\documentclass[10pt,twoside]{amundi_article}

\usepackage[table]{xcolor}
\usepackage{lmodern}
\usepackage{amssymb}
\usepackage{amsfonts}
\usepackage{amsmath}
\usepackage{amsbsy}
\usepackage{fancybox}
\usepackage{fancyhdr}
\usepackage{graphicx}
\usepackage[hyphens]{url}
\usepackage{arydshln}
\usepackage{multirow}
\usepackage{pdflscape}
\usepackage{tikz}
\usepackage{comment}
\usepackage{nicefrac}
\usepackage{dsfont}
\usepackage{algorithmic}
\usepackage{algorithm}
\usepackage{lipsum}

\newlength{\figurewidth}
\newlength{\figureheight}
\def\tableskip{\vskip 10pt plus 2pt minus 2pt\relax}
\def\figureskip{\vskip 10pt plus 2pt minus 2pt\relax}

\newtheorem{remark}{Remark}

\def\limfunc#1{\mathop{\rm #1}}
\def\func#1{\mathop{\rm #1}}

\DeclareMathAlphabet\mathbfcal{OMS}{cmsy}{b}{n}
\DeclareFontFamily{OT1}{pzc}{}
\DeclareFontShape{OT1}{pzc}{m}{it}{<-> s * [1.3] pzcmi7t}{}
\DeclareMathAlphabet{\mathpzc}{OT1}{pzc}{m}{it}

\newcommand{\cost}{\pmb{c}}
\newcommand{\turnover}{\boldsymbol{\tau}}

\usetikzlibrary{positioning}
\usetikzlibrary{patterns,arrows,decorations.pathreplacing}
\usetikzlibrary{backgrounds}

\newcommand{\TsV}{\hspace{5pt}}
\newcommand{\TsVII}{\hspace{7pt}}

\begin{document}

\setcounter{page}{1}

\title{\textbf{\color{amundi_blue}Machine Learning Optimization\\Algorithms \& Portfolio Allocation}%
\footnote{This survey article has been prepared for the book Machine Learning
and Asset Management edited by Emmanuel Jurczenko. We would like to thank
Mohammed El Mendili, Edmond Lezmi, Lina Mezghani, Jean-Charles Richard, Jules
Roche and Jiali Xu for their helpful comments.}}


\author{{\color{amundi_dark_blue} Sarah Perrin} \\
Artificial Intelligence \& Advanced \\
Visual Computing Master \\
Ecole Polytechnique \\
\texttt{sarah.perrin@polytechnique.edu} \and
{\color{amundi_dark_blue} Thierry Roncalli} \\
Quantitative Research \\
Amundi Asset Management \\
Paris \\
\texttt{thierry.roncalli@amundi.com}}

\date{\color{amundi_dark_blue}June 2019}

\maketitle

\begin{abstract}
Portfolio optimization emerged with the seminal paper of Markowitz (1952).
The original mean-variance framework is appealing because it is very
efficient from a computational point of view. However, it also has one
well-established failing since it can lead to portfolios that are not optimal
from a financial point of view (Michaud, 1989). Nevertheless, very few models
have succeeded in providing a real alternative solution to the Markowitz model.
The main reason lies in the fact that most academic portfolio optimization
models are intractable in real life although they present solid
theoretical properties. By intractable we mean that they can be implemented
for an investment universe with a small number of assets using a lot of
computational resources and skills, but they are unable to manage a universe
with dozens or hundreds of assets. However, the emergence and the
rapid development of robo-advisors means that we need to rethink portfolio optimization
and go beyond the traditional mean-variance optimization approach.\smallskip

Another industry and branch of science has faced similar issues concerning
large-scale optimization problems. Machine learning and applied statistics
have long been associated with linear and logistic regression models. Again,
the reason was the inability of optimization algorithms to solve
high-dimensional industrial problems. Nevertheless, the end of the 1990s
marked an important turning point with the development and the rediscovery of
several methods that have since produced impressive results. The goal of this
paper is to show how portfolio allocation can benefit from the development of
these large-scale optimization algorithms. Not all of these algorithms are
useful in our case, but four of them are essential when solving complex
portfolio optimization problems. These four algorithms are the coordinate
descent, the alternating direction method of multipliers, the proximal
gradient method and the Dykstra's algorithm.  This paper reviews them and
shows how they can be implemented in portfolio allocation.
\end{abstract}

\noindent \textbf{Keywords:} Portfolio allocation, mean-variance
optimization, risk budgeting optimization, quadratic programming,
coordinate descent, alternating direction method of multipliers,
proximal gradient method, Dykstra's algorithm.\medskip

\noindent \textbf{JEL classification:} C61, G11.

\section{Introduction}

The contribution of Harry Markowitz to economics is considerable.
The mean-variance optimization framework marks the beginning of
portfolio allocation in finance. In addition to the seminal paper of
1952, Harry Markowitz proposed an algorithm for solving quadratic
programming problems in 1956. At that time, very few people were aware of
this optimization framework. We can cite Mann (1943) and Martin
(1955), but it is widely accepted that Harry Markowitz is the
\textquotedblleft \textsl{father of quadratic
programming}\textquotedblright\ (Cottle and Infanger, 2010). This is
not the first time that economists are participating in the development of
mathematics\footnote{For example, Leonid Kantorovich made major
contributions to the success of linear programming.}, but this is
certainly the first time that mathematicians will explore a field of
research, whose main application during the first years of
research is exclusively an economic problem\footnote{If we consider
the first publications on quadratic programming, most of them were
published in \textit{Econometrica} or illustrated the Markowitz
problem (see \cite{qp-list-1}, \cite{qp-list-2}, \cite{qp-list-3},
\cite{qp-list-4}, \cite{qp-list-5}, \cite{qp-list-6},
\cite{qp-list-7}, \cite{qp-list-8} and
\cite{qp-list-9}).}.\smallskip

The success of mean-variance optimization (MVO) is due to the
appealing properties of the quadratic utility function, but it
should also be assessed in light of the success of quadratic
programming (QP). Because it is easy to solve QP problems and
because QP problems are available in mathematical software, solving
MVO problems is straightforward and does not require a specific
skill. This is why the mean-variance optimization is a universal
method which is used by all portfolio managers. However, this
approach has been widely criticized by academics and professionals.
Indeed, mean-variance optimization is very sensitive to input
parameters and produces corner solutions. Moreover, the concept of
mean-variance diversification is confused with the concept of hedging
(Bourgeron \textsl{et al.}, 2018). These different issues make the
practice of mean-variance optimization less attractive than the
theory (Michaud, 1989). In fact, solving MVO allocation problems
requires the right weight constraints to be specified in order to obtain
acceptable solutions. It follows that designing the constraints
is the most important component of mean-variance optimization. In this
case, MVO appears to be a trial-and-error process, not a
systematic solution.\smallskip

The success of the MVO framework is also explained by the fact that there are
very few competing portfolio allocation models that can be implemented from an
industrial point of view. There are generally two reasons for this. The first
one is that some models use input parameters that are difficult to estimate or
understand, making these models definitively unusable. The second reason is
that other models use a more complex objective function than the simple
quadratic utility function. In this case, the computational complexity makes
these models less attractive than the standard MVO model. Among these models,
some of them are based on the mean-variance objective function, but introduce
regularization penalty functions in order to improve the robustness of the
portfolio allocation. Again, these models have little chance of being used if
they cannot be cast into a QP problem. However, new optimization algorithms
have emerged for solving large-scale machine learning problems. The purpose of
this article is to present these new mathematical methods and show that they
can be easily applied to portfolio allocation in order to go beyond the MVO/QP
model.\smallskip

This survey article is based on several previous research papers
(\cite{roncalli1}, \cite{roncalli2}, \cite{roncalli3} and
\cite{roncalli4}) and extensively uses four leading references
(Beck, 2017; Boyd \textsl{et al.}, 2010; Combettes and Pesquet,
2011; Tibshirani, 2017). It is organized as follows. In section two,
we present the mean-variance approach and how it is related to the
QP framework. The third section is dedicated to large-scale
optimization algorithms that have been used in machine learning:
coordinate descent, alternating direction method of
multipliers, proximal gradient and Dykstra's algorithm.
Section four shows how these algorithms can be implemented in order to
solve portfolio optimization problems and build a more robust
asset allocation. Finally, section five offers some concluding
remarks.

\section{The quadratic programming world of portfolio optimization}

\subsection{Quadratic programming}

\subsubsection{Primal formulation}

A quadratic programming (QP) problem is an optimization problem with a
quadratic objective function and linear inequality constraints:
\begin{eqnarray}
x^{\star } &=&\arg \min_{x} \frac{1}{2}x^{\top }Qx-x^{\top }R  \label{eq:qp1} \\
&\text{s.t.}&Sx\leq T \notag
\end{eqnarray}%
where $x$ is a $n \times 1$ vector, $Q$ is a $n \times n$ matrix and $R$ is a
$n \times 1$ vector. We note that the system of constraints $Sx\leq T$ allows
us to specify linear equality constraints\footnote{This is equivalent to impose
that $Ax\geq B$ and $Ax\leq B$.} $Ax=B$ or box constraints $x^{-}\leq x\leq
x^{+}$. Most numerical packages then consider the following formulation:
\begin{eqnarray}
x^{\star } &=&\arg \min_{x} \frac{1}{2}x^{\top }Qx-x^{\top }R  \label{eq:qp2} \\
&\text{s.t.}&\left\{
\begin{array}{l}
Ax=B \\
Cx\leq D \\
x^{-}\leq x\leq x^{+}%
\end{array}%
\right. \notag
\end{eqnarray}
because the problem (\ref{eq:qp2}) is equivalent to the canonical problem
(\ref{eq:qp1}) with the following system of linear inequalities:
\begin{equation*}
\left[
\begin{array}{c}
-A{\TsV} \\
{\TsVII}A{\TsV} \\
{\TsVII}C{\TsV} \\
-I_n \\
{\TsVII}I_n%
\end{array}%
\right] x\leq \left[
\begin{array}{c}
-B{\TsV} \\
{\TsVII}B{\TsV} \\
{\TsVII}D{\TsV} \\
-x^{-} \\
{\TsVII}x^{+}%
\end{array}%
\right]
\end{equation*}%
If the space $\Omega $ defined by $Sx\leq T$ is non-empty and if $Q$ is a
symmetric positive definite matrix, the solution exists because the function
$f\left( x\right) =\dfrac{1}{2}x^{\top }Qx-x^{\top }R$ is convex. In the
general case where $Q$ is a square matrix, the solution may not exist.

\subsubsection{Dual formulation}

The Lagrange function is equal to:
\begin{equation*}
\mathcal{L}\left( x;\lambda \right) =\frac{1}{2}x^{\top }Qx-x^{\top
}R+\lambda ^{\top }\left( Sx-T\right)
\end{equation*}%
We deduce that the dual problem is defined by:
\begin{eqnarray*}
\lambda ^{\star } &=&\arg \max_{\lambda} \left\{ \inf_{x}\mathcal{L}\left( x;\lambda
\right) \right\}  \\
& \text{s.t. } & \lambda \geq 0
\end{eqnarray*}%
We note that $\partial _{x}\,\mathcal{L}\left( x;\lambda \right) =Qx-R+S^{\top
}\lambda $. The solution to the equation $\partial _{x}\,\mathcal{L}\left(
x;\lambda \right) =0$ is then $x=Q^{-1}\left( R-S^{\top }\lambda \right) $. We
finally obtain:
\begin{eqnarray*}
\inf_{x}\mathcal{L}\left( x;\lambda \right)  &=&\frac{1}{2}\left( R^{\top
}-\lambda ^{\top }S\right) Q^{-1}\left( R-S^{\top }\lambda \right) -\left(
R^{\top }-\lambda ^{\top }S\right) Q^{-1}R+ \\
&&\lambda ^{\top }\left( SQ^{-1}\left( R-S^{\top }\lambda \right) -T\right) \\
&=&\frac{1}{2}R^{\top }Q^{-1}R-\lambda ^{\top }SQ^{-1}R+\frac{1}{2}\lambda
^{\top }SQ^{-1}S^{\top }\lambda -R^{\top }Q^{-1}R+ \\
&&2\lambda ^{\top }SQ^{-1}R-\lambda ^{\top }SQ^{-1}S^{\top }\lambda -\lambda
^{\top }T \\
&=&-\frac{1}{2}\lambda ^{\top }SQ^{-1}S^{\top }\lambda +\lambda ^{\top
}\left( SQ^{-1}R-T\right) -\frac{1}{2}R^{\top }Q^{-1}R
\end{eqnarray*}%
We deduce that the dual program is another quadratic programming problem:%
\begin{eqnarray}
\lambda ^{\star } &=&\arg \min_{\lambda} \frac{1}{2}\lambda ^{\top }\bar{Q}\lambda
-\lambda ^{\top }\bar{R} \label{eq:qp-dual} \\
& \text{s.t.} & \lambda \geq 0 \notag
\end{eqnarray}%
where $\bar{Q}=SQ^{-1}S^{\top }$ and $\bar{R}=SQ^{-1}R-T$.

\begin{remark}
This duality property is very important for some machine learning methods. For
example, this is the case of support vector machines and kernel methods that
extensively use the duality for defining the solution (Cortes and Vapnik,
1995).
\end{remark}

\subsubsection{Numerical algorithms}

There is a substantial literature on the methods for solving quadratic
programming problems (Gould and Toint, 2000). The research begins in the 1950s
with different key contributions: Frank and Wolfe (1956), Markowitz (1956),
Beale (1959) and Wolfe (1959). Nowadays, QP problems are generally solved using
three approaches: active set methods, gradient projection methods and interior
point methods. All these algorithms are implemented in standard mathematical
programming languages (Matlab, Matematica, Python, Gauss, R, etc.). This
explains the success of QP problems since 2000s, because they can be easily and
rapidly solved.

\subsection{Mean-variance optimized portfolios}

The concept of portfolio allocation has a long history and dates back to the
seminal work of Markowitz (1952). In his paper, Markowitz defined precisely
what \textit{portfolio selection} means: \textquotedblleft\textsl{the investor
does (or should) consider expected return a desirable thing and variance of
return an undesirable thing}\textquotedblright. Indeed, Markowitz showed that
an efficient portfolio is the portfolio that maximizes the expected return for
a given level of risk (corresponding to the variance of portfolio return) or a
portfolio that minimizes the risk for a given level of expected return. Even if
this framework has been extended to many other allocation problems (index
sampling, turnover management, etc.), the mean-variance model remains the
optimization approach that is the most widely used in finance.

\subsubsection{The Markowitz framework}

We consider a universe of $n$ assets. Let $x=\left( x_{1},\ldots ,x_{n}\right)
$ be the vector of weights in the portfolio. We assume that the portfolio is
fully invested meaning that $\sum_{i=1}^{n}x_{i}=\mathbf{1}_n^{\top }x=1$. We
denote $\mathfrak{R}=\left( \mathfrak{R}_{1},\ldots ,\mathfrak{R}_{n}\right) $
as the vector of asset returns where $\mathfrak{R}_{i}$ is the return of asset
$i$. The return of the portfolio is then equal to $\mathfrak{R}\left( x\right)
=\sum_{i=1}^{n}x_{i}\mathfrak{R}_{i}=x^{\top }\mathfrak{R}$. Let $\mu
=\mathbb{E}\left[ \mathfrak{R}\right] $ and $\Sigma =\mathbb{E}\left[ \left(
\mathfrak{R}-\mu \right) \left( \mathfrak{R}-\mu \right) ^{\top }\right] $ be
the vector of expected returns and the covariance matrix of asset returns. The
expected return of the portfolio is equal to:
\begin{equation*}
\mu \left( x\right) =\mathbb{E}\left[ \mathfrak{R}\left( x\right) \right] =x^{\top }\mu
\end{equation*}%
whereas its variance is equal to:%
\begin{equation*}
\sigma ^{2}\left( x\right) =\mathbb{E}\left[ \left( \mathfrak{R}\left( x\right) -\mu
\left( x\right) \right) \left( \mathfrak{R}\left( x\right) -\mu \left( x\right) \right)
^{\top }\right] =x^{\top }\Sigma x
\end{equation*}%
Markowitz (1952) formulated the investor's financial problem as follows:
\begin{enumerate}
\item Maximizing the expected return of the portfolio under a volatility
constraint ($\sigma $-problem):%
\begin{equation}
\max \mu \left( x\right) \quad \text{s.t.}\quad \sigma \left( x\right) \leq
\sigma ^{\star }  \label{eq:markowitz1}
\end{equation}

\item Or minimizing the volatility of the portfolio under a return
constraint ($\mu $-problem):%
\begin{equation}
\min \sigma \left( x\right) \quad \text{s.t.}\quad \mu \left( x\right) \geq
\mu ^{\star }  \label{eq:markowitz2}
\end{equation}
\end{enumerate}
Markowitz's bright idea was to consider a quadratic utility function:
\begin{equation*}
\mathcal{U}\left( x\right) =x^{\top }\mu -\frac{\phi }{2}x^{\top }\Sigma x
\end{equation*}%
where $\phi \geq 0$ is the risk aversion. Since maximizing $\mathcal{U}\left(
x\right) $ is equivalent to minimizing $-\mathcal{U}\left( x\right) $, the
Markowitz problems (\ref{eq:markowitz1}) and (\ref{eq:markowitz2}) can be cast
into a QP problem\footnote{This transformation is called the QP trick.}:
\begin{eqnarray}
x^{\star }\left( \gamma \right)  &=&\arg \min_x \frac{1}{2}x^{\top }\Sigma
x-\gamma x^{\top }\mu   \label{eq:markowitz3} \\
&\text{s.t.}&\mathbf{1}_{n}^{\top }x=1  \notag
\end{eqnarray}%
where $\gamma =\phi ^{-1}$. Therefore, solving the $\mu $-problem or the
$\sigma $-problem is equivalent to finding the optimal value of $\gamma $ such
that $\mu \left( x^{\star }\left( \gamma \right) \right) =\mu ^{\star }$ or
$\sigma \left( x^{\star }\left( \gamma \right) \right) =\sigma ^{\star }$. We
know that the functions $\mu \left( x^{\star }\left( \gamma \right) \right) $
and $\sigma \left( x^{\star }\left( \gamma \right) \right) $ are increasing
with respect to $\gamma $ and are bounded. The optimal value of $\gamma $ can
then be easily computed using the bisection algorithm. It is obvious that a
large part of the success of the Markowitz framework lies on the QP trick.
Indeed, Problem (\ref{eq:markowitz3}) corresponds to the QP problem
(\ref{eq:qp2}) where $Q=\Sigma $, $R=\gamma \mu $, $A=\mathbf{1}_{n}^{\top }$
and $B=1$. Moreover, it is easy to include bounds on the weights, inequalities
between asset classes, etc.

\subsubsection{Solving complex MVO problems}

The previous framework can be extended to other portfolio allocation problems.
However, from a numerical point of view, the underlying idea is to always find
an equivalent QP formulation (Roncalli, 2013).

\paragraph{Portfolio optimization with a benchmark}

We now consider a benchmark $b$. We note $\mu \left( x\mid b\right) =\left(
x-b\right) ^{\top }\mu $ as the expected excess return and $\sigma \left( x\mid
b\right) =\sqrt{\left( x-b\right) ^{\top }\Sigma \left( x-b\right) }$ as the
tracking error volatility of Portfolio $x$ with respect to Benchmark $b$. The
objective function corresponds to a trade-off between minimizing the tracking
error volatility and maximizing the expected excess return (or the alpha):
\begin{equation*}
f\left( x\mid b\right) =\frac{1}{2}\sigma ^{2}\left( x\mid b\right) -\gamma
\mu \left( x\mid b\right)
\end{equation*}%
We can show that the equivalent QP problem is\footnote{See Appendix
\ref{appendix:qp-benchmark} on page \pageref{appendix:qp-benchmark}.}:
\begin{equation*}
x^{\star }\left( \gamma \right) =\arg \min_x \frac{1}{2}x^{\top }\Sigma
x-\gamma x^{\top }\tilde{\mu}
\end{equation*}%
where $\tilde{\mu}=\mu +\gamma ^{-1}\Sigma b$ is the regularized vector of
expected returns. Therefore, portfolio allocation with a benchmark can be
viewed as a regularization of the MVO problem and is solved using a QP
numerical algorithm.

\paragraph{Index sampling}

The goal of index sampling is to replicate an index portfolio with a smaller
number of assets than the index (or the benchmark) $b$. From a mathematical
point of view, index sampling could be written as follows:
\begin{eqnarray}
x^{\star } &=&\arg \min_x \frac{1}{2}\left( x-b\right) ^{\top }\Sigma \left(
x-b\right)  \label{eq:mvo2} \\
&\text{s.t.}&\left\{
\begin{array}{l}
\mathbf{1}_n^{\top }x=1 \\
x\geq \mathbf{0}_n \\
\sum_{i=1}^{n}\mathds{1}\left\{ x_{i}>0\right\} \leq n_{x}%
\end{array}%
\right. \notag
\end{eqnarray}%
The idea is to minimize the volatility of the tracking error such that the
number of stocks $n_{x}$ in the portfolio is smaller than the number of stocks
$n_{b}$ in the benchmark. For example, one would like to replicate the S\&P 500
index with only 50 stocks and not the entire 500 stocks that compose this
index. Professionals generally solve Problem (\ref{eq:mvo2}) with the following
heuristic algorithm:

\begin{enumerate}
\item We set $x_{\left( 0\right) }^{+}=\mathbf{1}_{n}$. At the iteration
    $k+1$, we solve the QP problem:
\begin{eqnarray*}
x^{\star } &=&\arg \min_x \frac{1}{2}\left(x-b\right)^{\top }\Sigma \left(x-b\right)  \\
&\text{s.t.}&\left\{
\begin{array}{l}
\mathbf{1}_n^{\top }x = 1 \\
\mathbf{0}_n\leq x \leq x_{\left( k\right) }^{+}%
\end{array}%
\right.
\end{eqnarray*}

\item We then update the upper bound $x_{\left( k\right) }^{+}$ of the QP
    problem by deleting the asset $i^{\star }$ with the lowest non-zero
    optimized weight\footnote{We have $i^{\star }=\left\{ \left. i:\arg \inf
    x_{i}^{\star }\right\vert x_{i}^{\star }>0\right\} $.}:
\begin{equation*}
x_{\left( k+1\right) ,i}^{+}\leftarrow x_{\left( k\right) ,i}^{+}
\quad \text{if }i\neq i^{\star }
\qquad \text{and}\qquad
x_{\left( k+1\right) ,i^{\star}}^{+}\leftarrow 0
\end{equation*}

\item We iterate the two steps until $\sum_{i=1}^{n}\mathds{1}\left\{
    x_{i}^{\star }>0\right\} = n_{x}$.
\end{enumerate}
The purpose of the heuristic algorithm is to delete one asset at each iteration
in order to obtain an invested portfolio, which is exactly composed of $n_{x}$
assets and has a low tracking error volatility. Again, we notice that solving
the index sampling problem is equivalent to solving $\left( n_{b}-n_{x}\right)
$ QP problems.

\paragraph{Turnover management}
\label{section:mvo-turnover}

If we note $\bar{x}$ as the current portfolio and $x$ as the new portfolio, the
turnover of Portfolio $x$ with respect to Portfolio $\bar{x}$ is the sum of
purchases and sales:%
\begin{equation*}
\turnover \left( x\mid \bar{x}\right) =\sum_{i=1}^{n}\left( x_{i}-\bar{x}%
_{i}\right) ^{+}+\sum_{i=1}^{n}\left( \bar{x}_{i}-x_{i}\right)
^{+}=\sum_{i=1}^{n}\left\vert x_{i}-\bar{x}_{i}\right\vert
\end{equation*}
Adding a turnover constraint in long-only MVO portfolios leads to the following
problem:
\begin{eqnarray*}
x^{\star } &=&\arg \min_x \frac{1}{2}x^{\top }\Sigma x-\gamma x^{\top }\mu  \\
&\text{s.t.}&\left\{
\begin{array}{l}
\sum_{i=1}^{n}x_{i}=1 \\
\sum_{i=1}^{n}\left\vert x_{i}-\bar{x}_{i}\right\vert \leq \turnover ^{+} \\
0\leq x_{i}\leq 1%
\end{array}%
\right.
\end{eqnarray*}%
where $\turnover ^{+}$ is the maximum turnover with respect to the current
portfolio $\bar{x}$. Scherer (2007) introduces the additional variables
$x_{i}^{-}$ and $x_{i}^{+}$ such that:
\begin{equation*}
x_{i} = \bar{x}_{i} + x_{i}^{+} - x_{i}^{-}
\end{equation*}%
with $x_{i}^{-}\geq 0$ indicates a negative weight change with respect to the
initial weight $\bar{x}_{i}$ and $x_{i}^{+}\geq 0$ indicates a positive weight
change. The expression of the turnover becomes:
\begin{equation*}
\sum_{i=1}^{n}\left\vert x_{i}- \bar{x}_{i}\right\vert
=\sum_{i=1}^{n}\left\vert x_{i}^{+}-x_{i}^{-}\right\vert
=\sum_{i=1}^{n}x_{i}^{+}+\sum_{i=1}^{n}x_{i}^{-}
\end{equation*}%
because one of the variables $x_{i}^{+}$ or $x_{i}^{-}$ is necessarily equal to
zero due to the minimization problem. The $\gamma $-problem of Markowitz
becomes:
\begin{eqnarray*}
x^{\star } &=&\arg \min_x \frac{1}{2}x^{\top }\Sigma x-\gamma x^{\top }\mu  \\
&\text{s.t.}&\left\{
\begin{array}{l}
\sum_{i=1}^{n}x_{i}=1 \\
x_{i} = \bar{x}_{i} + x_{i}^{+} - x_{i}^{-} \\
\sum_{i=1}^{n}x_{i}^{+}+\sum_{i=1}^{n}x_{i}^{-}\leq \turnover ^{+} \\
0\leq x_{i},x_{i}^{-},x_{i}^{+}\leq 1%
\end{array}%
\right.
\end{eqnarray*}%
We obtain an augmented QP problem of dimension $3n$ (see Appendix
\ref{appendix:augmented-qp-turnover} on page
\pageref{appendix:augmented-qp-turnover}).

\paragraph{Transaction costs}
\label{section:mvo-transaction-cost}

The previous analysis assumes that there is no transaction cost $\cost\left( x
\mid \bar{x}\right) $ when we rebalance the portfolio from the current
portfolio $\bar{x}$ to the new optimized portfolio $x$. If we note $c_{i}^{-}$
and $c_{i}^{+}$ as the bid and ask transaction costs, we have:
\begin{equation*}
\cost\left( x\mid \bar{x}\right) =\sum_{i=1}^{n} x_{i}^{-}c_{i}^{-} + \sum_{i=1}^{n}x_{i}^{+}c_{i}^{+}
\end{equation*}
The net expected return of Portfolio $x$ is then equal to $\mu \left( x\right)
-\cost\left( x \mid \bar{x}\right) $. It follows that the $\gamma $-problem of
Markowitz becomes\footnote{The equality constraint $\mathbf{1}_{n}^{\top }x=1$
becomes $\mathbf{1}_{n}^{\top }x + \cost\left( x\mid \bar{x}\right) =1$ because
the rebalancing process has to be financed.}:
\begin{eqnarray*}
x^{\star } &=&\arg \min_x \frac{1}{2}x^{\top }\Sigma x-\gamma \left( \sum_{i=1}^{n}
x_{i}\mu _{i}-\sum_{i=1}^{n} x_{i}^{-}c_{i}^{-}-\sum_{i=1}^{n} x_{i}^{+}c_{i}^{+}\right)  \\
&\text{s.t.}&\left\{
\begin{array}{l}
\sum_{i=1}^{n} x_{i} + \sum_{i=1}^{n} x_{i}^{-}c_{i}^{-} + \sum_{i=1}^{n} x_{i}^{+}c_{i}^{+}=1 \\
x_{i} = \bar{x}_{i} + x_{i}^{+} - x_{i}^{-} \\
0\leq x_{i},x_{i}^{-},x_{i}^{+}\leq 1%
\end{array}%
\right.
\end{eqnarray*}%
Once again, we obtain a QP problem (see Appendix
\ref{appendix:augmented-qp-transaction-cost} on page
\pageref{appendix:augmented-qp-transaction-cost}).

\subsection{Issues with QP optimization}

The concurrent model of the Markowitz framework is the risk budgeting approach
(Qian, 2005; Maillard \textsl{et al.}, 2010; Roncalli, 2013). The goal is to
define a convex risk measure $\mathcal{R}\left( x\right) $ and to allocate the
risk according to some specified risk budgets $\mathcal{RB}=\left(
\mathcal{RB}_{1},\ldots ,\mathcal{RB}_{n}\right) $ where $\mathcal{RB}_{i}>0$.
This approach exploits the Euler decomposition property of the risk measure:
\begin{equation*}
\mathcal{R}\left( x\right) =\sum_{i=1}^{n}x_{i}\frac{\partial \,\mathcal{R}%
\left( x\right) }{\partial \,x_{i}}
\end{equation*}%
By noting $\mathcal{RC}_{i}\left( x\right) =x_{i}\cdot \partial
_{x_{i}}\,\mathcal{R}\left( x\right) $ as the risk contribution of Asset $i$
with respect to portfolio $x$, the risk budgeting (RB) portfolio is defined by
the following set of equations:
\begin{equation*}
x^{\star }=\left\{ x\in \left[ 0,1\right] ^{n}:\mathcal{RC}_{i}\left(
x\right) =\mathcal{RB}_{i}\right\}
\end{equation*}%
Roncalli (2013) showed that it is equivalent to solving the following
non-linear optimization problem\footnote{In fact, the solution $x^{\star }$
must be rescaled after the optimization step.}:
\begin{eqnarray}
x^{\star } &=&\arg \min_x \mathcal{R}\left( x\right) -\lambda \sum_{i=1}^{n}%
\mathcal{RB}_{i}\cdot \ln x_{i} \label{eq:rb1} \\
\text{} &\text{s.t.}&x_{i}>0 \notag
\end{eqnarray}%
where $\lambda >0$ is an arbitrary positive constant. Generally, the most
frequently used risk measures are the volatility risk measure (Maillard
\textsl{et al.}, 2010):
\begin{equation*}
\mathcal{R}\left( x\right) =\sqrt{x^{\top }\Sigma x}
\end{equation*}%
and the standard deviation-based risk measure (Roncalli, 2015):%
\begin{equation*}
\mathcal{R}\left( x\right) =-x^{\top }\left( \mu -r\right) +\xi \sqrt{%
x^{\top }\Sigma x}
\end{equation*}%
where $r$ is the risk-free rate and $\xi $ is a positive scalar. In particular,
this last one encompasses the Gaussian value-at-risk --- $\xi =\Phi ^{-1}\left(
\alpha \right) $ --- and the Gaussian expected shortfall
--- $\xi =\left( 1-\alpha \right) ^{-1}\phi \left( \Phi ^{-1}\left( \alpha
\right) \right) $.\smallskip

The risk budgeting approach has displaced the MVO approach in many fields of
asset management, in particular in the case of factor investing and alternative
risk premia. Nevertheless, we notice that Problem (\ref{eq:rb1}) is a not a
quadratic programming problem, but a logarithmic barrier problem. Therefore,
the risk budgeting framework opens a new world of portfolio optimization that
is not necessarily QP! That is all the more true since MVO portfolios face
robustness issues (Bourgeron \textsl{et al.}, 2018). Regularization of
portfolio allocation has then become the industry standard. Indeed, it is
frequent to add a $\boldsymbol{\ell}_{1}$-norm or $\boldsymbol{\ell}_{2}$-norm
penalty functions to the MVO objective function. This type of penalty is,
however, tractable in a quadratic programming setting. With the development of
robo-advisors, non-linear penalty functions have emerged, in particular the
logarithmic barrier penalty function. And these regularization techniques
result in a non-quadratic programming world of portfolio
optimization.\smallskip

The success of this non-QP financial world will depend on how quickly and
easily these complex optimization problems can be solved. Griveau-Billon
\textsl{at al.} (2013), Bourgeron \textsl{et al.} (2018), and Richard and
Roncalli (2019) have already proposed numerical algorithms that are doing the
work in some special cases. The next section reviews the candidate algorithms
that may compete with QP numerical algorithms.

\section{Machine learning optimization algorithms}

The machine learning industry has experienced a similar trajectory to portfolio
optimization. Before the 1990s, statistical learning focused mainly on models
that were easy to solve from a numerical point of view. For instance, the
linear (and the ridge) regression has an analytical solution, we can solve
logistic regression with the Newton-Raphson algorithm whereas supervised and
unsupervised classification models\footnote{e.g. principal component analysis
(PCA), linear/quadratic discriminant analysis (LDA/QDA), Fisher classification
method, etc.} consist in performing a singular value decomposition or a
generalized eigenvalue decomposition. The 1990s saw the emergence of three
models that have deeply changed the machine learning approach: neural networks,
support vector machines and lasso regression.\smallskip

Neural networks have been extensively studied since the seminal work of
Rosenblatt (1958). However, the first industrial application dates back to the
publication of LeCun \textsl{et al.} (1989) on handwritten zip code
recognition. At the beginning of 1990s, a fresh craze then emerged with the
writing of many handbooks that were appropriate for students, the most popular
of which was Bishop (1995). With neural networks, two main issues arise
concerning calibration: the large number of parameters to estimate and the
absence of a global maximum. The traditional numerical optimization
algorithms\footnote{For example, we can cite the quasi-Newton BFGS
(Broyden-Fletcher-Goldfarb-Shanno) and DFP (Davidon-Fletcher-Powell) methods,
and the Fletcher-Reeves and Polak-Ribiere conjugate gradient methods.} that
were popular in the 1980s cannot be applied to neural networks. New
optimization approaches are then proposed. First, researchers have considered
more complex learning rules than the steepest descent (Jacobs, 1988), for
example the momentum method of Polyak (1964) or the Nesterov accelerated
gradient approach (Nesterov, 1983). Second, the descent method is generally not
performed on the full sample of observations, but on a subset of observations
that changes at each iteration. This is the underlying idea behind batch
gradient descent (BGD), stochastic descent gradient (SGD) and mini-batch
gradient descent (MGD). We notice that adaptive learning methods and batch
optimization techniques have marked the revival of the gradient descend
method.\smallskip

The development of support vector machines is another important step in the
development of machine learning techniques. Like neural networks, they can be
seen as an extension of the perceptron. However, they present nice theoretical
properties and a strong geometrical framework. Once SVMs have been first
developed for linear classification, they have been extended for non-linear
classification and regression. A support vector machine consists in separating
hyperplanes and finding the optimal separation by maximizing the margin. The
original problem called the hard margin classification can be formulated as a
quadratic programming problem. However, the dual problem, which is also a QP
problem, is generally preferred to the primal problem for solving SVM
classification, because of the sparse property of the solution (Cortes and
Vapnik, 1995). Over the years, the original hard margin classification has been
extended to other problems: soft margin classification with binary hinge loss,
soft margin classification with squared hinge loss, least squares SVM
regression, $\varepsilon$-SVM regression, kernel machines (Vapnik, 1998). All
these statistical problems share the same calibration framework. The primal
problem can be cast into a QP problem, implying that the corresponding dual
problem is also a QP problem. Again, we notice that the success and the
prominence of statistical methods are related to the efficiency of the
optimization algorithms, and it is obvious that support vector machines have
substantially benefited from the QP formulation. From an industrial point of
view, support vector machines present however some limitations. Indeed, if the
dimension of the primal QP problem is the number $p$ of features (or
parameters), the dimension of the dual QP problem is the number $n$ of
observations. It is becoming absolutely impossible to solve the dual problem
when the number of observations is larger than $100\,000$ and sometimes as high
as several millions. This implies that new algorithms that are more appropriate
for large-scale optimization problems need to be developed.\smallskip

Lasso regression is the third disruptive approach that put machine learning in
the spotlight in the 1990s. Like the ridge regression, lasso regression is a
regularized linear regression where the $\boldsymbol{\ell}_2$-norm penalty is
replaced by the $\boldsymbol{\ell}_1$-norm penalty (Tibshirani, 1996). Since
the $\boldsymbol{\ell}_1$ regularization forces the solution to be sparse, it
has been first largely used for variable selection, and then for pattern
recognition and robust estimation of linear models. For finding the lasso
solution, the technique of augmented QP problems is widely used since it is
easy to implement. The extension of the lasso-ridge regularization to the other
$\boldsymbol{\ell}_p$ norms is straightforward, but these approaches have never
been popular. The main reason is that existing numerical algorithms are not
sufficient to make these models tractable.\smallskip

Therefore, the success of a quantitative model may be explained by two
conditions. First, the model must be obviously appealing. Second, the model
must be solved by an efficient numerical algorithm that is easy to implement or
available in mathematical programming software. As shown previously, quadratic
programming and gradient descent methods have been key for many statistical and
financial models. In what follows, we consider four algorithms and techniques
that have been popularized by their use in machine learning: coordinate
descent, alternating direction method of multipliers, proximal operators and
Dykstra's algorithm. In particular, we illustrate how they can be used for
solving complex optimization problems.

\subsection{Coordinate descent}

\subsubsection{Definition}

We consider the following unconstrained minimization problem:%
\begin{equation}
x^{\star }=\arg \min_{x} f\left( x\right)   \label{eq:ccd1}
\end{equation}%
where $x\in \mathbb{R}^{n}$ and $f\left( x\right) $ is a continuous, smooth and
convex function. A popular method to find the solution $x^{\star }$ is to
consider the descent algorithm, which is defined by the following rule:
\begin{equation*}
x^{\left( k+1\right) }=x^{\left( k\right) }+\Delta x^{\left( k\right)
}=x^{\left( k\right) }-\eta D^{\left( k\right) }
\end{equation*}%
where $x^{\left( k\right) }$ is the approximated solution of Problem (\ref{eq:ccd1})
at the $k^{\mathrm{th}}$ Iteration, $\eta >0$ is a scalar that
determines the step size and $D^{\left( k\right) }$ is the direction. We notice
that the current solution $x^{\left( k\right) }$ is updated by going
in the opposite direction to $D^{\left( k\right) }$ in order to obtain $%
x^{\left( k+1\right) }$. In the case of the gradient descent, the direction is
equal to the gradient vector of $f\left( x\right) $ at the current point:
$D^{\left( k\right) }=\nabla f\left( x^{\left( k\right) }\right) $. Coordinate
descent (CD) is a variant of the gradient descent and minimizes
the function along one coordinate at each step:%
\begin{equation*}
x_{i}^{\left( k+1\right) }=x_{i}^{\left( k\right) }+\Delta x_{i}^{\left(
k\right) }=x_{i}^{\left( k\right) }-\eta D_{i}^{\left( k\right) }
\end{equation*}%
where $D_{i}^{\left( k\right) }=\nabla _{i}f\left( x^{\left( k\right) }\right)
$ is the $i^{\mathrm{th}}$ element of the gradient vector. At each iteration, a
coordinate $i$ is then chosen via a certain rule, while the other coordinates
are assumed to be fixed. Coordinate descent is an appealing algorithm, because
it transforms a vector-valued problem into a scalar-valued problem that is
easier to implement. Algorithm (\ref{alg:ccd1}) summarizes the CD algorithm.
The convergence criterion can be a predefined number of iterations or an error
rule between two iterations. The step size $\eta$ can be either a given
parameter or computed with a line search, implying that the parameter
$\eta^{\left(k\right)}$ changes at each iteration.

\begin{algorithm}[tbph]
\begin{algorithmic}
\STATE The goal is to find the solution $x^{\star }=\arg \min f\left( x\right)$
\STATE We initialize the vector $x^{\left(0\right)}$ and we note $\eta$ the step size
\STATE Set $k \leftarrow 0$
\REPEAT
    \STATE Choose a coordinate $i \in \left\{1,n\right\}$
    \STATE $x_{i}^{\left( k+1\right) }\leftarrow x_{i}^{\left( k\right) }-\eta \nabla_{i}f\left( x^{\left( k\right) }\right) $
    \STATE $x_{j}^{\left( k+1\right) }\leftarrow x_{j}^{\left( k\right) } \quad \text{if } j \neq i$
    \STATE $k \leftarrow k + 1 $
\UNTIL{convergence}
\RETURN $x^{\star } \leftarrow x^{\left(k\right)}$
\end{algorithmic}
\caption{Coordinate descent algorithm (gradient formulation)}
\label{alg:ccd1}
\end{algorithm}

Another formulation of the coordinate descent method is given in Algorithm
(\ref{alg:ccd2}). The underlying idea is to replace the descent approximation
by the exact problem. Indeed, the objective of the descend step is to minimize
the scalar-valued problem:
\begin{equation}
x_{i}^{\star} =\arg \min_{\varkappa} f\left( x_{1}^{\left( k\right)},\ldots
,x_{i-1}^{\left( k\right) },\varkappa, x_{i+1}^{\left( k\right) },\ldots,x_{n}^{\left( k\right) }\right)
\label{eq:ccd2}
\end{equation}

\begin{algorithm}[tbph]
\begin{algorithmic}
\STATE The goal is to find the solution $x^{\star }=\arg \min f\left( x\right)$
\STATE We initialize the vector $x^{\left(0\right)}$
\STATE Set $k \leftarrow 0$
\REPEAT
    \STATE Choose a coordinate $i \in \left\{1,n\right\}$
    \STATE $x_{i}^{\left( k+1\right) }=\arg \min_{\varkappa}f\left( x_{1}^{\left( k\right)},\ldots ,x_{i-1}^{\left( k\right) },\varkappa,
    x_{i+1}^{\left( k\right) },\ldots,x_{n}^{\left( k\right) }\right)$
    \STATE $x_{j}^{\left( k+1\right) }\leftarrow x_{j}^{\left( k\right) } \quad \text{if } j \neq i$
    \STATE $k \leftarrow k + 1 $
\UNTIL{convergence}
\RETURN $x^{\star } \leftarrow x^{\left(k\right)}$
\end{algorithmic}
\caption{Coordinate descent algorithm (exact formulation)}
\label{alg:ccd2}
\end{algorithm}

Coordinate descent is efficient in large-scale optimization problems, in
particular when there is a solution to the scalar-valued problem
(\ref{eq:ccd2}). Furthermore, convergence is guaranteed when $f\left(x\right)$
is convex and differentiable (Luo and Tseng, 1992; Luo and Tseng, 1993).

\begin{remark}
Coordinate descent methods have been introduced in several handbooks on
numerical optimization in the 1980s and 1990s (Wright, 1985). However, the most
important step is the contribution of Tseng (2001), who studied the
block-coordinate descent method and extended CD algorithms in the case of a
non-differentiable and non-convex function $f\left( x\right) $.
\end{remark}

\subsubsection{Cyclic or random coordinates?}

There are several options for choosing the coordinate of the $k^{\mathrm{th}} $
iteration. A natural choice could be to choose the coordinate which minimizes
the function:
\begin{equation*}
i^{\star }=\arg \inf \left\{ f_{i}^{\star }:i\in \left\{ 1,n\right\}
,f_{i}^{\star }=\min_{\varkappa}f\left( \left( 1-e_{i}\right) x^{\left( k\right)
}+e_{i}\varkappa\right) \right\}
\end{equation*}%
However, it is obvious that choosing the optimal coordinate $i^{\star }$ would
require the gradient along each coordinate to be calculated. This causes the
coordinate descent to be no longer efficient, since a classic gradient descent
would then be of equivalent cost at each iteration and would converge faster
because it requires fewer iterations.\smallskip

The simplest way to implement the CD algorithm is to consider cyclic
coordinates, meaning that we cyclically iterate through the coordinates (Tseng,
2001):
\begin{equation*}
i=k\func{mod}n
\end{equation*}%
This ensures that all the coordinates are selected during one cycle $\left\{
k-n+1,\ldots, k\right\} $ in the same order. This approach, called cyclical
coordinate descent (CCD), is the most popular and used method, even if it is
difficult to estimate the rate of convergence.\smallskip

The second way is to consider random coordinates. Let $\pi_i$ be the
probability of choosing the coordinate $i$ at the iteration $k$. The simplest
approach is to consider uniform probabilities: $\pi_i =1/n $. A better approach
consists in pre-specifying probabilities according to the Lipschitz
constants\footnote{Nesterov (2012) assumes that $f\left( x\right) $ is convex,
differentiable and Lipschitz-smooth for each coordinate:
\begin{equation*}
\left\Vert \nabla _{i}f\left( x+e_{i}h\right) -\nabla _{i}f\left(
x\right) \right\Vert \leq \mathfrak{L}_{i}\left\Vert h\right\Vert
\end{equation*}%
where $h\in \mathbb{R}$.}:%
\begin{equation}
\pi_i =\frac{\mathfrak{L}_{i}^{\alpha }}{\sum_{j=1}^{n}\mathfrak{L%
}_{j}^{\alpha }}  \label{eq:rcd1}
\end{equation}%
Nesterov (2012) considers three schemes: $\alpha =0$, $\alpha =1$ and $\alpha
=\infty $ --- in this last case, we have $i=\arg \max \left\{
\mathfrak{L}_{1},\ldots ,\mathfrak{L}_{n}\right\} $. From a theoretical point
of view, the random coordinate descent (RCD) method based on the probability
distribution (\ref{eq:rcd1}) leads to a faster convergence, since coordinates
that have a large Lipschitz constant $\mathfrak{L}_{i}$ are more likely to be
chosen. However, it requires additional calculus to compute the Lipschitz
constants and CCD is often preferred from a practical point of view. In what
follows, we only use the CCD algorithm described below. In Algorithm
(\ref{alg:ccd3}), the variable $k$ represents the number of cycles whereas the
number of iterations is equal to $k \cdot n$. For the coordinate $i$, the lower
coordinates $j < i$ correspond to the current cycle $k+1$ while the upper
coordinates $j > i$ correspond to the previous cycle $k$.

\begin{algorithm}[tbph]
\begin{algorithmic}
\STATE The goal is to find the solution $x^{\star }=\arg \min f\left( x\right)$
\STATE We initialize the vector $x^{\left(0\right)}$
\STATE Set $k \leftarrow 0$
\REPEAT
    \FOR {$i=1:n$}
        \STATE $x_{i}^{\left( k+1\right) }=\arg \min_{\varkappa}f\left( x_{1}^{\left( k+1\right)},\ldots ,x_{i-1}^{\left( k+1\right) },\varkappa,
    x_{i+1}^{\left( k\right) },\ldots,x_{n}^{\left( k\right) }\right)$
    \ENDFOR
    \STATE $k \leftarrow k + 1 $
\UNTIL{convergence}
\RETURN $x^{\star } \leftarrow x^{\left(k\right)}$
\end{algorithmic}
\caption{Cyclical coordinate descent algorithm}
\label{alg:ccd3}
\end{algorithm}

\subsubsection{Application to the $\lambda$-problem of the lasso regression}

We consider the linear regression:
\begin{equation}
Y=X\beta +\varepsilon   \label{eq:ccd-ols1}
\end{equation}%
where $Y$ is the $n\times 1$ vector, $X$ is the $n\times p$ design matrix, $\beta $
is the $p\times 1$ vector of coefficients and $\varepsilon $ is the $n\times 1$ vector of residuals. In this
model, $n$ is the number of observations and $p$ is the number of parameters
(or the number of explanatory variables). The objective of the ordinary least
squares is to minimize the residual sum of squares:
\begin{equation*}
\hat{\beta}=\arg \min_{\beta} \frac{1}{2}\limfunc{RSS}\left( \beta \right)
\end{equation*}%
where $\limfunc{RSS}\left( \beta \right) =\sum_{i=1}^{n}\varepsilon _{i}^{2}$%
. Since we have:%
\begin{equation*}
\limfunc{RSS}\left( \beta \right) =\left( Y-X\beta \right) ^{\top }\left(
Y-X\beta \right)
\end{equation*}%
we obtain:%
\begin{equation*}
\frac{\partial \,f\left( \beta \right) }{\partial \,\beta _{j}}=-x_{j}^{\top
}\left( Y-X\beta \right)
\end{equation*}%
where $x_{j}$ is the $n\times 1$ design vector corresponding to the
$j^{\mathrm{th}}$ explanatory variable. Because we can write:
\begin{equation*}
X\beta =X_{\left( -j\right) }\beta _{\left( -j\right) }+x_{j}\beta _{j}
\end{equation*}%
where $X_{\left( -j\right) }$ and $\beta _{\left( -j\right) }$ are the design
matrix and the beta vector by excluding the $j^{\mathrm{th}}$ explanatory
variable, it follows that:
\begin{eqnarray*}
\frac{\partial \,f\left( \beta \right) }{\partial \,\beta _{j}}
&=&x_{j}^{\top }\left( X_{\left( -j\right) }\beta _{\left( -j\right)
}+x_{j}\beta _{j}-Y\right)  \\
&=&x_{j}^{\top }X_{\left( -j\right) }\beta _{\left( -j\right) }+x_{j}^{\top
}x_{j}\beta _{j}-x_{j}^{\top }Y
\end{eqnarray*}%
At the optimum, we have $\partial _{\beta _{j}}\,f\left( \beta \right) =0$ or:
\begin{equation}
\beta _{j}=\frac{x_{j}^{\top }\left( Y-X_{\left( -j\right) }\beta _{\left(
-j\right) }\right) }{x_{j}^{\top }x_{j}}  \label{eq:ccd-ols2}
\end{equation}%
The implementation of the coordinate descent algorithm is straightforward. It
suffices to iterate Equation (\ref{eq:ccd-ols2}) through the
coordinates.\smallskip

The lasso regression problem is a variant of the OLS regression by adding a
$\boldsymbol{\ell}_{1}$-norm regularization (Tibshirani, 1996):
\begin{equation}
\hat{\beta}\left( \lambda \right) =\arg \min_{\beta} \frac{1}{2}\left( Y-X\beta
\right) ^{\top }\left( Y-X\beta \right) +\lambda \left\Vert \beta
\right\Vert_1   \label{eq:ccd-lasso1}
\end{equation}%
In this formulation, the residual sum of squares of the linear regression is
penalized by a term that will force a sparse selection of the coordinates.
Since the objective function is the sum of two convex norms, the convergence is
guaranteed for the lasso problem. Because $\left\Vert \beta \right\Vert_1
=\sum_{j=1}^{n}\left\vert \beta _{j}\right\vert $, the first order condition
becomes:
\begin{eqnarray*}
0 &=&\nabla _{i}f\left( \beta \right)  \\
&=&x_{j}^{\top }x_{j}\beta _{j}-x_{j}^{\top }\left( Y-X_{\left( -j\right)
}\beta _{\left( -j\right) }\right) +\lambda \partial \,\left\vert \beta
_{j}\right\vert
\end{eqnarray*}%
In Appendix \ref{appendix:soft-thresholding}, we show that the solution is
given by:
\begin{equation}
\beta _{j}=\frac{\mathcal{S}\left( x_{j}^{\top }\left(
Y-X_{\left( -j\right) }\beta _{\left( -j\right) }\right); \lambda \right) }{%
x_{j}^{\top }x_{j}}  \label{eq:ccd-lasso2}
\end{equation}%
where $\mathcal{S}\left(v; \lambda\right)$ is the soft-thresholding operator:
\begin{equation*}
\mathcal{S}\left( v; \lambda\right) =\limfunc{sign}\left( v\right) \cdot
\left( \left\vert v\right\vert -\lambda \right) _{+}
\end{equation*}%
It follows that the lasso CD algorithm is a variation of the linear regression
CD algorithm by applying the soft-threshold operator to the residuals
$x_{j}^{\top }\left( Y-X_{\left( -j\right) }\beta _{\left( -j\right) }\right) $
at each iteration.\smallskip

\begin{figure}[tbh]
\centering
\caption{CCD algorithm applied to the lasso optimization problem}
\label{fig:lasso2}
\figureskip
\includegraphics[width = \figurewidth, height = \figureheight]{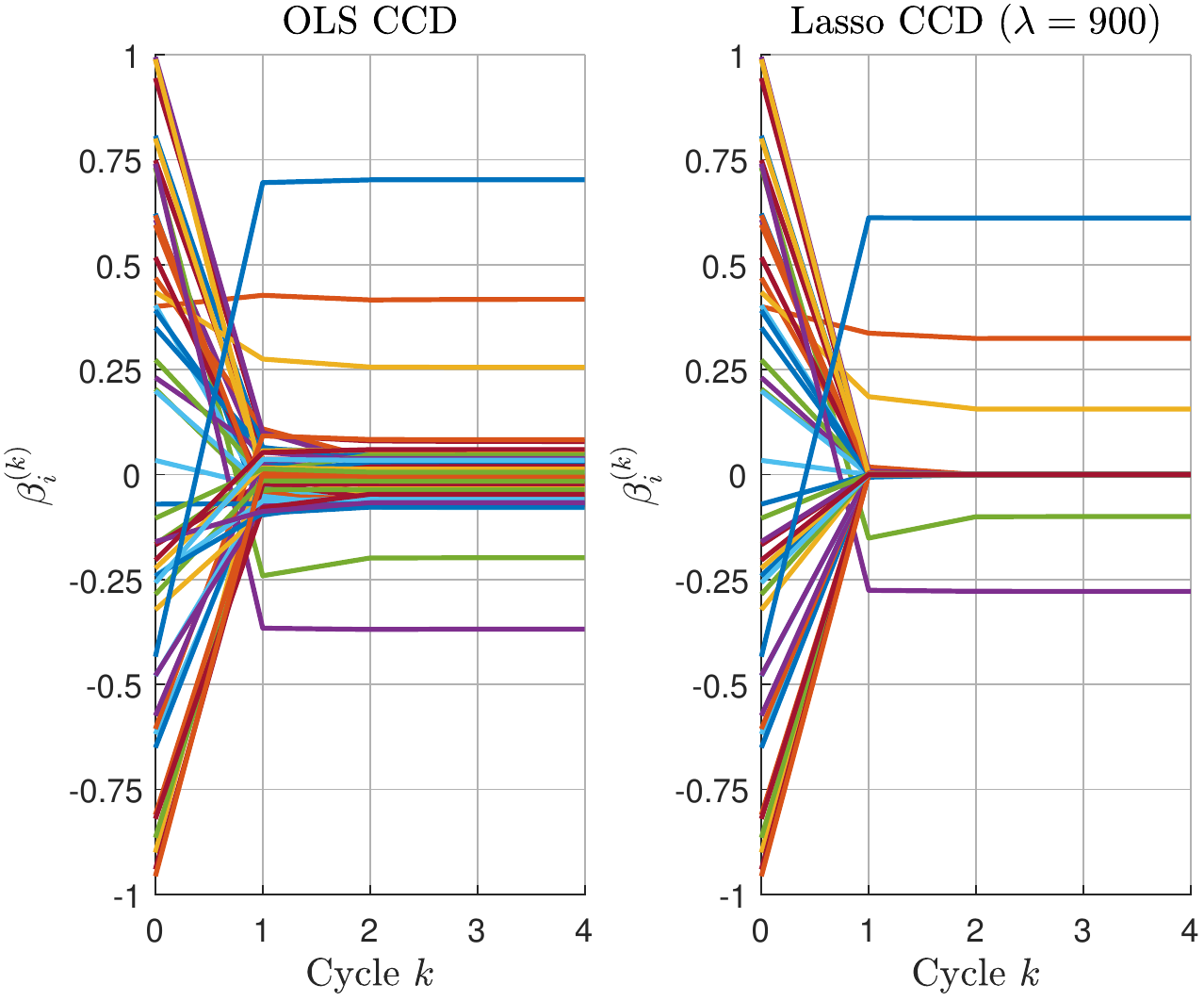}
\end{figure}

Let us consider an experiment with $n = 10\,0000$ and
$p=50$\label{example:ccd-lasso}. The design matrix $X$ is built using the
uniform distribution while the residuals are simulated using a Gaussian
distribution and a standard deviation of $20\%$. The beta coefficients are
distributed uniformly between $-3$ and $+3$ except four coefficients that take
a larger value. We then standardize the data of X and Y because the practice of
the lasso regression is to consider comparable beta coefficients. By
considering uniform numbers between $-1$ and $+1$ for initializing the
coordinates, results of the CCD algorithm are given in Figure \ref{fig:lasso2}.
We notice that the CCD algorithm converges quickly after three complete cycles.
In the case of a large-scale problem when $p \gg 1\,000$, it has been shown
that CCD may be faster for the lasso regression than for the OLS regression
because of the soft-thresholding operator. Indeed, we can initialize the
algorithm with the null vector $\mathbf{0}_p$. If $\lambda$ is large, a lot of
optimal coordinates are equal to zero and a few cycles are needed to find the
optimal values of non-zero coefficients.

\subsubsection{Application to the box-constrained QP problem}

Coordinate descent can also be applied to the box-constrained QP problem:
\begin{equation}
x^{\star }=\arg \min_x \frac{1}{2}x^{\top }Qx-x^{\top }R \quad  \text{s.t.} \quad
x^{-}\leq x\leq x^{+}  \label{eq:ccd-qp1}
\end{equation}%
In Appendix \ref{appendix:qp-ccd} on page \pageref{appendix:qp-ccd}, we show
that the coordinate update of the CCD algorithm is equal to:
\begin{equation*}
x_{i}^{\left( k+1\right) }=\mathcal{T}\left( \frac{R_{i}-\dfrac{1}{2}%
\sum_{j<i}x_{j}^{\left( k+1\right) }\left( Q_{i,j}+Q_{j,i}\right) -\dfrac{1}{2%
}\sum_{j>i}x_{j}^{\left( k\right) }\left( Q_{i,j}+Q_{j,i}\right) }{Q_{i,i}}%
;x_{i}^{-},x_{i}^{+}\right)
\end{equation*}%
where $\mathcal{T}\left( v;x^{-},x^{+}\right) $ is the truncation operator:%
\begin{eqnarray}
\mathcal{T}\left( v;x^{-},x^{+}\right)  &=&v\odot \mathds{1}\left\{
x^{-}<v<x^{+}\right\} +  \label{eq:truncation} \\
&&x^{-}\odot \mathds{1}\left\{ v\leq x^{-}\right\} +  \notag \\
&&x^{+}\odot \mathds{1}\left\{ v\geq x^{+}\right\}   \notag
\end{eqnarray}
Generally, we assume that $Q$ is a symmetric matrix, implying that the CCD
update reduces to:%
\begin{equation*}
x_{i}^{\left( k+1\right) }=\mathcal{T}\left( \frac{R_{i}-\sum_{j<i}x_{j}^{%
\left( k+1\right) }Q_{i,j}-\sum_{j>i}x_{j}^{\left( k\right) }Q_{i,j}}{Q_{i,i}%
};x_{i}^{-},x_{i}^{+}\right)
\end{equation*}

\begin{remark}
CCD can be applied to Problem (\ref{eq:ccd-qp1}) because the box
constraint $x^{-}\leq x\leq x^{+}$ is pointwise\footnote{%
See the discussion on page \pageref{section:proximal-ccd-pointwise}.}.
\end{remark}

\begin{figure}[tbph]
\centering
\caption{CCD algorithm applied to the box-constrained QP problem}
\label{fig:qp_box2}
\figureskip
\includegraphics[width = \figurewidth, height = \figureheight]{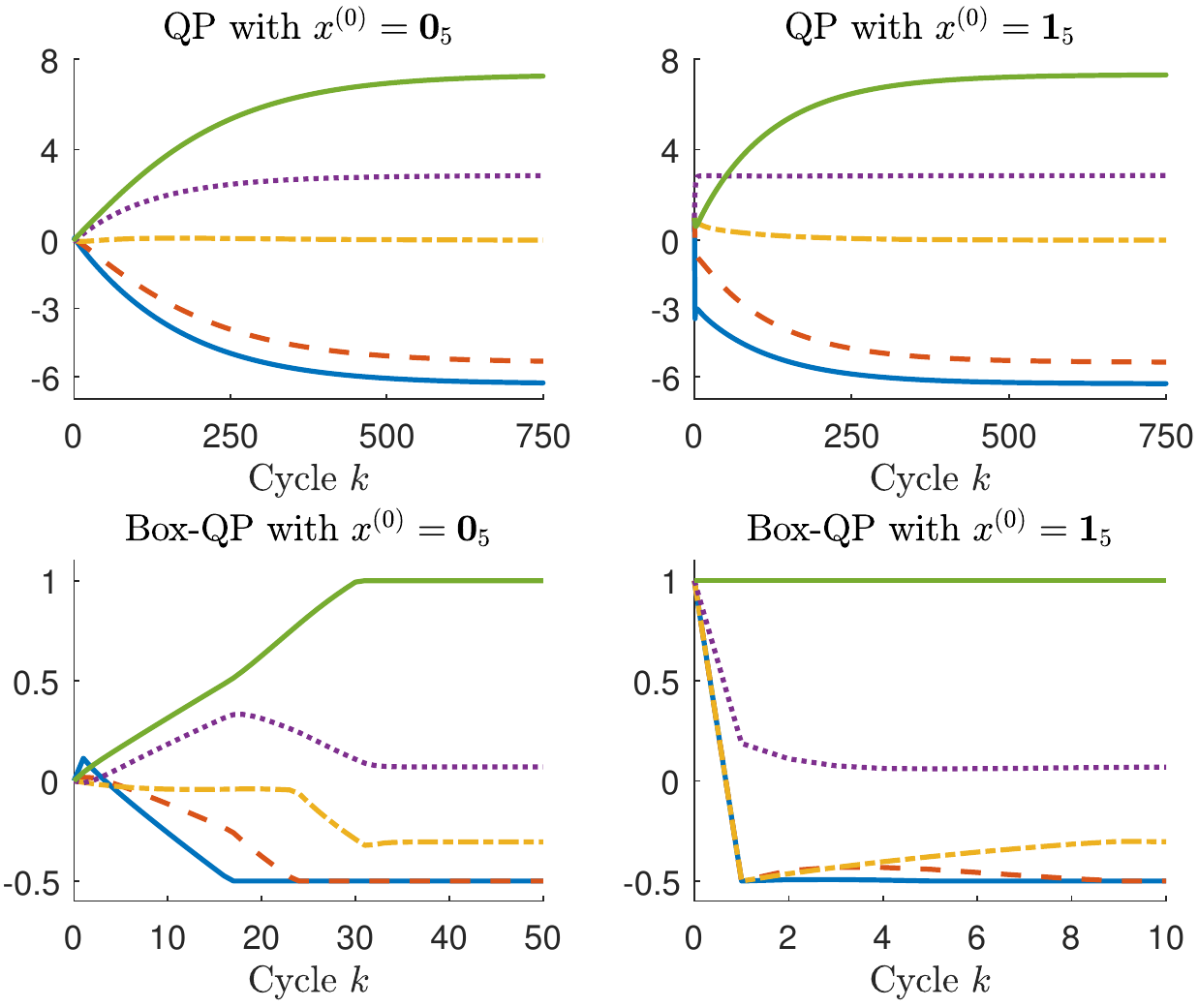}
\end{figure}

We consider the following example:%
\begin{equation*}
Q=\left(
\begin{array}{rrrrr}
5.76 & 5.11 & 3.47 & 5.13 &  6.82 \\
5.11 & 7.98 & 5.38 & 4.30 &  8.70 \\
3.47 & 5.38 & 4.01 & 2.83 &  5.91 \\
5.13 & 4.30 & 2.83 & 4.70 &  5.84 \\
6.82 & 8.70 & 5.91 & 5.84 & 10.18
\end{array}%
\right) \quad \text{and} \quad R=\left(
\begin{array}{r}
0.65 \\
0.72 \\
0.46 \\
0.59 \\
1.26
\end{array}%
\right)
\end{equation*}
In Figure \ref{fig:qp_box2}, we have reported the solution obtained with the
CCD algorithm. The top panels correspond to the QP problem without any
constraints, whereas the bottom panel corresponds to the QP problem with the
box constraint $-0.5\leq x_{i}\leq 1$. We notice that we need more than $500$
cycles for the convergence of the CCD algorithm in the case of the
unconstrained QP problem, whereas CCD finds the solution of the constrained QP
problem using less than $50$ cycles. We also observe that the convergence speed
is highly dependent on the starting values. In the case of the box-constrained
QP problem, we need $40$ cyclical iterations if the starting value is the
vector $x^{\left(0\right)} = \textbf{0}_5$, whereas less than $10$ cyclical
iterations are sufficient if we consider the unit vector $x^{\left(0\right)} =
\textbf{1}_5$.

\subsection{Alternating direction method of multipliers}

\subsubsection{Definition}

The alternating direction method of multipliers (ADMM) is an algorithm
introduced by Gabay and Mercier (1976) to solve optimization problems which can
be expressed as:
\begin{eqnarray}
\left\{ x^{\star },y^{\star }\right\}  &=&\arg \min_{\left( x,y\right)
}f_{x}\left( x\right) +f_{y}\left( y\right)   \label{eq:admm1} \\
&\text{s.t.}&Ax+By=c  \notag
\end{eqnarray}%
where $A\in \mathbb{R}^{p\times n}$, $B\in \mathbb{R}^{p\times m}$, $c\in
\mathbb{R}^{p}$, and the functions $f_{x}:\mathbb{R}^{n}\rightarrow \mathbb{R%
}\cup \{+\infty \}$ and $f_{y}:\mathbb{R}^{m}\rightarrow \mathbb{R}\cup
\{+\infty \}$ are proper closed convex functions. Boyd \textsl{et al.} (2011)
show that the ADMM algorithm consists of the following three steps:
\begin{enumerate}
\item The $x$-update is:
\begin{equation}
x^{\left( k+1\right) }=\arg \min_{x}\left\{ f_{x}\left( x\right) +\frac{%
\varphi }{2}\left\Vert Ax+By^{\left( k\right) }-c+u^{\left( k\right)
}\right\Vert _{2}^{2}\right\}   \label{eq:admm2a}
\end{equation}

\item The $y$-update is:%
\begin{equation}
y^{\left( k+1\right) }=\arg \min_{y}\left\{ f_{y}\left( y\right) +\frac{%
\varphi }{2}\left\Vert Ax^{\left( k+1\right) }+By-c+u^{\left( k\right)
}\right\Vert _{2}^{2}\right\}   \label{eq:admm2b}
\end{equation}

\item The $u$-update is:%
\begin{equation}
u^{\left( k+1\right) }=u^{\left( k\right) }+\left( Ax^{\left( k+1\right)
}+By^{\left( k+1\right) }-c\right)   \label{eq:admm2c}
\end{equation}
\end{enumerate}
In this approach, $u^{\left( k\right) }$ is the dual variable of the primal
residual $r=Ax+By-c$ and $\varphi $ is the $\boldsymbol{\ell }_{2} $-norm
penalty variable. The parameter $\varphi $ can be constant or may change at
each iteration\footnote{See Appendix \ref{appendix:admm} on page
\pageref{appendix:admm-varphi} for a discussion about the convergence of the
ADMM algorithm.}. The ADMM algorithm benefits from the dual ascent principle
and the method of multipliers. The difference with the latter is that the $x$-
and $y$-updates are performed in an alternating way. Therefore, it is more
flexible because the updates are equivalent to computing proximal operators for
$f_{x}$ and $f_{y}$ independently. In practice, ADMM may be slow to converge
with high accuracy, but is fast to converge if we consider modest accuracy.
This is why ADMM is a good candidate for solving large-scale machine learning
problems, where high accuracy does not necessarily lead to a better solution.

\begin{remark}
In this paper, we use the notations $f_x^{\left( k+1\right) }\left( x\right) $
and $f_y^{\left( k+1\right) }\left( y\right) $ when referring to the objective
functions that are defined in the $x$- and $y$-updates. Algorithm
(\ref{alg:admm1}) summarizes the different ADMM steps.
\end{remark}

\begin{algorithm}[tbph]
\begin{algorithmic}
\STATE The goal is to compute the solution $\left( x^{\star },y^{\star }\right) $
\STATE We initialize the vectors $x^{\left( 0\right) }$ and $y^{\left(0\right) }$ and we choose a value for the parameter $\varphi$
\STATE We set $u^{\left( 0\right) }=\mathbf{0}_{n}$
\STATE $k \leftarrow 0$
\REPEAT
\STATE $x^{\left( k+1\right) } =\arg \min_{x}\left\{ f_{x}^{\left( k+1\right)
}\left( x\right) =f_{x}\left( x\right) +\dfrac{\varphi }{2}\left\Vert
Ax+By^{\left( k\right) }-c+u^{\left( k\right) }\right\Vert _{2}^{2}\right\}$
\STATE $y^{\left( k+1\right) } =\arg \min_{y}\left\{ f_{y}^{\left( k+1\right)
}\left( y\right) =f_{y}\left( y\right) +\dfrac{\varphi }{2}\left\Vert
Ax^{\left( k+1\right) }+By-c+u^{\left( k\right) }\right\Vert_{2}^{2}\right\} $
\STATE $u^{\left( k+1\right) } = u^{\left( k\right) }+\left( Ax^{\left( k+1\right)
}+By^{\left( k+1\right) }-c\right)$
\STATE $k \leftarrow k+1$
\UNTIL{convergence}
\RETURN $x^{\star }\leftarrow x^{\left( k\right) }$ and $y^{\star}\leftarrow y^{\left( k\right) }$
\end{algorithmic}
\caption{ADMM algorithm}
\label{alg:admm1}
\end{algorithm}

\subsubsection{ADMM tricks}

The appeal of ADMM is that it can separate a complex problem into two
sub-problems that are easier to solve. However, most of the time, the
optimization problem is not formulated using a separable objective function.
The question is then how to formulate the initial problem as a separable
problem. We now list some tricks that show how ADMM may be used in practice.

\paragraph{First trick}

We consider a problem of the form $x^{\star }=\arg \min_x g\left( x\right) $. The
idea is then to write $g\left( x\right) $ as a separable function $g\left(
x\right) =g_{1}\left( x\right) +g_{2}\left( x\right) $ and to consider the
following equivalent ADMM problem:
\begin{eqnarray}
\left\{ x^{\star },y^{\star }\right\}  &=&\arg \min_{\left(x,y\right)} f_{x}\left( x\right)
+f_{y}\left( y\right)   \label{eq:admm-trick1} \\
&\text{s.t.}&x=y  \notag
\end{eqnarray}%
where $f_{x}\left( x\right) =g_{1}\left( x\right) $ and $f_{y}\left( y\right)
=g_{2}\left( y\right) $. Usually, the smooth part of $g\left( x\right) $ will
correspond to $g_{1}\left( x\right) $ while the non-smooth part will be
included in $g_{2}\left( x\right) $. The underlying idea is that the $x$-update
is straightforward, whereas the $y$-update deals with the tricky part of
$g\left( x\right) $.

\paragraph{Second trick}

If we want to minimize the function $g\left( x\right) $ where $x\in \Omega $ is
a set of constraints, the optimization problem can be cast into the ADMM form
(\ref{eq:admm-trick1}) where $f_{x}\left( x\right) =g\left( x\right) $,
$f_{y}\left( y\right) =\mathds{1}_{\Omega }\left( y\right) $ and
$\mathds{1}_{\Omega }\left( x\right) $ is the convex indicator function of
$\Omega $:
\begin{equation}
\mathds{1}_{\Omega }\left( x\right) =\left\{
\begin{array}{lll}
0 & \text{if} & x\in \Omega  \\
+\infty  & \text{if} & x\notin \Omega
\end{array}%
\right.   \label{eq:admm-trick2}
\end{equation}%
For example, if we want to solve the QP problem (\ref{eq:qp2}) given on page
\pageref{eq:qp2}, we have:
\begin{equation*}
f_{x}\left( x\right) =\frac{1}{2}x^{\top }Qx-x^{\top }R
\end{equation*}%
and:%
\begin{equation*}
\Omega =\left\{ x\in \mathbb{R}^{n}:Ax=B,Cx\leq D,x^{-}\leq x\leq
x^{+}\right\}
\end{equation*}

\paragraph{Third trick}

We can combine the first and second tricks. For instance, if we consider the
following optimization problem:
\begin{eqnarray*}
x^{\star } &=&\arg \min_x g_{1}\left( x\right) +g_{2}\left( x\right)  \\
&\text{s.t.}&x\in \Omega _{1}\cap \Omega _{2}
\end{eqnarray*}%
the equivalent ADMM form is:
\begin{eqnarray*}
\left\{ x^{\star },y^{\star }\right\}  &=&\arg \min_{\left(x,y\right)}\, \underset{f_{x}\left(
x\right) }{\underbrace{\left( g_{1}\left( x\right) +\mathds{1}_{\Omega
_{1}}\left( x\right) \right) }}+\underset{f_{y}\left( y\right) }{\underbrace{%
\left( g_{2}\left( y\right) +\mathds{1}_{\Omega _{2}}\left( y\right) \right)
}} \\
&\text{s.t.}&x=y
\end{eqnarray*}%
Let us consider a variant of the QP problem where we add a non-linear
constraint $h\left( x\right) =0$. In this case, we can write the set of
constraints as $\Omega =\Omega _{1}\cap \Omega _{2}$ where:
\begin{equation*}
\Omega _{1}=\left\{ x\in \mathbb{R}^{n}:Ax=B,Cx\leq D,x^{-}\leq x\leq
x^{+}\right\}
\end{equation*}%
and:%
\begin{equation*}
\Omega _{2}=\left\{ x\in \mathbb{R}^{n}:h\left( x\right) =0\right\}
\end{equation*}

\paragraph{Fourth trick}

Finally, if we want to minimize the function $g\left( x\right) =g\left(
x,Ax+b\right) =g_{1}\left( x\right) +g_{2}\left( Ax+b\right) $, we can write:
\begin{eqnarray*}
\left\{ x^{\star },y^{\star }\right\}  &=&\arg \min_{\left(x,y\right)} g_{1}\left( x\right)
+g_{2}\left( y\right)  \\
&\text{s.t.}&y=Ax+b
\end{eqnarray*}%
For instance, this trick can be used for a QP problem with a non-linear part:
\begin{equation*}
g\left( x\right) =\frac{1}{2}x^{\top }Qx - x^{\top} R + h\left( x\right)
\end{equation*}%
If we assume that $Q$ is a symmetric positive-definite matrix, we set $x=Ly$
where $L$ is the lower Cholesky matrix such that $LL^{\top }=Q$. It follows
that the ADMM form is equal to\footnote{This Cholesky trick has been used by
Gonzalvez \textsl{et al.} (2019) to solve trend-following strategies using the
ADMM algorithm in the context of Bayesian learning.}:
\begin{eqnarray*}
\left\{ x^{\star },y^{\star }\right\}  &=&\arg \min_{\left(x,y\right)}\, \underset{f_{x}\left(
x\right) }{\underbrace{\qquad \frac{1}{2}x^{\top }x\qquad }}+
\underset{f_{y}\left( y\right) }{\underbrace{\quad
\vphantom{\frac{1}{2}} h\left( y\right) -y^{\top }R\quad }} \\
&\text{s.t.}&x-Ly=\mathbf{0}_n
\end{eqnarray*}%
We notice that the $x$-update is straightforward because it corresponds to a
standard QP problem. If we add a set $\Omega $ of constraints, we specify:
\begin{equation*}
f_{y}\left( y\right) =h\left( y\right) -y^{\top }R+\mathds{1}_{\Omega
}\left( y\right)
\end{equation*}

\begin{remark}
\label{remark:admm-trick-qp} In the previous cases, we have seen that when the
function $g\left( x\right) $ may contain a QP problem, it is convenient to isolate
this QP problem into the $x$-update:
\begin{equation*}
x^{\left( k+1\right) }=\arg \min_x \left\{ \frac{1}{2}x^{\top }Qx-x^{\top }R+%
\mathds{1}_{\Omega }\left( x\right) +\frac{\varphi }{2}\left\Vert
x-y^{\left( k\right) }+u^{\left( k\right) }\right\Vert _{2}^{2}\right\}
\end{equation*}%
Since we have:
\begin{equation*}
\frac{\varphi }{2}\left\Vert x-y^{\left( k\right) }+u^{\left( k\right)
}\right\Vert _{2}^{2} =\frac{\varphi }{2}x^{\top }x-\varphi x^{\top
}\left( y^{\left( k\right) }-u^{\left( k\right) }\right) +
\frac{\varphi }{2}\left( y^{\left( k\right) }-u^{\left( k\right) }\right)
^{\top }\left( y^{\left( k\right) }-u^{\left( k\right) }\right)
\end{equation*}%
we deduce that the $x$-update is a standard QP problem where:
\begin{equation}
f_{x}^{\left( k+1\right) }\left( x\right) =\frac{1}{2}x^{\top }\left(
Q+\varphi I_{n}\right) x-x^{\top }\left( R+\varphi \left( y^{\left( k\right)
}-u^{\left( k\right) }\right) \right) +\mathds{1}_{\Omega }\left( x\right)
\label{eq:admm-trick-qp}
\end{equation}
\end{remark}

\subsubsection{Application to the $\lambda$-problem of the lasso regression}
\label{section:admm-lasso-lambda}

The $\lambda $-problem of the lasso regression (\ref{eq:ccd-lasso1}) has the
following ADMM formulation:
\begin{eqnarray*}
\left\{ \beta ^{\star },\bar{\beta}^{\star }\right\}  &=&\arg \min \frac{1}{2%
}(Y-X\beta )^{\top }(Y-X\beta )+\lambda \Vert \bar{\beta}\Vert _{1} \\
&\text{s.t.}&\beta -\bar{\beta}=\mathbf{0}_{p}
\end{eqnarray*}%
Since the $x$-step corresponds to a QP problem\footnote{We have $Q=X^{\top }X$
and $R=X^{\top }Y$.}, we use the results given in Remark
\ref{remark:admm-trick-qp} to find the value of $\beta ^{\left( k+1\right) }$:
\begin{eqnarray*}
\beta ^{\left( k+1\right) } &=&\left( Q+\varphi I_{p}\right) ^{-1}\left(
R+\varphi \left( \bar{\beta}^{\left( k\right) }-u^{\left( k\right) }\right)
\right)  \\
&=&\left( X^{\top }X+\varphi I_{p}\right) ^{-1}\left( X^{\top
}Y+\varphi \left( \bar{\beta}^{\left( k\right) }-u^{\left( k\right)
}\right) \right)
\end{eqnarray*}%
The $y$-step is:%
\begin{eqnarray*}
\bar{\beta}^{\left( k+1\right) } &=&\arg \min_{\bar{\beta}} \left\{ \lambda \Vert \bar{%
\beta}\Vert _{1}+\frac{\varphi }{2}\left\Vert \beta ^{\left( k+1\right) }-%
\bar{\beta}+u^{\left( k\right) }\right\Vert _{2}^{2}\right\}  \\
&=&\arg \min \left\{ \frac{1}{2}\left\Vert \bar{\beta}-\left( \beta ^{\left(
k+1\right) }+u^{\left( k\right) }\right) \right\Vert _{2}^{2}+\frac{\lambda
}{\varphi }\Vert \bar{\beta}\Vert _{1}\right\}
\end{eqnarray*}%
We recognize the soft-thresholding problem with $v=\beta ^{\left( k+1\right)
}+u^{\left( k\right) }$. Finally, the ADMM algorithm is made up of the
following steps (Boyd \textsl{et al.}, 2011):
\begin{equation*}
\left\{
\begin{array}{l}
\beta ^{\left( k+1\right) }=\left( X^{\top }X+\varphi I_{p}\right)
^{-1}\left( X^{\top }Y+\varphi \left( \bar{\beta}^{\left( k\right)
}-u^{\left( k\right) }\right) )\right)  \\
\bar{\beta}^{\left( k+1\right) }=\mathcal{S}\left( \beta ^{\left( k+1\right)
}+u^{\left( k\right) };\varphi ^{-1}\lambda \right)  \\
u^{\left( k+1\right) }=u^{\left( k\right) }+\left( \beta ^{\left( k+1\right)
}-\bar{\beta}^{\left( k+1\right) }\right)
\end{array}%
\right.
\end{equation*}

\begin{figure}[tbph]
\centering
\caption{ADMM algorithm applied to the lasso optimization problem}
\label{fig:lasso5}
\figureskip
\includegraphics[width = \figurewidth, height = \figureheight]{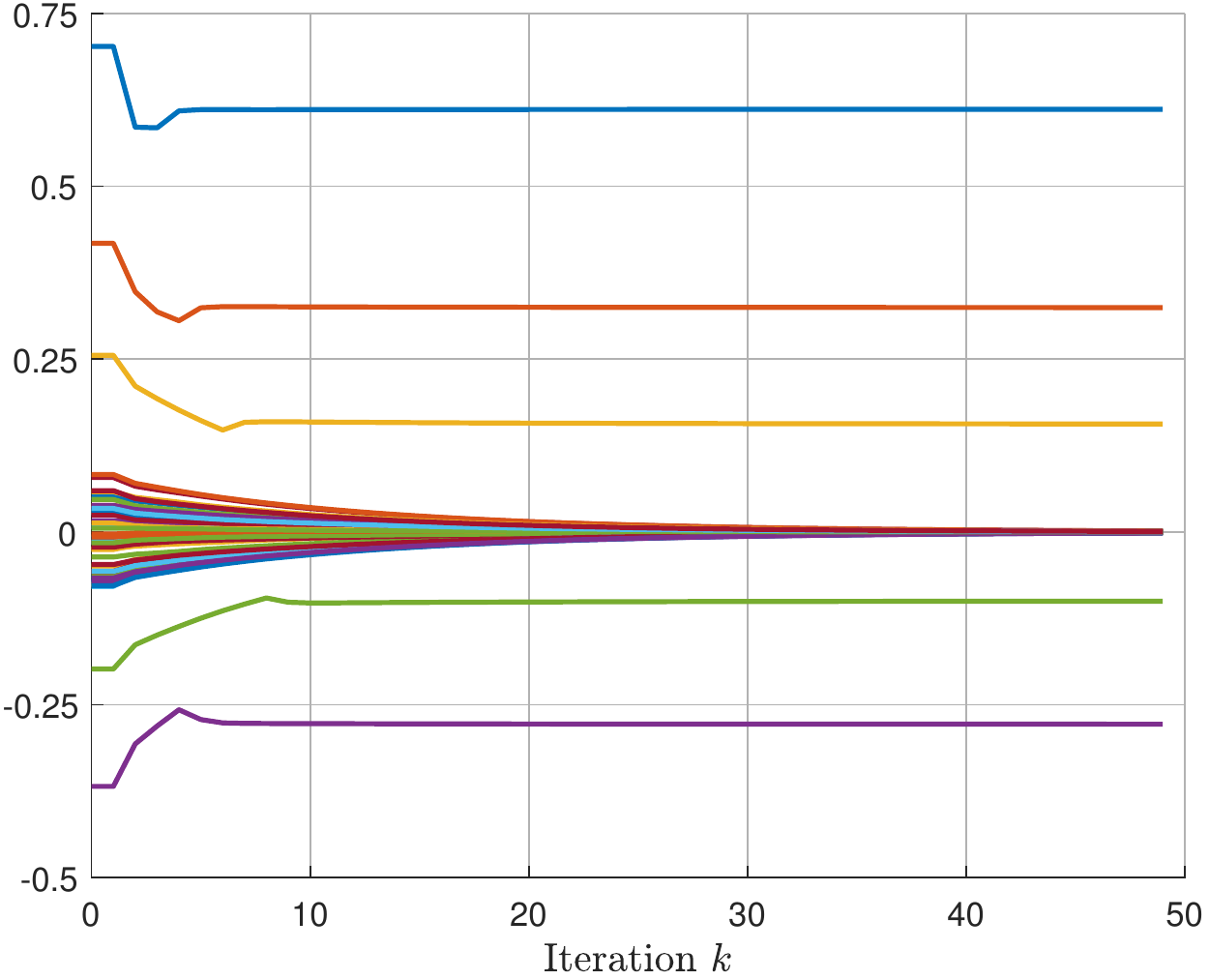}
\end{figure}

We consider the example of the lasso regression with $\lambda = 900$ on page
\pageref{example:ccd-lasso}. By setting $\varphi = \lambda$ and by initialing
the algorithm with the OLS estimates, we obtain the convergence given in Figure
\ref{fig:lasso5}. We notice that the ADMM algorithm converges more slowly than
the CCD algorithm for this example. In practice, we generally observe that the
convergence is poor for low and very high values of $\varphi$. However, finding
an optimal value of $\varphi$ is difficult. A better approach involves using a
varying parameter $\varphi^{\left(k\right)}$ such as the method described on
page \pageref{appendix:admm-varphi}.

\subsection{Proximal operators}

The $x$- and $y$-update steps of the ADMM algorithm require a
$\boldsymbol{\ell}_{2}$-norm penalized optimization problem to be solved.
Proximal operators are special cases of this type of problem when the matrices
$A$ or $B$ correspond to the identity matrix $I_{n}$ or its opposite $-I_{n}$.

\subsubsection{Definition}

Let $f:\mathbb{R}^{n}\rightarrow \mathbb{R}\cup \left\{ +\infty \right\} $ be a
proper closed convex function. The proximal operator $\mathbf{prox}_{f}\left(
v\right) :\mathbb{R}^{n}\rightarrow \mathbb{R}^{n}$ is defined by:
\begin{equation}
\mathbf{prox}_{f}\left( v\right) =x^{\star }=\arg \min\limits_{x}\left\{
f_v\left( x\right) = f\left( x\right) +\frac{1}{2}\left\Vert x-v\right\Vert _{2}^{2}\right\}
\label{eq:proximal1}
\end{equation}%
Since the function $f_{v}\left( x\right) =f\left( x\right)
+\dfrac{1}{2}\left\Vert x-v\right\Vert _{2}^{2}$ is strongly convex, it has a
unique minimum for every $v\in \mathbb{R}^{n}$ (Parikh and Boyd, 2014). By
construction, the proximal operator defines a point $x^{\star }$ which is a
trade-off between minimizing $f\left( x\right) $ and being close to
$v$.\smallskip

In many situations, we need to calculate the proximal of the scaled function
$\lambda f\left( x\right) $ where $\lambda >0$. In this case, we use the
notation $\mathbf{prox}_{\lambda f}\left( v\right) $ and we have:
\begin{eqnarray*}
\mathbf{prox}_{\lambda f}\left( v\right)  &=&\arg \min\limits_{x}\left\{
\lambda f\left( x\right) +\frac{1}{2}\left\Vert x-v\right\Vert
_{2}^{2}\right\}  \\
&=&\arg \min\limits_{x}\left\{ f\left( x\right) +\frac{1}{2\lambda }
\left\Vert x-v\right\Vert _{2}^{2}\right\}
\end{eqnarray*}%
For instance, if we consider the $y$-update of the ADMM\ algorithm with
$B=-I_{n}$, we have:
\begin{eqnarray*}
y^{\left( k+1\right) } &=&\arg \min_{y}\left\{ f_{y}\left( y\right) +\frac{%
\varphi }{2}\left\Vert y-v_{y}^{\left( k+1\right) }\right\Vert
_{2}^{2}\right\}  \\
&=&\arg \min_{y}\left\{ \varphi ^{-1}f_{y}\left( y\right) +\frac{1}{2}%
\left\Vert y-v_{y}^{\left( k+1\right) }\right\Vert _{2}^{2}\right\}  \\
&=&\mathbf{prox}_{\varphi ^{-1}f_{y}}\left( v_{y}^{\left( k+1\right)
}\right)
\end{eqnarray*}%
where $v_{y}^{\left( k\right) }=Ax^{\left( k+1\right) }-c+u^{\left( k\right)
}$. ADMM is then given by Algorithm (\ref{alg:admm2}). The interest of this
mathematical formulation is to write the ADMM algorithm in a convenient form
such that the $x$-update corresponds to the tricky part of the optimization
while the $y$-update is reduced to an analytical formula.

\begin{algorithm}[tbph]
\begin{algorithmic}
\STATE The goal is to compute the solution $\left( x^{\star },y^{\star }\right) $
\STATE We initialize the vectors $x^{\left( 0\right) }$ and $y^{\left(0\right) }$ and we choose a value for the parameter $\varphi$
\STATE We set $u^{\left( 0\right) }=\mathbf{0}_{n}$
\STATE $k \leftarrow 0$
\REPEAT
\STATE $x^{\left( k+1\right) } =\arg \min_{x}\left\{ f_{x}^{\left( k+1\right)
}\left( x\right) =f_{x}\left( x\right) +\dfrac{\varphi }{2}\left\Vert
Ax-y^{\left( k\right) }-c+u^{\left( k\right) }\right\Vert _{2}^{2}\right\}$
\STATE $v_{y}^{\left( k+1\right) }=Ax^{\left( k+1\right) }-c+u^{\left( k\right) }$
\STATE $y^{\left( k+1\right) }=\mathbf{prox}_{\varphi ^{-1}f_{y}}\left(
v_{y}^{\left( k+1\right) }\right) $
\STATE $u^{\left( k+1\right) } = u^{\left( k\right) }+\left( Ax^{\left( k+1\right)
}-y^{\left( k+1\right) }-c\right)$
\STATE $k \leftarrow k+1$
\UNTIL{convergence}
\RETURN $x^{\star }\leftarrow x^{\left( k\right) }$ and $y^{\star}\leftarrow y^{\left( k\right) }$
\end{algorithmic}
\caption{ADMM algorithm in the case $Ax - y = c$}
\label{alg:admm2}
\end{algorithm}

\subsubsection{Proximal operators and generalized projections}

In the case where $f\left( x\right) =\mathds{1}_{\Omega }\left( x\right) $ is
the indicator function, the proximal operator is then the Euclidean projection
onto $\Omega $:
\begin{eqnarray*}
\mathbf{prox}_{f}\left( v\right)  &=&\arg \min\limits_{x}\left\{ \mathds{1}%
_{\Omega }\left( x\right) +\frac{1}{2}\left\Vert x-v\right\Vert
_{2}^{2}\right\}  \\
&=&\arg \min\limits_{x\in \Omega }\left\{ \left\Vert x-v\right\Vert
_{2}^{2}\right\}  \\
&=&\mathcal{P}_{\Omega }\left( v\right)
\end{eqnarray*}%
where $\mathcal{P}_{\Omega }\left( v\right) $ is the standard projection of $%
v$ onto $\Omega $. Parikh and Boyd (2014) interpret then proximal operators as
a generalization of the Euclidean projection.\smallskip

Let us consider the constrained optimization problem $x^{\star }=\arg \min
f\left( x\right) $ subject to $x\in \Omega $. Using the second ADMM trick, we
have $f_{x}\left( x\right) =f\left( x\right) $, $f_{y}\left( y\right) =
\mathds{1}_{\Omega }\left( y\right) $ and $x-y=\mathbf{0}_{n}$. Therefore, we
can use Algorithm (\ref{alg:admm2}) since the $v$- and $y$-steps become
$v_{y}^{\left( k+1\right) }=x^{\left( k+1\right) }+u^{\left( k\right) }$
and\footnote{We notice that the parameter $\varphi $ has no impact on the
$y$-update because $\varphi ^{-1}f_{y}\left( y\right) =f_{y}\left( y\right)
=\mathds{1}_{\Omega }\left( y\right) $. We then deduce that:
\begin{equation*}
\mathbf{prox}_{\varphi ^{-1}f_{y}}\left( v_{y}^{\left( k+1\right) }\right) =
\mathbf{prox}_{f_{y}}\left( v_{y}^{\left( k+1\right) }\right) =
\mathcal{P}_{\Omega }\left( v_{y}^{\left( k+1\right) }\right)
\end{equation*}%
} $y^{\left( k+1\right) }=\mathcal{P}_{\Omega }\left( v_{y}^{\left( k+1\right)
}\right) $.\smallskip

Here, we give the results of Parikh and Boyd (2014) for some simple polyhedra:
\begin{equation*}
\label{tab:basic-projection}
\begin{tabular}{ccc}
\hline
Notation                                      & $\Omega $                         & $\mathcal{P}_{\Omega }\left( v\right) $                                         \\ \hline
$\mathcal{A}_{ffineset}\left[ A,B\right] $   & $Ax = B$                           & $v - A^{\dagger }\left( Av-B\right) $                                           \\
$\mathcal{H}_{yperplane}\left[ a,b\right] $   & $a^{\top }x = b$                  & $v -\dfrac{\left( a^{\top }v-b\right) }{\left\Vert a\right\Vert _{2}^{2}}a$     \\
$\mathcal{H}_{alfspace}\left[ c,d\right] $   & $c^{\top }x\leq d$                 & $v -\dfrac{\left( c^{\top }v-d\right) _{+}}{\left\Vert c\right\Vert _{2}^{2}}c$ \\
$\mathcal{B}_{ox}\left[ x^{-},x^{+}\right] $  & $x^{-}\leq x\leq x^{+}$           & $\mathcal{T}\left( v;x^{-},x^{+}\right) $                                       \\ \hline
\end{tabular}%
\end{equation*}%
where $A^{\dagger }$ is the Moore-Penrose pseudo-inverse of $A$, and
$\mathcal{T}\left( v;x^{-},x^{+}\right) $ is the truncation operator.

\subsubsection{Main properties}

There are many properties that are useful for finding the analytical expression
of the proximal operator. In what follows, we consider three main properties,
but the reader may refer to Combettes and Pesquet (2011), Parikh and Boyd
(2014) and Beck (2017) for a more exhaustive list.

\paragraph{Separable sum}

Let us assume that $f\left( x\right) =\sum_{i=1}^{n}f_{i}\left( x_{i}\right) $
is fully separable, then the proximal of $f\left( v\right) $ is the vector of
the proximal operators applied to each scalar-valued function $f_{i}\left(
x_{i}\right) $:%
\begin{equation*}
\mathbf{prox}_{f}\left( v\right) =\left(
\begin{array}{c}
\mathbf{prox}_{f_{1}}\left( v_{1}\right)  \\
\vdots  \\
\mathbf{prox}_{f_{n}}\left( v_{n}\right)
\end{array}%
\right)
\end{equation*}%
For example, if $f\left( x\right) =\lambda \left\Vert x\right\Vert _{1}$, we
have $f\left( x\right) =\lambda \sum_{i=1}^{n}\left\vert x_{i}\right\vert $ and
$f_{i}\left( x_{i}\right) =\lambda \left\vert x_{i}\right\vert $. We deduce
that the proximal operator of $f\left( x\right) $ is the vector formulation of
the soft-thresholding operator:
\begin{equation*}
\mathbf{prox}_{\lambda \left\Vert x\right\Vert _{1}}\left( v\right)
=\left(
\begin{array}{c}
\limfunc{sign}\left( v_{1}\right) \cdot \left( \left\vert v_{1}\right\vert
-\lambda \right) _{+} \\
\vdots  \\
\limfunc{sign}\left( v_{n}\right) \cdot \left( \left\vert v_{n}\right\vert
-\lambda \right) _{+}%
\end{array}%
\right) =\limfunc{sign}\left( v\right) \odot \left( \left\vert v\right\vert
-\lambda \mathbf{1}_{n}\right) _{+}
\end{equation*}%
This result has been used to solve the $\lambda $-problem of the lasso
regression on page \pageref{section:admm-lasso-lambda}.\smallskip

If we consider the scalar-valued logarithmic barrier function $f\left( x\right)
=-\lambda \ln x$, we have:
\begin{eqnarray*}
f_v\left( x\right) &=&-\lambda
\ln x+\frac{1}{2}\left( x-v\right) ^{2} \\
&=&-\lambda \ln x+\frac{1}{2}x^{2}-xv+\frac{1}{2}v^{2}
\end{eqnarray*}%
The first-order condition is $-\lambda x^{-1}+x-v=0$. We obtain two roots
with opposite signs:%
\begin{equation*}
x^{\star }=\frac{v\pm \sqrt{v^{2}+4\lambda }}{2}
\end{equation*}%
Since the logarithmic function is defined for $x>0$, we deduce that the
proximal operator is the positive root. In the case of the vector-valued
logarithmic barrier $f\left( x\right) =-\lambda \sum_{i=1}^{n}\ln x_{i}$, it
follows that:
\begin{equation*}
\mathbf{prox}_{f}(v)=\frac{v+\sqrt{v\odot v+4\lambda }}{2}
\end{equation*}

\paragraph{Moreau decomposition}

An important property of the proximal operator is the Moreau decomposition
theorem:%
\begin{equation*}
\mathbf{prox}_{f}\left( v\right) +\mathbf{prox}_{f^{\ast }}\left( v\right) =v
\end{equation*}%
where $f^{\ast }$ is the convex conjugate of $f$. This result is
used extensively to find the proximal of norms, the max function, the sum-of-$k$%
-largest-values function, etc. (Beck, 2017).\smallskip

In the case of the pointwise maximum function $f\left( x\right) =\max x$, we
can show that:%
\begin{equation*}
\mathbf{prox}_{\lambda \max x }\left( v\right) =\min \left( v,s^{\star
}\right)
\end{equation*}%
where $s^{\star }$ is the solution of the following equation (see
Appendix \ref{appendix:proximal-max-function} on page
\pageref{appendix:proximal-max-function}):
\begin{equation*}
s^{\star }=\left\{ s\in \mathbb{R}:\sum_{i=1}^{n}\left( v_{i}-s\right)
_{+}=\lambda \right\}
\end{equation*}%
If we assume that $f\left( x\right) =\left\Vert x\right\Vert _{p}$, we obtain:
\begin{equation*}
\begin{tabular}{cc}
\hline
$p$         & $\mathbf{prox}_{\lambda f}\left( v\right) $ \\ \hline
$p=1$       & $\mathcal{S}\left( v;\lambda\right) = \limfunc{sign}\left( v\right) \odot \left( \left\vert v\right\vert -\lambda \mathbf{1}_n\right)_{+}$ \\
$p=2$       & $\left( 1-\dfrac{\lambda }{\max \left( \lambda ,\left\Vert v\right\Vert _{2}\right) }\right) v$ \\
$p=\infty $ & $\limfunc{sign}\left( v\right) \odot \mathbf{prox}_{\lambda \max x}\left( \left\vert v\right\vert \right)$ \\ \hline
\end{tabular}%
\end{equation*}
\smallskip

If $f\left( x\right) $ is a $\boldsymbol{\ell }_{q}$-norm function,
then $f^{\ast }\left( x\right) =\mathds{1}_{\mathcal{B}_{p}}\left(
x\right) $ where $\mathcal{B}_{p}$ is the $\boldsymbol{\ell }_{p}$
unit ball and $p^{-1} + q^{-1} = 1$. Since we have
$\mathbf{prox}_{f^{\ast }}\left( v\right)
=\mathcal{P}_{\mathcal{B}_{p}}\left( v\right) $, we deduce that:
\begin{equation*}
\mathbf{prox}_{f}\left( v\right) +\mathcal{P}_{\mathcal{B}_{p}}\left(
v\right) =v
\end{equation*}%
More generally, we have:%
\begin{equation*}
\mathbf{prox}_{\lambda f}\left( v\right) +\lambda \mathcal{P}_{\mathcal{B}%
_{p}}\left( \frac{v}{\lambda }\right) =v
\end{equation*}%
It follows that the projection onto the $\boldsymbol{\ell }_{p}$
ball can be deduced from the proximal operator of the
$\boldsymbol{\ell }_{q}$-norm
function. Let $\mathcal{B}_{p}\left( c,\lambda \right) =\left\{ x\in \mathbb{%
R}^{n}:\left\Vert x-c\right\Vert _{p}\leq \lambda \right\} $ be the $%
\boldsymbol{\ell }_{p}$ ball with center $c$ and radius $\lambda $. We obtain:
\begin{equation*}
\begin{tabular}{ccc}
\hline
$p$         & $\mathcal{P}_{\mathcal{B}_{p}\left( \mathbf{0}_n,\lambda \right) }\left(v\right) $                              & $q$          \\ \hline
$p=1$       & $v - \limfunc{sign}\left( v\right) \odot \mathbf{prox}_{\lambda \max x}\left( \left\vert v\right\vert \right) $ & $q=\infty$   \\
$p=2$       & $v - \mathbf{prox}_{\lambda \left\Vert x\right\Vert _{2}}\left( v \right) $                                     & $q=2$        \\
$p=\infty $ & $\mathcal{T}\left( v;-\lambda ,\lambda \right) $                                                                & $q = 1$      \\ \hline
\end{tabular}%
\end{equation*}

\paragraph{Scaling and translation}

Let us define $g\left( x\right) =f\left( ax+b\right) $ where $a\neq 0$. We
have\footnote{The proof can be found in Beck (2017) on page 138. We have
reported it in Appendix \ref{appendix:proximal-scaling} on page
\pageref{appendix:proximal-scaling}.}:
\begin{equation*}
\mathbf{prox}_{g}\left( v\right) =\frac{\mathbf{prox}_{a^{2}f}\left(
av+b\right) -b}{a}
\end{equation*}%
We can use this property when the center $c$ of the $\boldsymbol{\ell }_{p}$
ball is not equal to $\mathbf{0}_n$. Since we have $\mathbf{prox}_{g}\left(
v\right) =\mathbf{prox}_{f}\left( v-c\right) +c$ where $g\left( x\right)
=f\left( x-c\right) $ and the equivalence $\mathcal{B}_{p}\left( \mathbf{0}_n%
,\lambda \right) =\left\{ x\in \mathbb{R}^{n}:f\left( x\right) \leq \lambda
\right\} $ where $f\left( x\right) =\left\Vert x\right\Vert _{p}$, we deduce
that:%
\begin{equation*}
\mathcal{P}_{\mathcal{B}_{p}\left( c,\lambda \right) }\left( v\right) =%
\mathcal{P}_{\mathcal{B}_{p}\left( \mathbf{0}_n,\lambda \right) }\left(
v-c\right) +c
\end{equation*}

\subsubsection{Application to the $\tau$-problem of the lasso regression}
\label{section:admm-lasso-tau}

We have previously presented the lasso regression problem by considering the
Lagrange formulation ($\lambda $-problem). We now consider the original $\tau
$-problem:
\begin{eqnarray*}
\hat{\beta}\left( \tau \right)  &=&\arg \min_{\beta} \frac{1}{2}\left( Y-X\beta
\right) ^{\top }\left( Y-X\beta \right)  \\
&\text{s.t.}&\left\Vert \beta \right\Vert _{1}\leq \tau
\end{eqnarray*}%
The ADMM formulation is:
\begin{eqnarray*}
\left\{ \beta ^{\star },\bar{\beta}^{\star }\right\}  &=&\arg \min_{\left(\beta,\bar{\beta}\right)} \, \frac{1}{2%
}\left( Y-X\beta \right) ^{\top }\left( Y-X\beta \right) +\mathds{1}_{\Omega
}\left( \bar{\beta}\right)  \\
&\text{s.t.}&\beta = \bar{\beta}
\end{eqnarray*}%
where $\Omega =\mathcal{B}_{1}\left( \mathbf{0}_n,\tau \right) $ is the
centered $\boldsymbol{\ell} _{1}$ ball with radius $\tau $. We notice that the
$x$-update is:
\begin{eqnarray*}
\beta ^{\left( k+1\right) } &=&\arg \min_{\beta }\left\{ \frac{1}{2}\left(
Y-X\beta \right) ^{\top }\left( Y-X\beta \right) +\frac{\varphi }{2}%
\left\Vert \beta -\bar{\beta}^{\left( k\right) }+u^{\left( k\right)
}\right\Vert _{2}^{2}\right\}  \\
&=&\left( X^{\top }X+\varphi I_{p}\right) ^{-1}\left( X^{\top }Y+\varphi
\left( \bar{\beta}^{\left( k\right) }-u^{\left( k\right) }\right) \right)
\end{eqnarray*}%
where $v_{x}^{\left( k+1\right) }=\bar{\beta}^{\left( k\right) }-u^{\left(
k\right) }$. For the $y$-update, we deduce that:
\begin{eqnarray*}
\bar{\beta}^{\left( k+1\right) } &=&\arg \min_{\bar{\beta}}\left\{ \mathds{1}%
_{\Omega }\left( \bar{\beta}\right) +\frac{\varphi }{2}\left\Vert \beta
^{\left( k+1\right) }-\bar{\beta}+u^{\left( k\right) }\right\Vert
_{2}^{2}\right\}  \\
&=&\mathbf{prox}_{f_{y}}\left( \beta ^{\left( k+1\right) }+u^{\left(
k\right) }\right)  \\
&=&\mathcal{P}_{\Omega }\left( v_{y}^{\left( k+1\right) }\right)  \\
&=&v_{y}^{\left( k+1\right) }-\limfunc{sign}\left( v_{y}^{\left( k+1\right)
}\right) \odot \mathbf{prox}_{\tau \max x}\left( \left\Vert v_{y}^{\left(
k+1\right) }\right\Vert \right)
\end{eqnarray*}%
where $v_{y}^{\left( k+1\right) }=\beta ^{\left( k+1\right) }+u^{\left(
k\right) }$. Finally, the $u$-update is defined by $u^{\left( k+1\right)
}=u^{\left( k\right) }+\beta ^{\left( k+1\right) }-\bar{\beta}^{\left(
k+1\right) }$.

\begin{remark}
The ADMM algorithm is similar for $\lambda$- and $\tau$-problems since the only
difference concerns the $y$-step. For the $\lambda$-problem, we apply the
soft-thresholding operator while we use the $\boldsymbol{\ell}_1$ projection in
the case of the $\tau$-problem. However, our experience shows that the
$\tau$-problem is easier to solve with the ADMM algorithm from a practical
point of view. The reason is that the $y$-update of the $\tau$-problem is
independent of the penalization parameter $\varphi$. This is not the case for
the $\lambda$-problem, because the soft-thresholding depends on the value taken
by $\varphi^{-1}\lambda$.
\end{remark}

\subsubsection{Application to the CD algorithm with pointwise constraints}
\label{section:proximal-ccd-pointwise}

We consider the following constrained minimization problem:
\begin{equation*}
x^{\star } = \arg \min_{x}f\left( x\right) \quad \text{s.t.} \quad x\in \Omega
\end{equation*}
where the set $\Omega $ of constraints is fully separable:
\begin{equation*}
\mathds{1}_{\Omega }\left( x\right) =\sum_{i=1}^{n}\mathds{1}_{\Omega
_{i}}\left( x_{i}\right)
\end{equation*}%
The scalar-valued problem of the CD algorithm becomes:
\begin{equation*}
x_{i}^{\star }=\arg \min_{\varkappa }f\left( x_{1},\ldots ,x_{i-1},\varkappa
,x_{i+1},\ldots ,x_{n}\right) +\lambda \sum_{i=1}^{n}\mathds{1}_{\Omega
_{i}}\left( x_{i}\right)
\end{equation*}%
Nesterov (2012) and Wright (2015) propose the following coordinate update:
\begin{equation*}
x_{i}^{\star }=\arg \min_{\varkappa }\left( \varkappa -x_{i}\right) g_{i}+%
\frac{1}{2\eta }\left( \varkappa -x_{i}\right) ^{2}+\lambda \cdot \mathds{1}%
_{\Omega _{i}}\left( \varkappa \right)
\end{equation*}%
where $g_{i}=\nabla _{i}f\left( x\right) $ is the first-derivative of the
function with respect to $x_{i}$, $\eta >0$ controls the quadratic penalization
term and $\lambda $ is a positive scalar. The objective function is equivalent
to:
\begin{eqnarray*}
(\ast ) &=&\left( \varkappa -x_{i}\right) g_{i}+\frac{1}{2\eta }\left(
\varkappa -x_{i}\right) ^{2}+\lambda \cdot \mathds{1}_{\Omega _{i}}\left(
\varkappa \right)  \\
&=&\frac{1}{2\eta }\left( \left( \varkappa -x_{i}\right) ^{2}+2\left(
\varkappa -x_{i}\right) \eta g_{i}\right) +\lambda \cdot \mathds{1}_{\Omega
_{i}}\left( \varkappa \right)  \\
&=&\frac{1}{2\eta }\left( \varkappa -x_{i}+\eta g_{i}\right) ^{2}+\lambda
\cdot \mathds{1}_{\Omega _{i}}\left( \varkappa \right) -\frac{\eta }{2}%
g_{i}^{2}
\end{eqnarray*}%
By taking $\lambda =\eta ^{-1}$, we deduce that:%
\begin{eqnarray*}
x_{i}^{\star } &=&\arg \min_{\varkappa }\mathds{1}_{\Omega _{i}}\left(
\varkappa \right) +\frac{1}{2}\left\Vert \varkappa -\left( x_{i}-\eta
g_{i}\right) \right\Vert ^{2} \\
&=&\mathbf{prox}_{\psi }\left( x_{i}-\eta g_{i}\right)  \\
&=&\mathcal{P}_{\Omega _{i}}\left( x_{i}-\eta g_{i}\right)
\end{eqnarray*}%
where $\psi \left( \varkappa\right) =\mathds{1}_{\Omega _{i}}\left(
\varkappa\right) $. Extending the CD algorithm in the case of pointwise
constraints is then equivalent to implement a standard CD algorithm and apply
the projection onto the $i^{\mathrm{th}}$ coordinate at each iteration%
\footnote{This method corresponds to the proximal gradient algorithm.}. For
instance, this algorithm is particularly efficient when we consider box
constraints.

\subsection{Dykstra's algorithm}

We now consider the proximal optimization problem where the function $ f\left(
x\right) $ is the convex sum of basic functions $f_{j}\left( x\right) $:
\begin{equation*}
x^{\star }=\arg \min_{x}\left\{ \sum_{j=1}^{m}f_{j}\left( x\right) +\frac{1}{%
2}\left\Vert x-v\right\Vert _{2}^{2}\right\}
\end{equation*}%
and the proximal of each basic function is known.

\subsubsection{The $m=2$ case}

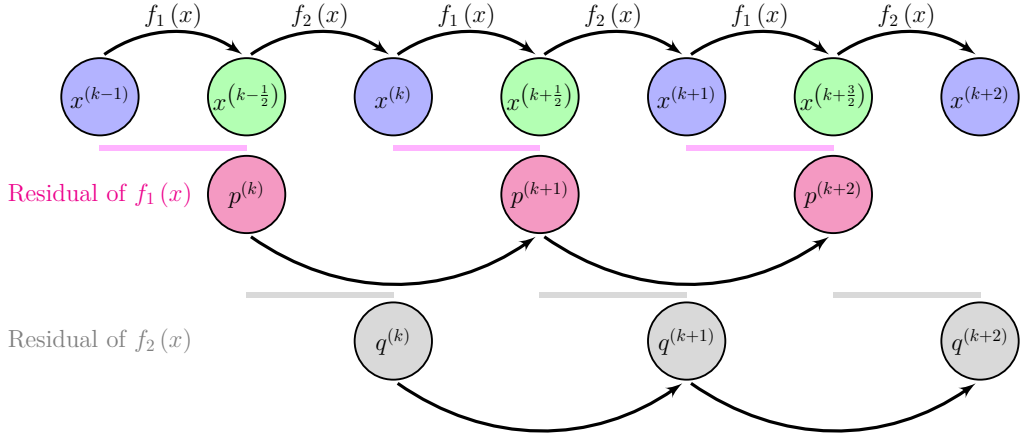
\begin{figure}[tbph]
\centering
\caption{Splitting method of the Dykstra's algorithm}
\figureskip
\label{fig:dykstra}
\begin{tikzpicture}[scale=0.65, transform shape]
    \tikzstyle{ball}=[circle,fill=black!25,minimum size=45pt,inner sep=0pt,draw=black,line width=0.25mm]
    \tikzstyle{blue ball}=[ball, fill=blue!30];
    \tikzstyle{green ball}=[ball, fill=green!30];
    \tikzstyle{magenta ball}=[ball, fill=magenta!50];
    \tikzstyle{gray ball}=[ball, fill=gray!30!white];
    \tikzstyle{every node}=[font=\Large]

    \node[blue ball] (X1) at (-6,0) {$x^{\left(k-1\right)}$};
    \node[blue ball] (X2) at  (0,0) {$x^{\left(k\right)}$};
    \node[blue ball] (X3) at  (6,0) {$x^{\left(k+1\right)}$};
    \node[blue ball] (X4) at  (12,0) {$x^{\left(k+2\right)}$};

    \node[green ball] (Y1) at (-3,0) {$x^{\left(k-\frac{1}{2}\right)}$};
    \node[green ball] (Y2) at  (3,0) {$x^{\left(k+\frac{1}{2}\right)}$};
    \node[green ball] (Y3) at  (9,0) {$x^{\left(k+\frac{3}{2}\right)}$};

    \draw[->, >=latex', shorten >=2pt, shorten <=2pt, bend left=35, very thick] (X1.north) to (Y1.north);
    \draw[->, >=latex', shorten >=2pt, shorten <=2pt, bend left=35, very thick] (X2.north) to (Y2.north);
    \draw[->, >=latex', shorten >=2pt, shorten <=2pt, bend left=35, very thick] (X3.north) to (Y3.north);

    \draw[->, >=latex', shorten >=2pt, shorten <=2pt, bend left=35, very thick] (Y1.north) to (X2.north);
    \draw[->, >=latex', shorten >=2pt, shorten <=2pt, bend left=35, very thick] (Y2.north) to (X3.north);
    \draw[->, >=latex', shorten >=2pt, shorten <=2pt, bend left=35, very thick] (Y3.north) to (X4.north);

    \node at (-4.5,1.65) {$f_1\left(x\right)$};
    \node at  (1.5,1.65) {$f_1\left(x\right)$};
    \node at  (7.5,1.65) {$f_1\left(x\right)$};

    \node at (-1.5,1.65) {$f_2\left(x\right)$};
    \node at  (4.5,1.65) {$f_2\left(x\right)$};
    \node at (10.5,1.65) {$f_2\left(x\right)$};

    \shade [top color=magenta!30, bottom color=magenta!30] (-6,-1.0) rectangle (-3,-1.1);
    \shade [top color=magenta!30, bottom color=magenta!30] (0,-1.0) rectangle (3,-1.1);
    \shade [top color=magenta!30, bottom color=magenta!30] (6,-1.0) rectangle (9,-1.1);

    \node[magenta ball] (P1) at (-3,-2.0) {$p^{\left(k\right)}$};
    \node[magenta ball] (P2) at  (3,-2.0) {$p^{\left(k+1\right)}$};
    \node[magenta ball] (P3) at  (9,-2.0) {$p^{\left(k+2\right)}$};

    \draw[->, >=latex', shorten >=2pt, shorten <=2pt, bend right=35, very thick] (P1.south) to (P2.south);
    \draw[->, >=latex', shorten >=2pt, shorten <=2pt, bend right=35, very thick] (P2.south) to (P3.south);

    \shade [top color=gray!30, bottom color=gray!30] (-3,-4.0) rectangle (0,-4.1);
    \shade [top color=gray!30, bottom color=gray!30] (3,-4.0) rectangle (6,-4.1);
    \shade [top color=gray!30, bottom color=gray!30] (9,-4.0) rectangle (12,-4.1);

    \node[gray ball] (Q1) at  (0,-5) {$q^{\left(k\right)}$};
    \node[gray ball] (Q2) at  (6,-5) {$q^{\left(k+1\right)}$};
    \node[gray ball] (Q3) at (12,-5) {$q^{\left(k+2\right)}$};

    \draw[->, >=latex', shorten >=2pt, shorten <=2pt, bend right=35, very thick] (Q1.south) to (Q2.south);
    \draw[->, >=latex', shorten >=2pt, shorten <=2pt, bend right=35, very thick] (Q2.south) to (Q3.south);

    \node[color=magenta!100,thick] at (-6,-2.0) {Residual of $f_1\left(x\right)$};
    \node[color=gray!100,thick] at (-6,-5.0) {Residual of $f_2\left(x\right)$};

\end{tikzpicture}
\end{figure}

In the previous section, we listed some analytical solutions of the proximal
problem when the function $f\left( x\right) $ is basic. For instance, we know
the proximal solution of the $\ell _{1}$-norm function $f_{1}\left( x\right)
=\lambda _{1}\left\Vert x\right\Vert _{1}$ or the proximal solution of the
logarithmic barrier function $f_{2}\left( x\right) =\lambda
_{2}\sum_{i=1}^{n}\ln x_{i}$. However, we don't know how to compute the
proximal operator of $f\left( x\right) =f_{1}\left( x\right) +f_{2}\left(
x\right) $:
\begin{eqnarray*}
x^{\star } &=&\arg \min_{x}f_{1}\left( x\right) +f_{2}\left( x\right) +
\frac{1}{2}\left\Vert x-v\right\Vert _{2}^{2} \\
&=&\mathbf{prox}_{f}\left( v\right)
\end{eqnarray*}%
Nevertheless, an elegant solution is provided by the Dykstra's algorithm
(Dykstra, 1983; Bauschke and Borwein, 1994; Combettes and Pesquet, 2011), which
is defined by the following iterations:
\begin{equation}
\left\{
\begin{array}{l}
x^{\left( k+1\right) }=\mathbf{prox}_{f_{1}}\left( y^{\left( k\right)
}+p^{\left( k\right) }\right)  \\
p^{\left( k+1\right) }=y^{\left( k\right) }+p^{\left( k\right) }-x^{\left(
k+1\right) } \\
y^{\left( k+1\right) }=\mathbf{prox}_{f_{2}}\left( x^{\left( k+1\right)
}+q^{\left( k\right) }\right)  \\
q^{\left( k+1\right) }=x^{\left( k+1\right) }+q^{\left( k\right) }-y^{\left(
k+1\right) }%
\end{array}%
\right.   \label{eq:douglas1}
\end{equation}%
where $x^{\left( 0\right) }=y^{\left( 0\right) }=v$ and $p^{\left( 0\right)
}=q^{\left( 0\right) }=\mathbf{0}_{n}$. This algorithm is obviously related to
the Douglas-Rachford splitting framework\footnote{See Douglas and Rachford
(1956), Combettes and Pesquet (2011), and Lindstrom and Sims (2018).} where
$x^{\left( k\right) }$ and $p^{\left( k\right) }$ are the variable and the
residual associated to $f_{1}\left( x\right) $, and $y^{\left( k\right) }$ and
$q^{\left( k\right) }$ are the variable and the residual associated to
$f_{2}\left( x\right) $. Algorithm (\ref{eq:douglas1}) can be reformulated by
introducing the intermediary step $k+\frac{1}{2}$:
\begin{equation}
\left\{
\begin{array}{l}
x^{\left( k+\frac{1}{2}\right) }=\mathbf{prox}_{f_{1}}\left( x^{\left(
k\right) }+p^{\left( k\right) }\right)  \\
p^{\left( k+1\right) }=p^{\left( k\right) }-\Delta _{1/2}x^{\left( k+\frac{1%
}{2}\right) } \\
x^{\left( k+1\right) }=\mathbf{prox}_{f_{2}}\left( x^{\left( k+\frac{1}{2}%
\right) }+q^{\left( k\right) }\right)  \\
q^{\left( k+1\right) }=q^{\left( k\right) }-\Delta _{1/2}x^{\left(
k+1\right) }%
\end{array}%
\right.   \label{eq:douglas2}
\end{equation}%
where $\Delta _{h}x^{\left( k\right) }=x^{\left( k\right) }-x^{\left(
k-h\right) }$. Figure \ref{fig:dykstra} illustrates the splitting method used
by the Dykstra's algorithm and clearly shows the relationship with the
Douglas-Rachford algorithm.

\subsubsection{The $m>2$ case}

The case $m>2$ is a generalization of the previous algorithm by considering $m$
residuals:
\begin{enumerate}
\item The $x$-update is:%
\begin{equation*}
x^{\left( k+1\right) }=\mathbf{prox}_{f_{j\left( k\right) }}\left( x^{\left(
k\right) }+z^{\left( k+1-m\right) }\right)
\end{equation*}

\item The $z$-update is:%
\begin{equation*}
z^{\left( k+1\right) }=x^{\left( k\right) }+z^{\left( k+1-m\right)
}-x^{\left( k+1\right) }
\end{equation*}
\end{enumerate}
where $x^{\left( 0\right) }=v$, $z^{\left( k\right) }=\mathbf{0}_{n}$ for $k<0$
and $j\left( k\right) =\limfunc{mod}\left( k+1,m\right) $ denotes the modulo
operator taking values in $\left\{ 1,\ldots ,m\right\} $. The variable
$x^{\left( k\right) }$ is updated at each iteration while the residual
$z^{\left( k\right) }$ is updated every $m$ iterations. This implies that the
basic function $f_{j}\left( x\right) $ is related to the residuals $z^{\left(
j\right) }$, $z^{\left( j+m\right) }$, $z^{\left( j+2m\right) }$, etc.
Following Tibshirani (2017), it is better to write the Dykstra's algorithm by
using two iteration indices $k$ and $j$. The main index $k$ refers to the
cycle\footnote{Exactly like the coordinate descent algorithm.}, whereas the
sub-index $j$ refers to the constraint number:
\begin{enumerate}
\item The $x$-update is:%
\begin{equation}
x^{\left( k+1,j\right) }=\mathbf{prox}_{f_{j}}\left( x^{\left(
k+1,j-1\right) }+z^{\left( k,j\right) }\right)   \label{eq:Dykstra1}
\end{equation}

\item The $z$-update is:%
\begin{equation}
z^{\left( k+1,j\right) }=x^{\left( k+1,j-1\right) }+z^{\left( k,j\right)
}-x^{\left( k+1,j\right) }  \label{eq:Dykstra2}
\end{equation}
\end{enumerate}
where $x^{\left( 1,0\right) }=v$, $z^{\left( k,j\right) }=\mathbf{0}_{n}$ for
$k=0$ and $x^{\left( k+1,0\right) }=x^{\left( k,m\right) }$.\smallskip

The Dykstra's algorithm is particularly efficient when we consider the
projection problem:%
\begin{equation*}
x^{\star }=\mathcal{P}_{\Omega }\left( v\right)
\end{equation*}%
where:%
\begin{equation*}
\Omega =\Omega _{1}\cap \Omega _{2}\cap \cdots \cap \Omega _{m}
\end{equation*}%
Indeed, the solution is found by replacing Equation (\ref{eq:Dykstra1}) with:%
\begin{equation}
x^{\left( k+1,j\right) }=\mathcal{P}_{\Omega_j }\left( x^{\left(
k+1,j-1\right) }+z^{\left( k,j\right) }\right)   \label{eq:Dykstra3}
\end{equation}

\subsubsection{Application to general linear constraints}

Let us consider the case $\Omega =\left\{ x\in \mathbb{R}^{n}:Cx\leq D\right\}
$ where the number of inequality constraints is equal to $m$. We can write:
\begin{equation*}
\Omega =\Omega _{1}\cap \Omega _{2}\cap \cdots \cap \Omega _{m}
\end{equation*}%
where $\Omega _{j}=\left\{ x\in \mathbb{R}^{n}:c_{\left( j\right) }^{\top
}x\leq d_{\left( j\right) }\right\} $, $c_{\left( j\right) }^{\top }$
corresponds to the $j^{\mathrm{th}}$ row of $C$ and $d_{\left( j\right) }$
is the $j^{\mathrm{th}}$ element of $D$. Since the projection $\mathcal{P}%
_{\Omega _{j}}$ is known and has been given on page
\pageref{tab:basic-projection}, we can find the projection
$\mathcal{P}_{\Omega}$ using Algorithm (\ref{alg:dykstra1}).

\begin{algorithm}[tbph]
\begin{algorithmic}
\STATE  The goal is to compute the solution $x^{\star }=\mathbf{prox}_{f}\left(
v\right) $ where $f\left( x\right) =\mathds{1}_{\Omega }\left( x\right) $ and $\Omega =\left\{ x\in \mathbb{R}^{n}:Cx\leq D\right\} $
\STATE  We initialize $x^{\left( 0,m\right) }\leftarrow v$
\STATE  We set $z^{\left( 0,1\right) }\leftarrow \mathbf{0}_n, \ldots, z^{\left( 0,m\right)
}\leftarrow \mathbf{0}_n$
\STATE $k \leftarrow 0$
\REPEAT
    \STATE $x^{\left( k+1,0\right) }\leftarrow x^{\left( k,m\right) }$
    \FOR {$j=1:m$}
        \STATE  The $x$-update is:%
            \begin{equation*}
                x^{\left( k+1,j\right) } = x^{\left( k+1,j-1\right) }+z^{\left( k,j\right) }-\frac{\left( c_{\left(
                j\right) }^{\top }x^{\left( k+1;j-1\right) }+c_{\left( j\right) }^{\top
                }z^{\left( k,j\right) }-d_{\left( j\right) }\right) _{+}}{\left\Vert
                c_{\left( j\right) }\right\Vert _{2}^{2}}c_{\left( j\right) }
            \end{equation*}
        \STATE The $z$-update is:%
            \begin{equation*}
                z^{\left( k+1,j\right) }=x^{\left( k+1,j-1\right) }+z^{\left( k,j\right)
                }-x^{\left( k+1,j\right) }
            \end{equation*}
    \ENDFOR
    \STATE $k \leftarrow k+1$
\UNTIL{Convergence}
\RETURN $x^{\star }\leftarrow x^{\left( k,m\right) }$
\end{algorithmic}
\caption{Dykstra's algorithm for solving the proximal problem with linear inequality constraints}
\label{alg:dykstra1}
\end{algorithm}

If we define $\Omega $ as follows:%
\begin{equation*}
\Omega =\left\{ x\in \mathbb{R}^{n}:Ax=B,Cx\leq D,x^{-}\leq x\leq
x^{+}\right\}
\end{equation*}%
we decompose $\Omega $ as the intersection of three basic convex sets:%
\begin{equation*}
\Omega =\Omega _{1}\cap \Omega _{2}\cap \Omega _{3}
\end{equation*}%
where $\Omega _{1}=\left\{ x\in \mathbb{R}^{n}:Ax=B\right\} $, $\Omega
_{2}=\left\{ x\in \mathbb{R}^{n}:Cx\leq D\right\} $ and $\Omega _{3}=\left\{
x\in \mathbb{R}^{n}:x^{-}\leq x\leq x^{+}\right\} $. Using Dykstra's
algorithm is equivalent to formulating Algorithm (\ref{alg:dykstra2}).

\begin{algorithm}[tbph]
\begin{algorithmic}
\STATE  The goal is to compute the solution $x^{\star }=\mathbf{prox}%
_{f}\left( v\right) $ where $f\left( x\right) =\mathds{1}_{\Omega }\left(
x\right) $ and $\Omega =\left\{ x\in \mathbb{R}^{n}:Ax=B,Cx\leq D,x^{-}\leq
x\leq x^{+}\right\} $
\STATE  We initialize $x_{m}^{\left( 0\right) }\leftarrow v$
\STATE  We set $z_{1}^{\left( 0\right) }\leftarrow \mathbf{0}_n$, $z_{2}^{\left( 0\right) }\leftarrow \mathbf{0}_n$ and
        $z_{3}^{\left( 0\right) }\leftarrow \mathbf{0}_n$
\STATE $k \leftarrow 0$
\REPEAT
    \STATE $x_{0}^{\left( k+1\right) }\leftarrow x_{m}^{\left( k\right) }$
    \STATE $x_{1}^{\left( k+1\right) }\leftarrow x_{0}^{\left( k+1\right)
    }+z_{1}^{\left( k\right) }-A^{\dag }\left( Ax_{0}^{\left( k+1\right)
    }+Az_{1}^{\left( k\right) }-B\right) $
    \STATE $z_{1}^{\left( k+1\right) }\leftarrow x_{0}^{\left( k+1\right)
    }+z_{1}^{\left( k\right) }-x_{1}^{\left( k+1\right) }$
    \STATE $x_{2}^{\left( k+1\right) }\leftarrow \mathcal{P}_{\Omega _{2}}\left(
    x_{1}^{\left( k+1\right) }+z_{2}^{\left( k\right) }\right) $ \hspace{4cm} $\blacktriangleright$ Algorithm (\ref{alg:dykstra1})
    \STATE $z_{2}^{\left( k+1\right) }\leftarrow x_{1}^{\left( k+1\right)
    }+z_{2}^{\left( k\right) }-x_{2}^{\left( k+1\right) }$
    \STATE $x_{3}^{\left( k+1\right) }\leftarrow \mathcal{T}\left( x_{2}^{\left(
    k+1\right) }+z_{3}^{\left( k\right) };x^{-},x^{+}\right) $
    \STATE $z_{3}^{\left( k+1\right) }\leftarrow x_{2}^{\left( k+1\right)
    }+z_{3}^{\left( k\right) }-x_{3}^{\left( k+1\right) }$
    \STATE $k \leftarrow k+1$
\UNTIL{Convergence}
\RETURN $x^{\star }\leftarrow x_{3}^{\left( k\right) }$
\end{algorithmic}
\caption{Dykstra's algorithm for solving the proximal problem with general linear constraints}
\label{alg:dykstra2}
\end{algorithm}

Since we have:%
\begin{equation*}
\frac{1}{2}\left\Vert x-v\right\Vert _{2}^{2}=\frac{1}{2}x^{\top }x-x^{\top
}v+\frac{1}{2}v^{\top }v
\end{equation*}%
we deduce that the two previous problems can be cast into a QP problem:
\begin{eqnarray*}
x^{\star } &=&\arg \min_x \frac{1}{2}x^{\top }I_{n}x-x^{\top }v \\
&\text{s.t.}&x\in \Omega
\end{eqnarray*}%
We can then compare the efficiency of Dykstra's algorithm with the QP
algorithm. Let us consider the proximal problem where the vector $v$ is defined
by the elements $v_{i}=\ln \left( 1+i^{2}\right) $ and the set of constraints
is:
\begin{equation*}
\Omega =\left\{ x\in \mathbb{R}^{n}:\sum_{i=1}^{n}x_{i}\leq \frac{1}{2}%
,\sum_{i=1}^{n}e^{-i}x_{i}\geq 0\right\}
\end{equation*}%
Using a Matlab implementation\footnote{%
The QP implementation corresponds to the \texttt{quadprog} function.}, we find
that the computational time of the Dykstra's algorithm when $n$ is equal to
$10$ million is equal to the QP algorithm when $n$ is equal to $12\,500$,
meaning that there is a factor of $800$ between the two methods!

\subsubsection{Application to the $\boldsymbol{\ell }_{2}$-penalized logarithmic barrier function}

We consider the following proximal problem:
\begin{eqnarray*}
x^{\star } &=&\arg \min_{x}-\lambda \sum_{i=1}^{n}b_{i}\ln x_{i}+\frac{1}{2}%
\left\Vert x-v\right\Vert _{2}^{2} \\
&\text{s.t.}&\left\Vert x-c\right\Vert _{2}\leq r
\end{eqnarray*}%
In Appendix \ref{appendix:proximal-penalized-logarithmic} on page
\pageref{appendix:proximal-penalized-logarithmic}, we show that the
corresponding Dykstra's algorithm is:
\begin{equation*}
\left\{
\begin{array}{l}
x^{\left( k+1\right) }=\dfrac{y^{\left( k\right) }+z_{1}^{\left( k\right) }+%
\sqrt{\left( y^{\left( k\right) }+z_{1}^{\left( k\right) }\right) \odot
\left( y^{\left( k\right) }+z_{1}^{\left( k\right) }\right) +4\lambda b}}{2}
\\
z_{1}^{\left( k+1\right) }=y^{\left( k\right) }+z_{1}^{\left( k\right)
}-x^{\left( k+1\right) } \\
y^{\left( k+1\right) }=c+\dfrac{r}{\max \left( r,\left\Vert x^{\left(
k+1\right) }+z_{2}^{\left( k\right) }-c\right\Vert _{2}\right) }\left(
x^{\left( k+1\right) }+z_{2}^{\left( k\right) }-c\right)  \\
z_{2}^{\left( k+1\right) }=x^{\left( k+1\right) }+z_{2}^{\left( k\right)
}-y^{\left( k+1\right) }%
\end{array}%
\right.
\end{equation*}

\section{Applications to portfolio optimization}

The development of the previous algorithms will fundamentally change the
practice of portfolio optimization. Until now, we have seen that portfolio
managers live in a quadratic programming world. With these new optimization
algorithms, we can consider more complex portfolio optimization programs with
non-quadratic objective function, regularization with penalty functions and
non-linear constraints.

\begin{table}[tbph]
\centering
\caption{Some objective functions used in portfolio optimization}
\label{tab:portfolio1}
\tableskip
\begin{tabular}{cccc}
\hline
Item & Portfolio & $f\left( x\right) $                                                           & Reference                         \\ \hline
(1) & MVO  & $\frac{1}{2}x^{\top }\Sigma x-\gamma x^{\top }\mu $                                 & Markowitz (1952)                  \\
(2) & GMV  & $\frac{1}{2}x^{\top }\Sigma x$                                                      & Jagganathan and Ma (2003)         \\
(3) & MDP  & $\ln \left( \sqrt{x^{\top }\Sigma x}\right) -\ln \left( x^{\top}\sigma \right) $    & Choueifaty and Coignard (2008)    \\
(4) & KL   & $\sum_{i=1}^{n}x_{i}\ln \left( x_{i}/\tilde{x}_{i}\right) $                         & Bera and Park (2008)              \\
(5) & ERC  & $\frac{1}{2}x^{\top }\Sigma x-\lambda \sum_{i=1}^{n}\ln x_{i}$                      & Maillard \textsl{et al.} (2010)   \\
(6) & RB   & $\mathcal{R}\left( x\right) -\lambda \sum_{i=1}^{n}\mathcal{RB}_{i}\cdot \ln x_{i}$ & Roncalli (2015)                   \\
(7) & RQE  & $\frac{1}{2}x^{\top }Dx$                                                            & Carmichael \textsl{et al.} (2018) \\
\hline
\end{tabular}
\end{table}

We consider a universe of $n$ assets. Let $x$ be the vector of weights in the
portfolio. We denote by $\mu $ and $\Sigma $ the vector of expected returns and
the covariance matrix of asset returns\footnote{The vector of volatilities is
defined by $\sigma =\left( \sigma _{1},\ldots ,\sigma _{n}\right) $.}. Some
models consider also a reference portfolio $\tilde{x}$. In Table
\ref{tab:portfolio1}, we report the main objective functions that are used by
professionals\footnote{For each model, we write the optimization
problem as a minimization problem.}. Besides the mean-variance optimized
portfolio (MVO) and the global minimum variance portfolio (GMV), we find the
equal risk contribution portfolio (ERC), the risk budgeting portfolio (RB) and
the most diversified portfolio (MDP). According to Choueifaty and Coignard
(2008), the MDP is defined as the portfolio which maximizes the diversification
ratio $\mathcal{DR}\left( x\right) =\dfrac{x^{\top }\sigma }{\sqrt{x^{\top
}\Sigma x}}$. We also include in the list two \textquoteleft
\textit{academic}\textquoteright\ portfolios, which are based on the
Kullback-Leibler (KL) information criteria and the Rao's quadratic entropy
(RQE) measure\footnote{$D$ is the dissimilarity matrix satisfying $D_{i,j}\geq
0$ and $D_{i,j}=D_{j,i}$.}.

\begin{table}[tbph]
\centering
\caption{Some regularization penalties used in portfolio optimization}
\label{tab:portfolio2}
\tableskip
\begin{tabular}{cccc}
\hline
Item & Regularization & $\mathfrak{R}\left( x\right) $ & Reference \\ \hline
(8)  & Ridge & $\lambda \left\Vert x-\tilde{x}\right\Vert _{2}^{2}$ & DeMiguel \textsl{et al.} (2009) \\
(9)  & Lasso & $\lambda \left\Vert x-\tilde{x}\right\Vert _{1}$ & Brodie \textsl{at al.} (2009) \\
(10) & Log-barrier & $-\sum_{i=1}^{n}\lambda _{i}\ln x_{i}$ & Roncalli (2013) \\
(11) & Shannon's entropy & $\lambda \sum_{i=1}^{n}x_{i}\ln x_{i}$ & Yu \textsl{et al.} (2014) \\
\hline
\end{tabular}
\end{table}

In a similar way, we list in Table \ref{tab:portfolio2} some popular
regularization penalty functions that are used in the industry (Bruder
\textsl{et al.}, 2013; Bourgeron \textsl{et al.}, 2018). The ridge and lasso
regularization are well-known in statistics and machine learning (Hastie
\textsl{et al.}, 2009). The log-barrier penalty function comes from the risk
budgeting optimization problem, whereas Shannon's entropy is another
approach for imposing a sufficient weight diversification.

\begin{table}[tbph]
\centering
\caption{Some constraints used in portfolio optimization}
\label{tab:portfolio3}
\tableskip
\begin{tabular}{ccc}
\hline
(12) & No cash and leverage  & $\sum_{i=1}^{n}x_{i}=1$ \\
(13) & No short selling      & $x_{i}\geq 0$ \\
(14) & Weight bounds         & $x_{i}^{-}\leq x_{i}\leq x_{i}^{+}$ \\
(15) & Asset class limits    & $c_{j}^{-}\leq \sum_{i\in \mathcal{C}_{j}}x_{i}\leq c_{j}^{+}$ \\
(16) & Turnover              & $\sum_{i=1}^{n}\left\vert x_{i}-\tilde{x}_{i}\right\vert \leq  \turnover^{+} $ \\
(17) & Transaction costs     & $\sum_{i=1}^{n}\left( c_{i}^{-}\left( \tilde{x}_{i}-x_{i}\right)_{+}+c_{i}^{+}\left( x_{i}-\tilde{x}_{i}\right) _{+}\right) \leq  \cost^{+}$ \\
(18) & Leverage limit        & $\sum_{i=1}^{n}\left\vert x_{i}\right\vert \leq \mathcal{L}^{+}$ \\
(19) & Long/short exposure   & $-\mathcal{LS}^{-}\leq \sum_{i=1}^{n}x_{i} \leq \mathcal{LS}^{+}$ \\
(20) & Benchmarking          & $\sqrt{\left( x-\tilde{x}\right) ^{\top }\Sigma \left( x-\tilde{x}\right) }\leq \sigma^{+}$ \\
(21) & Tracking error floor  & $\sqrt{\left( x-\tilde{x}\right) ^{\top }\Sigma \left( x-\tilde{x}\right) }\geq \sigma^{-}$ \\
(22) & Active share floor    & $\frac{1}{2}\sum_{i=1}^{n}\left\vert x_{i}-\tilde{x}_{i}\right\vert \geq \mathcal{AS}^{-}$ \\
(23) & Number of active bets & $\left( x^{\top }x\right) ^{-1}\geq \mathcal{N}^{-}$ \\
\hline
\end{tabular}
\end{table}

Concerning the constraints, the most famous are the no cash/no leverage and no
short selling restrictions. Weight bounds and asset class limits are also
extensively used by practitioners. Turnover and transaction cost
management may be an important topic when rebalancing a current portfolio
$\tilde{x}$. When managing long/short portfolios, we generally impose leverage
or long/short exposure limits. In the case of a benchmarked strategy, we might
also want to have a tracking error limit with respect to the benchmark
$\tilde{x}$. On the contrary, we can impose a minimum tracking error or active
share in the case of active management. Finally, the Herfindahl constraint is
used for some smart beta portfolios.\smallskip

In what follows, we consider several portfolio optimization problems. Most of
them are a combination of an objective function, one or two regularization
penalty functions and some constraints that have been listed above. From an
industrial point of view, it is interesting to implement the proximal operator
for each item. In this approach, solving any portfolio optimization problem is
equivalent to using CCD, ADMM, Dykstra and the appropriate proximal functions
as Lego bricks.

\subsection{Minimum variance optimization}

\subsubsection{Managing diversification}
\label{section:gmv-diversification}

The global minimum variance (GMV) portfolio corresponds to the following
optimization program:%
\begin{eqnarray*}
x^{\star } &=&\arg \min_x \frac{1}{2}x^{\top }\Sigma x \\
&\text{s.t.}&\mathbf{1}_{n}^{\top }x=1
\end{eqnarray*}%
We know that the solution is $x^{\star }=\left( \mathbf{1}_{n}^{\top }\Sigma
^{-1}\mathbf{1}_{n}\right) ^{-1}\Sigma ^{-1}\mathbf{1}_{n}$. In practice,
nobody implements the GMV portfolio because it is a long/short portfolio and it
is not robust. Most of the time, professionals impose weight bounds: $0\leq
x_{i}\leq x^{+}$. However, this approach generally leads to corner solutions,
meaning that a large number of optimized weights are equal to zero or the upper
bound and very few assets have a weight within the range. With the emergence of
smart beta portfolios, the minimum variance portfolio gained popularity among
institutional investors. For instance, we can find many passive indices based
on this framework. In order to increase the robustness of these portfolios, the
first generation of minimum variance strategies has used relative
weight bounds with respect to a benchmark $b$:%
\begin{equation}
\delta^{-}b_{i}\leq x_{i}\leq \delta^{+}b_{i}  \label{eq:gmv1}
\end{equation}%
where $0<\delta^{-} < 1$ and $\delta^{+} > 1$. For instance, the most popular
scheme is to take $\delta^{-}=0.5$ and  $\delta^{+}=2$. Nevertheless, the
constraint (\ref{eq:gmv1}) produces the illusion that the portfolio is
diversified, because the optimized weights are different. In fact, portfolio
weights are different because benchmark weights are different. The second
generation of minimum variance strategies imposes a global diversification
constraint. The most popular solution is based on the Herfindahl index
$\mathcal{H}\left( x\right) =\sum_{i=1}^{n}x_{i}^{2}$. This index takes the
value 1 for a pure concentrated portfolio ($\exists\, i:x_{i}=1 $) and $1/n$
for an equally-weighted portfolio. Therefore, we can define the number of
effective bets as the inverse of the Herfindahl index (Meucci, 2009):
$\mathcal{N}\left( x\right) =\mathcal{H}\left( x\right) ^{-1}$. The
optimization program becomes:
\begin{eqnarray}
x^{\star } &=&\arg \min_x \frac{1}{2}x^{\top }\Sigma x \label{eq:gmv2} \\
&\text{s.t.}&\left\{
\begin{array}{l}
\mathbf{1}_{n}^{\top }x=1 \\
\mathbf{0}_n\leq x\leq x^{+} \\
\mathcal{N}\left( x\right) \geq \mathcal{N}^{-}%
\end{array}%
\right. \notag
\end{eqnarray}%
where $\mathcal{N}^{-}$ is the minimum number of effective bets.\smallskip

The Herfindhal constraint is equivalent to:
\begin{eqnarray*}
\mathcal{N}\left( x\right) \geq \mathcal{N}^{-} &\Leftrightarrow &\left(
x^{\top }x\right) ^{-1}\geq \mathcal{N}^{-} \\
&\Leftrightarrow &x^{\top }x\leq \frac{1}{\mathcal{N}^{-}}
\end{eqnarray*}%
Therefore, a first solution to solve (\ref{eq:gmv2}) is to consider the
following QP problem\footnote{The objective function can be written as:
\begin{equation*}
\frac{1}{2}x^{\top }\Sigma x+\lambda x^{\top }x=\frac{1}{2}x^{\top }\left(
\Sigma +2I_{n}\right) x
\end{equation*}%
}:
\begin{eqnarray}
x^{\star }\left( \lambda \right)  &=&\arg \min_x \frac{1}{2}x^{\top }\Sigma
x+\lambda x^{\top }x  \label{eq:gmv3} \\
&\text{s.t.}&\left\{
\begin{array}{l}
\mathbf{1}_{n}^{\top }x=1 \\
\mathbf{0}_n\leq x\leq x^{+}%
\end{array}%
\right.   \notag
\end{eqnarray}%
where $\lambda \geq 0$ is a scalar. Since $\mathcal{N}\left( x^{\star }\left(
\infty \right) \right) $ is equal to the number $n$ of assets and
$\mathcal{N}\left( x^{\star }\left( \lambda \right) \right) $ is an increasing
function of $\lambda $, Problem (\ref{eq:gmv3}) has a unique solution if
$\mathcal{N}^{-}\in \left[ \mathcal{N}\left( x^{\star }\left( 0\right) \right)
,n\right] $. There is an optimal value $\lambda ^{\star }$ such that for each
$\lambda \geq \lambda ^{\star }$, we have $\mathcal{N}\left( x^{\star }\left(
\lambda \right) \right) \geq \mathcal{N}^{-}$. Computing the optimal portfolio
$x^{\star }\left( \lambda ^{\star }\right) $ therefore implies finding the solution
$\lambda ^{\star }$ of the non-linear equation\footnote{We generally use
the bisection algorithm to determine the optimal solution $\lambda ^{\star
}$.} $\mathcal{N}\left( x^{\star }\left( \lambda \right) \right)
=\mathcal{N}^{-}$.\smallskip

A second method is to consider the ADMM form:
\begin{eqnarray*}
\left\{ x^{\star },y^{\star }\right\}  &=&\arg \min_{\left(x,y\right)} \frac{1}{2}x^{\top
}\Sigma x+\mathbf{\mathds{1}}_{\Omega _{1}}\left( x\right) +\mathbf{%
\mathds{1}}_{\Omega _{2}}\left( y\right)  \\
&\text{s.t.}&x=y
\end{eqnarray*}%
where $\Omega _{1}=\left\{ x\in \mathbb{R}^{n}:\mathbf{1}_{n}^{\top }x=1,\mathbf{0}_n\leq
x\leq x^{+}\right\} $ and $\Omega _{2}=\mathcal{B}_{2}\left(
\mathbf{0}_n,\sqrt{\frac{1}{\mathcal{N}^{-}}}\right) $. We deduce that the $x$-update
is a QP problem:
\begin{equation*}
x^{\left( k+1\right) }=\arg \min_{x}\left\{ \frac{1}{2}x^{\top }\left(
\Sigma +\varphi I_{n}\right) x-\varphi x^{\top }\left( y^{\left( k\right)
}-u^{\left( k\right) }\right) +\mathbf{\mathds{1}}_{\Omega_1 }\left( x\right)
\right\}
\end{equation*}%
whereas the $y$-update is:
\begin{equation*}
y^{\left( k+1\right) }=\frac{x^{\left( k+1\right) }+u^{\left( k\right) }}{%
\max \left( 1,\sqrt{\mathcal{N}^{-}}\left\Vert x^{\left( k+1\right)
}+u^{\left( k\right) }\right\Vert _{2}\right) }
\end{equation*}%
A better approach is to write the problem as follows:%
\begin{eqnarray*}
\left\{ x^{\star },y^{\star }\right\}  &=&\arg \min_{\left(x,y\right)} \frac{1}{2}x^{\top
}\Sigma x+\mathbf{\mathds{1}}_{\Omega _{3}}\left( x\right) +\mathbf{%
\mathds{1}}_{\Omega _{4}}\left( y\right)  \\
&\text{s.t.}&x=y
\end{eqnarray*}%
where $\Omega _{3}=\mathcal{H}_{yperplane}\left[ \mathbf{1}_{n},1\right] $ and
$\Omega _{4}=\mathcal{B}_{ox}\left[ \mathbf{0}_{n},x^{+}\right] \cap \mathcal{
B}_{2}\left( \mathbf{0}_{n},\sqrt{\frac{1}{\mathcal{N}^{-}}}\right) $. In this
case, the $x$- and $y$-updates become\footnote{See Appendix
\ref{appendix:qp-hyperplane} on page \pageref{appendix:qp-hyperplane} for the
derivation of the $x$-update.}:
\begin{eqnarray*}
x^{\left( k+1\right) } &=&\arg \min_{x}\left\{ \frac{1}{2}x^{\top }\left(
\Sigma +\varphi I_{n}\right) x-\varphi x^{\top }\left( y^{\left( k\right)
}-u^{\left( k\right) }\right) +\mathbf{\mathds{1}}_{\Omega _{3}}\left(
x\right) \right\}  \\
&=&\left( \Sigma +\varphi I_{n}\right) ^{-1}\left( \varphi \left( y^{\left(
k\right) }-u^{\left( k\right) }\right) +\frac{1-\mathbf{1}_{n}^{\top }\left(
\Sigma +\varphi I_{n}\right) ^{-1}\varphi \left( y^{\left( k\right)
}-u^{\left( k\right) }\right) }{\mathbf{1}_{n}^{\top }\left( \Sigma +\varphi
I_{n}\right) ^{-1}\mathbf{1}_{n}}\mathbf{1}_{n}\right)
\end{eqnarray*}%
and:%
\begin{equation*}
y^{\left( k+1\right) }=\mathcal{P}_{\mathcal{B}\mathrm{ox}-\mathcal{B}%
\mathrm{all}}\left( x^{\left( k+1\right) }+u^{\left( k\right) };\mathbf{0}%
_{n},x^{+},\mathbf{0}_{n},\sqrt{\frac{1}{\mathcal{N}^{-}}}\right)
\end{equation*}%
where $\mathcal{P}_{\mathcal{B}\mathrm{ox}-\mathcal{B}\mathrm{all}}$
corresponds to the Dykstra's algorithm given in Appendix
\ref{appendix:proximal-ball-box} on page
\pageref{appendix:proximal-ball-box}.\smallskip

We consider the parameter set \#1 given in Appendix \ref{appendix:data1} on
page \pageref{appendix:data1}. The investment universe is made up of eight
stocks. We would like to build a diversified minimum variance long-only
portfolio without imposing an upper weight bound\footnote{This means that
$x^{+}$ is set to $\mathbf{1}_n$.}. In Table \ref{tab:herfindahl1}, we report
the solutions found by the ADMM algorithm for several values of
$\mathcal{N}^{-}$. When there is no Herfindahl constraint, the portfolio is
fully invested in the seventh stock, meaning that the asset diversification is
very poor. Then we increase the number of effective bets. If $\mathcal{N}^{-}$
is equal to the number $n$ of stocks, we verify that the solution corresponds
to the equally-weighted portfolio. Between these two limit cases, we see the
impact of the Herfindahl constraint on the portfolio diversification. The
parameter set \#1 is defined with respect to a capitalization-weighted index,
whose weights are equal to $23\%$, $19\%$, $17\%$, $9\%$, $8\%$, $6\%$ and
$5\%$. The number of effective bets of this benchmark is equal to $6.435$. If
we impose that the effective number of bets of the minimum variance portfolio
is at least equal to the effective number of bets of the benchmark, we find the
following solution: $14.74\%$, $15.45\%$, $1.79\%$, $15.49\%$, $6.17\%$,
$13.83\%$, $23.21\%$ and $9.31\%$.\smallskip

\begin{table}[tbh]
\centering
\caption{Minimum variance portfolios (in \%)}
\label{tab:herfindahl1}
\tableskip
\begin{tabular}{crrrrrrrrrrr}
\hline
$\mathcal{N}^{-}$    & $  1.00$ & $ 2.00$ & $ 3.00$ & $ 4.00$ & $ 5.00$ & $ 6.00$ & $ 6.50$ & $ 7.00$ & $ 7.50$ &  $ 8.00$ \\ \hline
$x_1^{\star }$       & $  0.00$ & $ 3.22$ & $ 9.60$ & $13.83$ & $15.18$ & $15.05$ & $14.69$ & $14.27$ & $13.75$ &  $12.50$ \\
$x_2^{\star }$       & $  0.00$ & $12.75$ & $14.14$ & $15.85$ & $16.19$ & $15.89$ & $15.39$ & $14.82$ & $14.13$ &  $12.50$ \\
$x_3^{\star }$       & $  0.00$ & $ 0.00$ & $ 0.00$ & $ 0.00$ & $ 0.00$ & $ 0.07$ & $ 2.05$ & $ 4.21$ & $ 6.79$ &  $12.50$ \\
$x_4^{\star }$       & $  0.00$ & $10.13$ & $15.01$ & $17.38$ & $17.21$ & $16.09$ & $15.40$ & $14.72$ & $13.97$ &  $12.50$ \\
$x_5^{\star }$       & $  0.00$ & $ 0.00$ & $ 0.00$ & $ 0.00$ & $ 0.71$ & $ 5.10$ & $ 6.33$ & $ 7.64$ & $ 9.17$ &  $12.50$ \\
$x_6^{\star }$       & $  0.00$ & $ 5.36$ & $ 8.95$ & $12.42$ & $13.68$ & $14.01$ & $13.80$ & $13.56$ & $13.25$ &  $12.50$ \\
$x_7^{\star }$       & $100.00$ & $68.53$ & $52.31$ & $40.01$ & $31.52$ & $25.13$ & $22.92$ & $20.63$ & $18.00$ &  $12.50$ \\
$x_8^{\star }$       & $  0.00$ & $ 0.00$ & $ 0.00$ & $ 0.50$ & $ 5.51$ & $ 8.66$ & $ 9.41$ & $10.14$ & $10.95$ &  $12.50$ \\ \hline
$\lambda ^{\star }$ (in \%) & $  0.00$ & $ 1.59$ & $ 3.10$ & $ 5.90$ & $10.38$ & $18.31$ & $23.45$ & $31.73$ & $49.79$ & $\infty$ \\ \hline
\end{tabular}
\end{table}

As explained before, we can also solve the optimization problem by combining
Problem (\ref{eq:gmv3}) and the bisection algorithm. This is why we have
reported the corresponding value $\lambda ^{\star }$ in the last row in Table
\ref{tab:herfindahl1}. However, this approach is no longer valid if we consider
diversification constraints that are not quadratic. For instance, let us
consider the generalized minimum variance problem:
\begin{eqnarray}
x^{\star } &=&\arg \min_{x}\frac{1}{2}x^{\top }\Sigma x  \label{eq:gmv4} \\
&\text{s.t.}&\left\{
\begin{array}{l}
\mathbf{1}_{n}^{\top }x=1 \\
\mathbf{0}_{n}\leq x\leq x^{+} \\
\mathcal{D}\left( x\right) \geq \mathcal{D}^{-}%
\end{array}%
\right.   \notag
\end{eqnarray}%
where $\mathcal{D}\left( x\right) $ is a weight diversification measure and
$\mathcal{D}^{-}$ is the minimum acceptable diversification. For example, we
can use Shannon's entropy, the Gini index or the diversification ratio. In
this case, it is not possible to obtain an equivalent QP problem, whereas the
ADMM algorithm is exactly the same as previously except for the
$y$-update\label{eq:admm-gmv-trick}:
\begin{equation*}
y^{\left( k+1\right) }=\mathcal{P}_{\mathcal{B}_{ox}\left[ \mathbf{0}_{n},x^{+}\right]\, \cap\, \mathfrak{D}}\left( x^{\left(
k+1\right) }+u^{\left( k\right) }\right)
\end{equation*}%
where $\mathfrak{D} =\left\{ x\in \mathbb{R}^{n}:\mathcal{D}\left( x\right)
\geq \mathcal{D}^{-}\right\} $. The projection onto $\mathfrak{D} $ can be
easily derived from the proximal operator of the dual function (see the
\textit{tips and tricks} on page \pageref{section:tips-and-tricks}).

\begin{remark}
If we compare the computational times, we observe that the best method is the
second version of the ADMM algorithm. In our example, the computational time is
divided by a factor of eight with respect to the bisection approach%
\footnote{In contrast, the first version of the ADMM algorithm is not
efficient since the computational time is multiply by a factor of five with
respect to the bisection approach.}. If we consider a large-scale problem with
$n$ larger than $1\,000$, the computational time is generally divided by a
factor greater than $50$!
\end{remark}

\subsubsection{Managing the portfolio rebalancing process}

Another big challenge of the minimum variance portfolio is the management of
the turnover between two rebalancing dates. Let $x_{t}$ be the current
portfolio. The optimization program for calibrating the optimal solution
$x_{t+1}$ for the next rebalancing date $t+1$ may include a penalty function
$\cost\left( x\mid x_{t}\right) $ and/or a weight constraint
$\mathfrak{C}\left( x\mid x_{t}\right) $ that are parameterized with respect to
the current portfolio $x_{t}$:
\begin{eqnarray}
x_{t+1} &=&\arg \min_{x}\frac{1}{2}x^{\top }\Sigma x+\cost\left( x\mid
x_{t}\right)  \label{eq:gmv5} \\
&\text{s.t.}&\left\{
\begin{array}{l}
\mathbf{1}_{n}^{\top }x=1 \\
\mathbf{0}_{n}\leq x\leq x^{+} \\
x\in \mathfrak{C}\left( x\mid x_{t}\right)%
\end{array}%
\right.  \notag
\end{eqnarray}%
Again, we can solve this problem using the ADMM algorithm. Thanks to the
Dykstra's algorithm, the only difficulty is finding the proximal operator of
$\cost\left( x\mid x_{t}\right) $ or $\mathfrak{C}\left( x\mid x_{t}\right) $
when performing the $y$-update.\smallskip

Let us define the cost function as:
\begin{equation*}
\cost\left( x\mid x_{t}\right) =\lambda \sum_{i=1}^{n}\left( c_{i}^{-}\left(
x_{i,t}-x_i\right) _{+}+c_{i}^{+}\left( x_i - x_{i,t}\right) _{+}\right)
\end{equation*}%
where $c_{i}^{-}$ and $c_{i}^{+}$ are the bid and ask transaction costs. In
Appendix \ref{appendix:linear-cost} on page \pageref{appendix:linear-cost}, we
show that the proximal operator is:
\begin{equation}
\mathbf{prox}_{\cost\left( x\mid x_{t}\right) }\left( v\right) =x_{t}+%
\mathcal{S}\left( v-x_{t};\lambda c^{-},\lambda c^{+}\right) \label{eq:gmv-cost}
\end{equation}%
where $\mathcal{S}\left( v;\lambda _{-},\lambda _{+}\right) =
\left( v-\lambda_{+}\right) _{+} - \left( v+\lambda
_{-}\right) _{-}$ is the two-sided soft-thresholding operator.\smallskip

If we define the cost constraint $\mathfrak{C}\left( x\mid x_{t}\right) $ as a
turnover constraint:
\begin{equation*}
\mathfrak{C}\left( x\mid x_{t}\right) =\left\{ x\in \mathbb{R}%
^{n}:\left\Vert x-x_{t}\right\Vert _{1}\leq \turnover^{+}\right\}
\end{equation*}%
the proximal operator is:
\begin{equation}
\mathcal{P}_{\mathfrak{C}}\left( v\right) =v-\limfunc{sign}\left(
v-x_{t}\right) \odot \min \left( \left\vert v-x_{t}\right\vert ,s^{\star
}\right) \label{eq:gmv-turnover}
\end{equation}%
where $s^{\star }=\left\{ s\in \mathbb{R}:\sum_{i=1}^{n}\left( \left\vert
v_{i}-x_{t,i}\right\vert -s\right) _{+}=\turnover^{+}\right\} $.

\begin{remark}
These two examples are very basic and show how we can easily introduce turnover
management using the ADMM framework. More sophisticated approaches are
presented in Section \ref{section:tips-and-tricks} on page
\pageref{section:tips-and-tricks}.
\end{remark}

\subsection{Smart beta portfolios}

In this section, we consider three main models of smart beta portfolios: the
equal risk contribution (ERC) portfolio, the risk budgeting (RB) portfolio and
the most diversified portfolio (MDP). Specific algorithms for these portfolios
based on the CCD method have already been presented in Griveau-Billion
\textsl{et al.} (2013) and Richard and Roncalli (2015, 2019). We extend these
results to the ADMM algorithm.

\subsubsection{The ERC portfolio}

The ERC portfolio uses the volatility risk measure $\sigma \left( x\right) =%
\sqrt{x^{\top }\Sigma x}$ and allocates the weights such that the risk
contribution is the same for all the assets of the portfolio (Maillard
\textsl{et al.}, 2010):%
\begin{equation*}
\mathcal{RC}_i\left(x\right) = x_{i}\frac{\partial \,\sigma \left( x\right) }{\partial \,x_{i}}=x_{j}\frac{%
\partial \,\sigma \left( x\right) }{\partial \,x_{j}} = \mathcal{RC}_j\left(x\right)
\end{equation*}
In this case, we can show that the ERC portfolio is the scaled solution $%
x^{\star }/\left( \mathbf{1}_{n}^{\top }x^{\star }\right) $ where $x^{\star }
$ is given by:%
\begin{equation*}
x^{\star }=\arg \min_x \frac{1}{2}x^{\top }\Sigma x-\lambda \sum_{i=1}^{n}\ln
x_{i}
\end{equation*}%
and $\lambda $ is any positive scalar. The first-order condition is $%
\left( \Sigma x\right) _{i}-\lambda x_{i}^{-1}=0$. It follows that $%
x_{i}\left( \Sigma x\right) _{i}-\lambda =0$ or:%
\begin{equation*}
x_{i}^{2}\sigma _{i}^{2}+x_{i}\sigma _{i}\sum_{j\neq i}x_{j}\rho
_{i,j}\sigma _{j}-\lambda =0
\end{equation*}%
We deduce that the solution is the positive root of the second-degree
equation. Finally, we obtain the following iteration for the CCD algorithm:%
\begin{equation}
x_{i}^{\left( k+1\right) }=\frac{-v_{i}^{\left( k+1\right) }+\sqrt{\left(
v_{i}^{\left( k+1\right) }\right) ^{2}+4\lambda \sigma _{i}^{2}}}{2\sigma
_{i}^{2}}  \label{eq:ccd-rb1}
\end{equation}%
where:%
\begin{equation*}
v_{i}^{\left( k+1\right) }=\sigma _{i}\sum_{j<i}x_{j}^{\left( k+1\right)
}\rho _{i,j}\sigma _{j}+\sigma _{i}\sum_{j>i}x_{j}^{\left( k\right) }\rho
_{i,j}\sigma _{j}
\end{equation*}%
\smallskip

The ADMM algorithm uses the first trick where $f_{x}\left( x\right) =
\frac{1}{2}x^{\top }\Sigma x$ and $f_{y}\left( y\right) =-\lambda
\sum_{i=1}^{n}\ln y_{i}$. It follows that the $x$- and $y$-update steps are:
\begin{equation*}
x^{\left( k+1\right) }=\left( \Sigma +\varphi I_{n}\right) ^{-1}\varphi
\left( y^{\left( k\right) }-u^{\left( k\right) }\right)
\end{equation*}%
and:%
\begin{equation*}
y_{i}^{\left( k+1\right) }=\frac{1}{2}\left( \left( x_{i}^{\left( k+1\right)
}+u_{i}^{\left( k\right) }\right) +\sqrt{\left( x_{i}^{\left( k+1\right)
}+u_{i}^{\left( k\right) }\right) ^{2}+4\lambda \varphi ^{-1}}\right)
\end{equation*}%
\smallskip

\begin{figure}[tbph]
\centering
\caption{Number of ADMM iterations for finding the ERC portfolio}
\label{fig:erc2}
\figureskip
\includegraphics[width = \figurewidth, height = \figureheight]{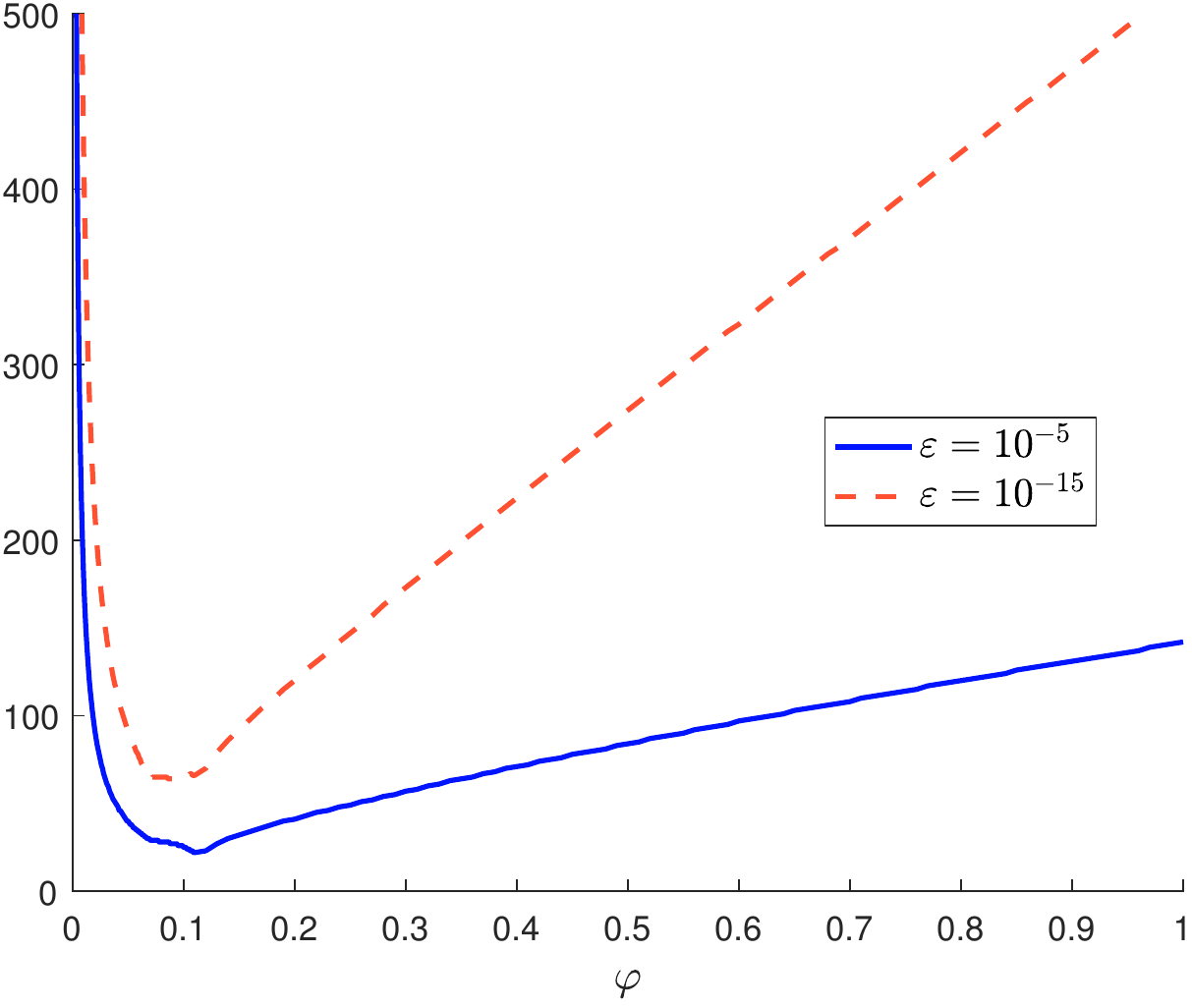}
\end{figure}

We apply the CCD and ADMM algorithms to the parameter set \#1. We find that the
ERC portfolio is equal to $11.40\%$, $12.29\%$, $5.49\%$, $11.91\%$, $6.65\%$,
$10.81\%$, $33.52\%$ and $7.93\%$. It appears that the CCD\ algorithm is much
more efficient than the ADMM algorithm. For instance, if we set $\lambda
=\sqrt{x^{\left( 0\right) ^{\top }}\Sigma x^{\left( 0\right) }}$, $x^{\left(
0\right) }=n^{-1}\mathbf{1}_{n}$ and $\varphi =1$, the CCD algorithm needs six
cycles to converge whereas the ADMM algorithm needs $156$ iterations if we set
the convergence criterion\footnote{The termination rule is defined as $\max
\left\vert x_{i}^{\left( k+1\right) }-x_{i}^{\left( k\right) }\right\vert \leq
\varepsilon $.} $\varepsilon =10^{-8}$. Whatever the values of $\lambda $,
$x^{\left( 0\right) }$ and $\varepsilon $, our experience is that the CCD
generally converges within less than $15$ cycles even if the number of assets
is greater than $1\,000 $. The convergence of the ADMM is more of an issue,
because it depends on the parameters $\lambda $ and $\varphi $. In Figure
\ref{fig:erc2}, we have reported the number of iterations of the ADMM with
respect to $\varphi $ for several values of $\varepsilon $ when $\lambda =1$
and $x^{\left( 0\right) }=\mathbf{1}_{n}$. We verify that it is very sensitive
to the value taken by $\varphi $. Curiously, the parameter $\lambda $ has
little influence, meaning that the convergence issue mainly concerns the
$x$-update step.

\subsubsection{Risk budgeting optimization}

The ERC portfolio has been extended by Roncalli (2013) when the risk budgets
are not equal and when the risk measure $\mathcal{R}\left( x\right) $ is convex
and coherent:
\begin{equation*}
\mathcal{RC}_{i}\left( x\right) =x_{i}\frac{\partial \,\mathcal{R}\left(
x\right) }{\partial \,x_{i}}=\mathcal{RB}_{i}
\end{equation*}%
where $\mathcal{RB}_{i}$ is the risk budget allocated to Asset $i$. In this
case, we can show that the risk budgeting portfolio is the scaled solution of
the following optimization problem:
\begin{equation*}
x^{\star }=\arg \min_{x}\mathcal{R}\left( x\right) -\lambda \sum_{i=1}^{n}\mathcal{RB}_{i}\cdot \ln x_{i}
\end{equation*}%
where $\lambda $ is any positive scalar. Depending on the risk measure, we can
use the CCD or the ADMM algorithm.\smallskip

For example, Roncalli (2015) proposes using the standard deviation-based risk
measure:
\begin{equation*}
\mathcal{R}\left( x\right) =-x^{\top }\left( \mu -r\right) +\xi \sqrt{
x^{\top }\Sigma x}
\end{equation*}%
In this case, the first-order condition for defining the CCD algorithm is:
\begin{equation*}
-\left( \mu _{i}-r\right) +\xi \frac{\left( \Sigma x\right) _{i}}{\sqrt{%
x^{\top }\Sigma x}}-\lambda \frac{\mathcal{RB}_{i}}{x_{i}}=0
\end{equation*}%
It follows that $\xi x_{i}\left( \Sigma x\right) _{i}-\left( \mu _{i}-r\right)
x_{i}\sigma \left( x\right) -\lambda \sigma \left( x\right) \cdot
\mathcal{RB}_{i}=0$ or equivalently:
\begin{equation*}
\alpha _{i}x_{i}^{2}+\beta _{i}x_{i}+\gamma _{i}=0
\end{equation*}%
where $\alpha _{i}=\xi \sigma _{i}^{2}$, $\beta _{i}=\xi \sigma _{i}\sum_{j\neq
i}x_{j}\rho _{i,j}\sigma _{j}-\left( \mu _{i}-r\right) \sigma \left( x\right) $
and $\gamma _{i}=-\lambda \sigma \left( x\right) \cdot \mathcal{RB}_{i}$. We
notice that the solution $x_{i}$ depends on the volatility $\sigma \left(
x\right) $. Here, we face an endogenous problem, because $\sigma \left(
x\right) $ depends on $x_{i}$. Griveau-Billon \textsl{et al.} (2015) notice
that this is not an issue, because we may assume that $\sigma \left( x\right) $
is almost constant between two coordinate iterations of the CCD algorithm. They
deduce that the coordinate solution is then the positive root of the
second-degree equation:
\begin{equation}
x_{i}^{\left( k+1\right) }=\frac{-\beta _{i}^{\left( k+1\right) }+\sqrt{%
\left( \beta _{i}^{\left( k+1\right) }\right) ^{2}-4\alpha _{i}^{\left(
k+1\right) }\gamma _{i}^{\left( k+1\right) }}}{2\alpha _{i}^{\left(
k+1\right) }}  \label{eq:ccd-rb2}
\end{equation}%
where:%
\begin{equation*}
\left\{
\begin{array}{l}
\alpha _{i}^{\left( k+1\right) }=\xi \sigma _{i}^{2} \\
\beta _{i}^{\left( k+1\right) }=\xi \sigma _{i}\left(
\sum_{j<i}x_{j}^{\left( k+1\right) }\rho _{i,j}\sigma
_{j}+\sum_{j>i}x_{j}^{\left( k\right) }\rho _{i,j}\sigma _{j}\right) -\left(
\mu _{i}-r\right) \sigma _{i}^{\left( k+1\right) }\left( x\right)  \\
\gamma _{i}^{\left( k+1\right) }=-\lambda \sigma _{i}^{\left( k+1\right)
}\left( x\right) \cdot \mathcal{RB}_{i} \\
\sigma _{i}^{\left( k+1\right) }\left( x\right) =\sqrt{\chi ^{\top }\Sigma
\chi } \\
\chi =\left( x_{1}^{\left( k+1\right) },\ldots ,x_{i-1}^{\left( k+1\right)
},x_{i}^{\left( k\right) },x_{i+1}^{\left( k\right) }\ldots ,x_{n}^{\left(
k\right) }\right)
\end{array}%
\right.
\end{equation*}
\smallskip

In the case of the volatility or the standard deviation-based risk measure, we
apply the exact formulation of the CCD algorithm because we have an analytical
solution of the first-order condition. This is not always the case, especially
when we consider skewness-based risk measure (Bruder \textsl{et al.}, 2016;
Lezmi \textsl{et al.}, 2018). In this case, we can use the gradient formulation
of the CCD algorithm or the ADMM algorithm, which is defined as follows:
\begin{equation*}
\left\{
\begin{array}{l}
x^{\left( k+1\right) }=\mathbf{prox}_{\varphi ^{-1}\mathcal{R}\left(
x\right) }\left( y^{\left( k\right) }-u^{\left( k\right) }\right)  \\
v_{y}^{\left( k+1\right) }=x^{\left( k+1\right) }+u^{\left( k\right) } \\
y^{\left( k+1\right) }=\frac{1}{2}\left( v_{y}^{\left( k+1\right) }+\sqrt{%
v_{y}^{\left( k+1\right) }\odot v_{y}^{\left( k+1\right) }+4\lambda \varphi
^{-1}\cdot \mathcal{RB}}\right)  \\
u^{\left( k+1\right) }=u^{\left( k\right) }+x^{\left( k+1\right) }-y^{\left(
k+1\right) }%
\end{array}%
\right.
\end{equation*}

\subsubsection{The most diversified portfolio}

Choueifaty and Coignard (2008) introduce the concept of diversification ratio,
which corresponds to the following expression:
\begin{equation*}
\mathcal{DR}\left( x\right) =\frac{\sum_{i=1}^{n}x_{i}\sigma _{i}}{\sigma
\left( x\right) }=\frac{x^{\top }\sigma }{\sqrt{x^{\top }\Sigma x}}
\end{equation*}%
By construction, the diversification ratio of a portfolio fully invested in one
asset is equal to one, whereas it is larger than one in the general case. The
authors then propose building the most diversified portfolio as the portfolio
which maximizes the diversification ratio. It is also the solution to the
following minimization problem\footnote{See Choueifaty \textsl{et al.}
(2013).}:
\begin{eqnarray*}
x^{\star } &=&\arg \min_{x}\frac{1}{2}\ln \left( x^{\top }\Sigma x\right)
-\ln \left( x^{\top }\sigma \right)  \\
&\text{s.t.}&\left\{
\begin{array}{l}
\mathbf{1}_{n}^{\top }x=1 \\
x\in \Omega
\end{array}%
\right.
\end{eqnarray*}%
This problem is relatively easy to solve using standard numerical algorithms if
$\Omega $ corresponds to linear constraints, for example weight constraints.
However, the optimal solution may face the same problem as the minimum variance
portfolio since most of the time it is concentrated on a small number of
assets. This is why it is interesting to add a weight diversification
constraint $\mathcal{D}\left( x\right) \geq \mathcal{D}^{-}$. For example, we
can assume that the number of effective bets $\mathcal{N}\left( x\right) $ is
larger than a minimum acceptable value $\mathcal{N}^{-}$. Contrary to the
minimum variance portfolio, we do not obtain a QP problem and we observe that
the optimization problem is tricky in practice. Thanks to the ADMM algorithm,
we can however simplify the optimization problem by splitting the constraints
and using the same approach that has been already described on page
\pageref{eq:admm-gmv-trick}. The $x$-update consists in finding the regularized
standard MDP:
\begin{equation*}
x^{\left( k+1\right) }=\arg \min_{x}\left\{ \frac{1}{2}\ln \left( x^{\top
}\Sigma x\right) -\ln \left( x^{\top }\sigma \right) +\frac{\varphi }{2}%
\left\Vert x-y^{\left( k\right) }+u^{\left( k\right) }\right\Vert
_{2}^{2}\quad \text{s.t. }\mathbf{1}_{n}^{\top }x=1\right\}
\end{equation*}%
whereas the $y$-update corresponds to the projection onto the intersection of
$\Omega $ and $\mathfrak{D}=\left\{ x\in \mathbb{R}^{n}:\mathcal{D}\left(
x\right) \geq \mathcal{D}^{-}\right\} $:
\begin{equation*}
y^{\left( k+1\right) }=\mathcal{P}_{\Omega\, \cap\, \mathfrak{D}}\left(
x^{\left( k+1\right) }+u^{\left( k\right) }\right)
\end{equation*}
\smallskip

We consider the parameter set \#2 given in Appendix \ref{appendix:data2} on
page \pageref{appendix:data2}. Results are reported in Table \ref{tab:mdp1}.
The second column corresponds to the long/short MDP portfolio (or $\Omega =%
\mathbb{R}^{n}$). By definition, we cannot compute the number of effective bets
because it contains short positions. The other columns correspond to the
long-only MDP portfolio (or $\Omega =\left[ 0,1\right] ^{n}$) when we impose a
sufficient number of effective bets $\mathcal{N}^{-}$. We notice that the
traditional long-only MDP is poorly diversified in terms of weights since we
have $\mathcal{N}\left( x\right) =2.30$. As for the minimum variance
portfolio, the MDP tends to the equally-weighted portfolio when $\mathcal{N}%
^{-}$ tends to the number of assets

\begin{table}[tbph]
\centering
\caption{MDP portfolios (in \%)}
\label{tab:mdp1}
\tableskip
\begin{tabular}{c:r:rrrrrr}
\hline
                     & L/S     & \multicolumn{6}{c}{Long-only}                             \\
$\mathcal{N}^{-}$    &         & $0.00$  & $ 3.00$ & $ 4.00$ & $ 5.00$ & $ 6.00$ & $ 7.00$ \\ \hline
$x_1^{\star }$       & $41.81$ & $41.04$ & $35.74$ & $30.29$ & $26.08$ & $22.44$ & $18.83$ \\
$x_2^{\star }$       & $51.88$ & $50.92$ & $43.91$ & $36.68$ & $31.05$ & $26.12$ & $21.19$ \\
$x_3^{\star }$       & $ 8.20$ & $ 8.05$ & $10.12$ & $11.52$ & $12.33$ & $12.80$ & $13.01$ \\
$x_4^{\star }$       & $-0.43$ & $ 0.00$ & $ 2.48$ & $ 5.12$ & $ 7.16$ & $ 8.90$ & $10.51$ \\
$x_5^{\star }$       & $-0.26$ & $ 0.00$ & $ 0.92$ & $ 2.28$ & $ 3.60$ & $ 5.02$ & $ 6.85$ \\
$x_6^{\star }$       & $-0.38$ & $ 0.00$ & $ 2.03$ & $ 4.36$ & $ 6.28$ & $ 8.02$ & $ 9.79$ \\
$x_7^{\star }$       & $-0.51$ & $ 0.00$ & $ 3.47$ & $ 6.68$ & $ 8.85$ & $10.44$ & $11.65$ \\
$x_8^{\star }$       & $-0.31$ & $ 0.00$ & $ 1.32$ & $ 3.07$ & $ 4.65$ & $ 6.27$ & $ 8.17$ \\ \hline
$\mathcal{N}\left(x\right)$
                     &         & $ 2.30$ & $ 3.00$ & $ 4.00$ & $ 5.00$ & $ 6.00$ & $ 7.00$ \\ \hline
\end{tabular}
\end{table}

\subsection{Robo-advisory optimization}

Today's financial industry is facing a digital revolution in all areas: payment
services, on-line banking, asset management, etc. This is particularly true for
the financial advisory industry, which has been impacted in the last few years
by the emergence of digitalization and robo-advisors. The demand for
robo-advisors is strong, which explains the growth of this
business\footnote{For instance, the growth was $60\%$ per year in the US over
the last five years. In Europe, the growth is also impressive, even though the
market is smaller. In the last two years, assets under management have
increased $14$-fold.}. How does one characterize a robo-advisor? This is not
simple, but the underlying idea is to build a systematic portfolio allocation
in order to provide a customized advisory service. A robo-advisor has two main
objectives. The first objective is to know the investor better than a
traditional asset manager. Because of this better knowledge, the robo-advisor
may propose a more appropriate asset allocation. The second objective is to
perform the task in a systematic way and to build an automated rebalancing
process. Ultimately, the goal is to offer a customized solution. In fact, the
reality is very different. We generally notice that many robo-advisors are more
a web or a digital application, but not really a robo-advisor. The reason is
that portfolio optimization is a very difficult task. In many robo-advisors,
asset allocation is then rather human-based or not completely systematic with
the aim to rectify the shortcomings of mean-variance optimization. Over the
next five years, the most important challenge for robo-advisors will be to
reduce these discretionary decisions and improve the robustness of their
systematic asset allocation process. But this means that robo-advisors must
give up the quadratic programming world of the portfolio allocation.

\subsubsection{Specification of the objective function}

In order to make mean-variance optimization more robust, two directions can be
followed. The first one has been largely explored and adds a penalty function
in order to regularize or sparsify the solution (Brodie \textsl{et al.} 2009;
DeMiguel \textsl{et al.}, 2009; Carrasco and Noumon, 2010; Bruder \textsl{et
al.}, 2013; Bourgeron \textsl{et al.}, 2018). The second one consists in
changing the objective function and considering risk budgeting portfolios
instead of mean-variance optimized portfolios (Maillard \textsl{et al.}, 2010;
Roncalli, 2013). Even if this second direction has encountered great success,
it presents a solution that is not sufficiently flexible in terms of active
management. Nevertheless, these two directions are not so divergent. Indeed,
Roncalli (2013) shows that the risk budgeting optimization can be viewed as a
non-linear shrinkage approach of the minimum variance optimization. Richard and
Roncalli (2015) propose then a unified approach of smart beta portfolios by
considering alternative allocation models as penalty functions of the minimum
variance optimization. In particular, they use the logarithmic barrier function
in order to regularize minimum variance portfolios. This idea has also been
reiterated by de Jong (2018), who considers a mean-variance
framework.\smallskip

Therefore, we propose defining the robo-advisor optimization problem as
follows:
\begin{eqnarray}
x_{t+1}^{\star } &=&\arg \min_{x}f_{\mathcal{R}obo}\left( x\right)
\label{eq:robo1} \\
&\text{s.t.}&\left\{
\begin{array}{l}
\mathbf{1}_{n}^{\top }x=1 \\
\mathbf{0}_{n}\leq x\leq \mathbf{1}_{n} \\
x\in \Omega
\end{array}%
\right.   \notag
\end{eqnarray}%
where:
\begin{eqnarray}
f_{\mathcal{R}obo}\left( x\right)  &=&\frac{1}{2}\left( x-b\right) ^{\top
}\Sigma _{t}\left( x-b\right) -\gamma \left( x-b\right) ^{\top }\mu _{t}+
\notag \\
&&\varrho _{1}\left\Vert \Gamma _{1}\left( x-x_{t}\right) \right\Vert _{1}+%
\frac{1}{2}\varrho _{2}\left\Vert \Gamma _{2}\left( x-x_{t}\right)
\right\Vert _{2}^{2}+  \notag \\
&&\tilde{\varrho}_{1}\left\Vert \tilde{\Gamma}_{1}\left( x-\tilde{x}\right)
\right\Vert _{1}+\frac{1}{2}\tilde{\varrho}_{2}\left\Vert \tilde{\Gamma}%
_{2}\left( x-\tilde{x}\right) \right\Vert _{2}^{2}-\lambda \sum_{i=1}^{n}%
\mathcal{RB}_{i}\cdot \ln x_{i}  \label{eq:robo2}
\end{eqnarray}%
$b$ is the benchmark portfolio, $\tilde{x}$ is the reference portfolio and
$x_{t}$ is the current portfolio.\smallskip

This specification is sufficiently broad that it encompasses most models used
by the industry. We notice that the objective function is made up of three
parts. The first part corresponds to the MVO objective function with a
benchmark. If we set $b$ equal to $\mathbf{0}_{n}$, we obtain the Markowitz
utility function. The second part contains $\boldsymbol{\ell }_{1}$- and
$\boldsymbol{\ell }_{2}$-norm penalty functions. The regularization can be done
with respect to the current allocation $x_{t}$ in order to control the
rebalancing process and the smoothness of the dynamic allocation. The
regularization can also be done with respect to a reference portfolio, which is
generally the strategic asset allocation of the fund. The idea is to control
the profile of the fund. For example, if the strategic asset allocation is an
80/20 asset mix policy, we do not want the portfolio to present a defensive or
balanced risk profile. Finally, the third part of the objective function
corresponds to the logarithmic barrier function, where the parameter $\lambda $
controls the trade-off between MVO optimization and RB optimization. This last
part is very important in order to make the dynamic asset allocation more
robust. The hyperparameters of the objective function are $\varrho _{1}$,
$\varrho _{2}$, $\tilde{\varrho}_{1}$, $\tilde{\varrho}_{2}$ and $\lambda $.
They are all positive and can also be set to zero in order to deactivate a
penalty function. For instance, $\varrho _{2}$ and $\tilde{\varrho}_{2}$ are
equal to zero if we don't want to have a shrinkage of the covariance matrix
$\Sigma _{t}$. The hyperparameters $\varrho _{1}$ and $\tilde{\varrho}_{1}$ can
also be equal to zero because the $\boldsymbol{\ell }_{1}$ regularization is
generally introduced when specifying the additional constraints $\Omega $. The
parameter $\gamma $ is not really a hyperparameter, because it is generally
calibrated to target volatility or an expected return. We also notice that this
model encompasses the Black-Litterman model thanks to the specification of $\mu
_{t}$ (Bourgeron \textsl{et al.}, 2018). Another important component of this
framework is the specification of the set $x\in \Omega $. It may include
traditional constraints such as weight bounds and/or asset class limits, but we
can also add non-linear constraints such as a turnover limit, an active share
floor or a weight diversification constraint.

\subsubsection{Derivation of the general algorithm}

Problem (\ref{eq:robo1}) is equivalent to solving:
\begin{equation*}
x_{t+1}^{\star }=\arg \min_{x}f_{\mathcal{R}obo}^{+}\left( x\right)
\end{equation*}%
where the objective function can be broken down as follows:
\begin{eqnarray*}
f_{\mathcal{R}obo}^{+}\left( x\right)  &=&f_{\mathrm{MVO}}\left(
x\right) +f_{\boldsymbol{\ell }_{1}}\left( x\right) +
f_{\boldsymbol{\ell }_{2}}\left( x\right) +
f_{\mathrm{RB}}\left( x\right) + \\
&&+\mathbf{\mathds{1}}_{\Omega _{0}}\left( x\right) +\mathbf{\mathds{1}}_{\Omega }\left( x\right)
\end{eqnarray*}%
where:%
\begin{eqnarray*}
f_{\mathrm{MVO}}\left( x\right)  &=&\frac{1}{2}\left( x-b\right)
^{\top }\Sigma _{t}\left( x-b\right) -\gamma \left( x-b\right) ^{\top }\mu_{t} \\
f_{\boldsymbol{\ell }_{1}}\left( x\right)  &=&
\varrho _{1}\left\Vert\Gamma _{1}\left( x-x_{t}\right) \right\Vert _{1}+
\tilde{\varrho}_{1}\left\Vert \tilde{\Gamma}_{1}\left( x-\tilde{x}\right) \right\Vert _{1} \\
f_{\boldsymbol{\ell }_{2}}\left( x\right)  &=&
\frac{1}{2}\varrho_{2}\left\Vert \Gamma _{2}\left( x-x_{t}\right) \right\Vert _{2}^{2}+
\frac{1}{2}\tilde{\varrho}_{2}\left\Vert \tilde{\Gamma}_{2}\left( x-\tilde{x}\right) \right\Vert _{2}^{2} \\
f_{\mathrm{RB}}\left( x\right)  &=&-\lambda \sum_{i=1}^{n}\mathcal{RB}_{i}\cdot \ln x_{i}
\end{eqnarray*}%
and $\Omega _{0}=\left\{ x\in \left[ 0,1\right] ^{n}:\mathbf{1}_{n}^{\top
}x=1\right\} $. The ADMM algorithm is implemented as follows:
\begin{eqnarray*}
\left\{ x^{\star },y^{\star }\right\}  &=&\arg \min f_{x}\left( x\right)
+f_{y}\left( y\right)  \\
&\text{s.t.}&x-y=\mathbf{0}_{n}
\end{eqnarray*}%
This is the general approach for solving the robo-advisor problem.\smallskip

The main task is then to split the function $f_{\mathcal{R}obo}^{+}$ into
$f_{x}$ and $f_{y}$. However, in order to be efficient, the $x$- and $y$-update
steps of the ADMM algorithm must be easy to compute. Therefore, we impose that
the $x$-step is solved using QP or CCD methods while the $y$-step is solved
using the Dykstra's algorithm, where each component corresponds to an
analytical proximal operator. Moreover, we also split the set of constraints
$\Omega $ into a set of linear constraints $\Omega _{\mathcal{L}inear}$ and a
set of non-linear constraints $\Omega _{\mathcal{N}onlinear}$. This lead
defining $f_{x}\left( x\right) $ and $f_{y}\left(y\right) $ as follows:
\begin{equation}
\left\{
\begin{array}{l}
f_{x}\left( x\right) =f_{\mathrm{MVO}}\left( x\right) +f_{%
\boldsymbol{\ell }_{2}}\left( x\right) +\mathbf{\mathds{1}}_{\Omega
_{0}}\left( x\right) +\mathbf{\mathds{1}}_{\Omega _{\mathcal{L}inear}}\left(
x\right)  \\
f_{y}\left( y\right) =f_{\boldsymbol{\ell }_{1}}\left( y\right)
+f_{\mathrm{RB}}\left( x\right) +\mathbf{\mathds{1}}%
_{\Omega _{\mathcal{N}onlinear}}\left( x\right)
\end{array}%
\right.   \label{eq:robo3}
\end{equation}%
We notice that $f_{x}\left( x\right) $ has a quadratic form, implying that the
$x$-step may be solved using a QP algorithm. Another formulation is:
\begin{equation}
\left\{
\begin{array}{l}
f_{x}\left( x\right) =f_{\mathrm{MVO}}\left( x\right) +f_{%
\boldsymbol{\ell }_{2}}\left( x\right) +f_{\mathrm{RB}}\left( x\right)  \\
f_{y}\left( y\right) =f_{\boldsymbol{\ell }_{1}}\left( y\right) +%
\mathbf{\mathds{1}}_{\Omega _{0}}\left( x\right) +\mathbf{\mathds{1}}%
_{\Omega _{\mathcal{L}inear}}\left( x\right) +\mathbf{\mathds{1}}_{\Omega _{%
\mathcal{N}onlinear}}\left( x\right)
\end{array}%
\right.   \label{eq:robo4}
\end{equation}%
In this case, the $x$-step is solved using the CCD algorithm.

\subsubsection{Specific algorithms}

\paragraph{The ADMM-QP formulation}

If we consider Formulation (\ref{eq:robo3}), we have:
\begin{eqnarray*}
f_{\mathrm{QP}}\left( x\right)  &=&f_{\mathrm{MVO}}\left( x\right)
+f_{\boldsymbol{\ell }_{2}}\left( x\right)  \\
&=&\frac{1}{2}\left( x-b\right) ^{\top }\Sigma _{t}\left( x-b\right) -\gamma
\left( x-b\right) ^{\top }\mu _{t}+\frac{1}{2}\varrho _{2}\left\Vert \Gamma
_{2}\left( x-x_{t}\right) \right\Vert _{2}^{2}+
\frac{1}{2}\tilde{\varrho}_{2}\left\Vert \tilde{\Gamma}_{2}\left( x-\tilde{x}\right) \right\Vert_{2}^{2} \\
&=&\frac{1}{2}x^{\top }Qx-x^{\top }R+C
\end{eqnarray*}%
where $Q=\Sigma _{t}+\varrho _{2}\Gamma _{2}^{\top }\Gamma
_{2}+\tilde{\varrho}_{2}\tilde{\Gamma}_{2}^{\top }\tilde{\Gamma}_{2}$,
$R=\gamma \mu _{t}+\Sigma _{t}b+\varrho _{2}\Gamma _{2}^{\top }\Gamma
_{2}x_{t}+\tilde{\varrho}_{2}\tilde{\Gamma}_{2}^{\top
}\tilde{\Gamma}_{2}\tilde{x}$ and $C$ is a constant\footnote{The expression of
$f_{\mathrm{QP}}\left( x\right) $ is computed in Appendix
\ref{appendix:admm-qp} on page \pageref{appendix:admm-qp}.}. Using the fourth
ADMM trick, we deduce that $x^{\left( k+1\right) }$ is the solution of the
following QP problem:
\begin{eqnarray*}
x^{\left( k+1\right) } &=&\arg \min_{x}\frac{1}{2}x^{\top }\left( Q+\varphi
I_{n}\right) x-x^{\top }\left( R+\varphi \left( y^{\left( k\right)
}-u^{\left( k\right) }\right) \right)  \\
&\text{s.t.}&\left\{
\begin{array}{l}
\mathbf{1}_{n}^{\top }x=1 \\
\mathbf{0}_{n}\leq x\leq \mathbf{1}_{n}%
\end{array}%
\right.
\end{eqnarray*}%
Since the proximal operators of $f_{\boldsymbol{\ell }_{1}}$ and
$f_{\mathrm{RB}}$ have been already computed, finding $y^{\left( k+1\right) }$
is straightforward with the Dykstra's algorithm as long as the proximal of each
non-linear constraint is known.

\paragraph{The ADMM-CCD formulation}

If we consider Formulation (\ref{eq:robo4}), we have:
\begin{equation*}
f_{x}\left( x\right) =f_{\mathrm{QP}}\left( x\right) -\lambda \sum_{i=1}^{n}%
\mathcal{RB}_{i}\cdot \ln x_{i}
\end{equation*}%
Using Appendix \ref{appendix:admm-ccd} on page \pageref{appendix:admm-ccd}, the
CCD algorithm applied to $x$-update is:
\begin{eqnarray*}
x_{i}^{\left( k+1\right) } &=&\frac{R_{i}-\sum_{j<i}x_{j}^{\left( k+1\right)
}Q_{i,j}-\sum_{j>i}x_{j}^{\left( k\right) }Q_{i,j}}{2Q_{i,i}}+ \\
&&\frac{\sqrt{\left( \sum_{j<i}x_{j}^{\left( k+1\right)
}Q_{i,j}+\sum_{j>i}x_{j}^{\left( k\right) }Q_{i,j}-R_{i}\right)
^{2}+4\lambda _{i}Q_{i,i}}}{2Q_{i,i}}
\end{eqnarray*}%
where the matrices $Q$ and $R$ are defined as:
\begin{equation*}
Q=\Sigma _{t}+\varrho _{2}\Gamma _{2}^{\top }\Gamma _{2}+\tilde{\varrho}_{2}%
\tilde{\Gamma}_{2}^{\top }\tilde{\Gamma}_{2}+\varphi I_{n}
\end{equation*}%
and:%
\begin{equation*}
R=\gamma \mu _{t}+\Sigma _{t}b+\varrho _{2}\Gamma _{2}^{\top }\Gamma
_{2}x_{t}+\tilde{\varrho}_{2}\tilde{\Gamma}_{2}^{\top }\tilde{\Gamma}_{2}%
\tilde{x}+\varphi \left( y^{\left( k\right) }-u^{\left( k\right) }\right)
\end{equation*}%
and $\lambda _{i}=\lambda \cdot \mathcal{RB}_{i}$. Like the ADMM-QP
formulation, the $y$-update step does not pose any particular difficulties.

\subsection{Tips and tricks}
\label{section:tips-and-tricks}

If we consider the different portfolio optimization approaches presented in
Table \ref{tab:portfolio1}, we have shown how to solve MVO (1), GMV (2), MDP
(3), ERC (4) and RB (5) models. The RQE (7) model is equivalent to the GMV (2)
model by replacing the covariance matrix $\Sigma $ by the dissimilarity matrix
$D$. Below, we implement the Kullback-Leibler model (4) of Bera and Park (2008)
using the ADMM framework. Concerning the regularization problems in Table
\ref{tab:portfolio2}, ridge (8), lasso (9) and log-barrier (10) penalty
functions have been already covered. Indeed, ridge and lasso penalizations
correspond to the proximal operator of $\boldsymbol{\ell }_{1}$- and
$\boldsymbol{\ell }_{2}$-norm functions by applying the translation $g\left(
x\right) =x-\tilde{x}$. Shannon's entropy (11) penalization is discussed below.
For the constraints that are considered in Table \ref{tab:portfolio3}, imposing
no cash and leverage (12) is done with the proximal of the hyperplane
$\mathcal{H}_{yperlane}\left[ \mathbf{1}_{n},1\right] $. No short selling (13)
and weight bounds (14) are equivalent to considering the box projections
$\mathcal{B}_{ox}\left[ \mathbf{0}_{n},\infty \right] $ and
$\mathcal{B}_{ox}\left[ x^{-},x^{+}\right] $. Asset class limits can be
implemented using the projection onto the intersection of two half-spaces
$\mathcal{H}_{alfspace}\left[ \mathbf{1}_{i\in C_{j}},c_{j}^{+}\right] $ and
$\mathcal{H}_{alfspace}\left[ -\mathbf{1}_{i\in C_{j}},-c_{j}^{-}\right] $. The
proximal of the turnover (16) had been already given in Equation
(\ref{eq:gmv-turnover}) on page \pageref{eq:gmv-turnover}. If we want to impose
an upper limit on transaction costs (17), we use the Moreau decomposition and
Equation (\ref{eq:gmv-cost}). Finally, Section
\ref{section:gmv-diversification} on page \pageref{section:gmv-diversification}
dealt with the weight diversification problem of the number of active bets.
Therefore, it remains to solve leverage limits (18), long/short exposure (19)
restrictions and active management constraints: benchmarking (20), tracking
error floor (21) and active share floor (22).

\subsubsection{Volatility and return targeting}

We first consider the $\mu $-problem and the $\sigma $-problem. Targeting a
minimum expected return $\mu \left( x\right) \geq \mu ^{\star }$ can be
implemented in the ADMM framework using the proximal operator of the
hyperplane\footnote{%
We have $\mu \left( x\right) \geq \mu ^{\star }\Leftrightarrow x^{\top }\mu
\geq \mu ^{\star }\Leftrightarrow -\mu ^{\top }x\leq -\mu ^{\star }$.} $%
\mathcal{H}_{yperlane}\left[ -\mu ,-\mu ^{\star }\right] $. In the
case of the $\sigma $-problem $\sigma \left( x\right) \leq \sigma
^{\star }$, we use
the fourth ADMM trick. Let $L$ be the lower Cholesky decomposition of $%
\Sigma $, we have:%
\begin{eqnarray*}
\sigma \left( x\right) \leq \sigma ^{\star } &\Leftrightarrow &\sqrt{x^{\top
}\Sigma x}\leq \sigma ^{\star } \\
&\Leftrightarrow &\sqrt{x^{\top }\left( LL^{\top }\right) x}\leq \sigma
^{\star } \\
&\Leftrightarrow &\left\Vert y^{\top }y\right\Vert _{2}\leq \sigma ^{\star }
\end{eqnarray*}%
where $y=L^{\top }x$. It follows that the proximal of the $y$-update is the
projection onto the $\boldsymbol{\ell }_{2}$ ball $\mathcal{B}_{2}\left(
\mathbf{0}_{n},\sigma ^{\star }\right) $.

\subsubsection{Leverage management}

If we impose a leverage limit $\sum_{i=1}^{n}\left\vert x_{i}\right\vert \leq
\mathcal{L}^{+}$, we have $\left\Vert x\right\Vert _{1}\leq \mathcal{L}^{+}$
and the proximal of the $y$-update is the projection onto the $\boldsymbol{\ell
}_{1}$ ball $\mathcal{B}_{1}\left( \mathbf{0}_{n},\mathcal{L}^{+}\right) $. If
the leverage constraint concerns the long/short limits $-\mathcal{LS}^{-}\leq
\sum_{i=1}^{n}x_{i}\leq \mathcal{LS}^{+}$, we consider the intersection of the
two half-spaces $\mathcal{H}_{alfspace}\left[
\mathbf{1}_{n},\mathcal{LS}^{+}\right] $ and $\mathcal{H}_{alfspace}\left[
-\mathbf{1}_{n},\mathcal{LS}^{-}\right] $. If we consider an absolute leverage
$\left\vert \sum_{i=1}^{n}x_{i}\right\vert \leq \mathcal{L}^{+}$, we obtain the
previous case with $\mathcal{LS}^{-}=\mathcal{LS}^{+} = \mathcal{L}^{+}$.
Portfolio managers can also use another constraint concerning the sum of the
$k$ largest values\footnote{An example is the 5/10/40 UCITS rule: A UCITS fund
may invest no more than $10\%$ of its net assets in transferable securities or
money market instruments issued by the same body, with a further aggregate
limitation of $40\%$ of net assets on exposures of greater than $5\%$ to single
issuers.}:
\begin{equation*}
f\left( x\right) =\sum_{i=n-k+1}^{n}x_{\left( i:n\right) }=x_{\left(
n:n\right) }+\ldots +x_{\left( n-k+1:n\right) }
\end{equation*}%
where $x_{\left( i:n\right) }$ is the order statistics of $x$: $x_{\left(
1:n\right) }\leq x_{\left( 2:n\right) }\leq \cdots \leq x_{\left( n:n\right)
}$. Beck (2017) shows that:
\begin{equation*}
\mathbf{prox}_{\lambda f\left( x\right) }\left( v\right) =v-\lambda \mathcal{%
P}_{\Omega }\left( \frac{v}{\lambda }\right)
\end{equation*}%
where:%
\begin{equation*}
\Omega =\left\{ x\in \left[ 0,1\right] ^{n}:\mathbf{1}_{n}^{\top
}x=k\right\} =\mathcal{B}_{ox}\left[ \mathbf{0}_{n},\mathbf{1}_{n}\right]
\cap \mathcal{H}_{yperlane}\left[ \mathbf{1}_{n},k\right]
\end{equation*}

\subsubsection{Entropy portfolio and diversification measure}

Bera and Park (2008) propose using a cross-entropy measure as the objective
function:
\begin{eqnarray*}
x^{\star } &=&\arg \min_{x}\limfunc{KL}\left( x\mid \tilde{x}\right)  \\
&\text{s.t.}&\left\{
\begin{array}{l}
\mathbf{1}_{n}^{\top }x=1 \\
\mathbf{0}_{n}\leq x\leq \mathbf{1}_{n} \\
\mu \left( x\right) \geq \mu ^{\star } \\
\sigma \left( x\right) \leq \sigma ^{\star }%
\end{array}%
\right.
\end{eqnarray*}%
where $\limfunc{KL}\left( x\mid \tilde{x}\right) =\sum_{i=1}^{n}x_{i}\ln \left(
x_{i}/\tilde{x}_{i}\right) $ and $\tilde{x}$ is a reference portfolio, which is
well-diversified (e.g. the EW\footnote{In this case, it is equivalent to
maximize Shannon's entropy because $\tilde{x}=\mathbf{1}_{n}$.} or ERC
portfolio). In Appendix \ref{appendix:proximal-kullblack_leibler} on page
\pageref{appendix:proximal-kullblack_leibler}, we show that the proximal
operator of $\lambda \limfunc{KL}\left( x\mid \tilde{x}\right) $ is equal to:
\begin{equation*}
\mathbf{prox}_{\lambda \limfunc{KL}\left( v\mid \tilde{x}\right) }\left(
v\right) =\lambda \left(
\begin{array}{c}
W\left( \lambda ^{-1}\tilde{x}_{1}e^{\lambda ^{-1}v_{1}-\tilde{x}%
_{1}^{-1}}\right)  \\
\vdots  \\
W\left( \lambda ^{-1}\tilde{x}_{n}e^{\lambda ^{-1}v_{n}-\tilde{x}%
_{n}^{-1}}\right)
\end{array}%
\right)
\end{equation*}%
where $W\left( x\right) $ is the Lambert $W$ function.

\begin{remark}
Using the previous result and the fact that $\limfunc{SE}\left( x\right)
=-\limfunc{KL}\left( x\mid \mathbf{1}_{n}\right) $, we can use Shannon's
entropy to define the diversification measure $\mathcal{D}\left( x\right)
=\limfunc{SE}\left( x\right) $. Therefore, solving Problem (\ref{eq:gmv4}) is
straightforward when we consider the following diversification set:
\begin{equation*}
\mathfrak{D}=\left\{ x\in \left[ 0,1\right] ^{n}:-\sum_{i=1}^{n}x_{i}\ln
x_{i}\mathbf{\geq }\limfunc{SE}\nolimits^{-}\right\}
\end{equation*}
\end{remark}

\subsubsection{Passive and active management}

In the case of the active share, we use the translation property:%
\begin{eqnarray*}
\mathcal{AS}\left( x\mid \tilde{x}\right)  &=&\frac{1}{2}\sum_{i=1}^{n}\left%
\vert x_{i}-\tilde{x}_{i}\right\vert  \\
&=&\frac{1}{2}\left\Vert x-\tilde{x}\right\Vert _{1}
\end{eqnarray*}
The proximal operator is given in Appendix
\ref{appendix:proximal-l1-ball-complement} on page
\pageref{appendix:proximal-l1-ball-complement}. It is interesting to notice
that this type of problem cannot be solved using an augmented QP algorithm
since it involves the complement of the $\boldsymbol{\ell}_1$ ball and not
directly the $\boldsymbol{\ell}_1$ ball itself. In this case, we face a
maximization problem and not a minimization problem, and the technique of
augmented variables does not work.\smallskip

For tracking error volatility, again we use the fourth ADMM trick:
\begin{eqnarray*}
\sigma \left( x\mid \tilde{x}\right)  &=&\sqrt{\left( x-\tilde{x}\right)
^{\top }\Sigma \left( x-\tilde{x}\right) } \\
&=&\left\Vert y\right\Vert _{2}
\end{eqnarray*}%
where $y=L^{\top }x-L^{\top }\tilde{x}$. Using our ADMM notations, we have $%
Ax+By=c$ where $A=L^{\top }$, $B=-I_{n}$ and $c=L^{\top }\tilde{x}$.

\subsubsection{Index sampling}

Index sampling is based on the cardinality constraint
$\sum_{i=1}^{n}\mathds{1}\left\{ x_{i}>0\right\} \leq n_{x}$. It is closed to
the $\boldsymbol{\ell }_{0}$-norm function $\left\Vert x\right\Vert
_{0}=\sum_{i=1}^{n}\mathds{1}\left\{ x_{i}\neq 0\right\} $. Beck (2017) derives
the proximal of $\lambda \left\Vert x\right\Vert _{0}$ on pages 137-138 of his
monograph. However, it does not help to solve the index sampling problem,
because we are interested in computing the projection onto the
$\boldsymbol{\ell }_{0}$ ball and not the proximal of the $\boldsymbol{\ell
}_{0}$-norm function\footnote{We cannot use the Moreau decomposition, because
the dual of $\lambda \left\Vert x\right\Vert _{p}$ is not necessarily the ball
$\mathcal{B}_{p}\left( \mathbf{0}_{n},\lambda \right) $. For example, the dual
of the $\boldsymbol{\ell }_{2}$-norm function is the $\boldsymbol{\ell }_{2}$
ball, but the dual of the $\boldsymbol{\ell }_{1}$-norm function is the
$\boldsymbol{\ell }_{\infty }$ ball.}. This is why index sampling remains an
open problem using the ADMM framework.

\section{Conclusion}

The aim of this paper is to propose an alternative solution to the quadratic
programming algorithm in the context of portfolio allocation. In numerical
analysis, the quadratic programming model is a powerful optimization tool,
which is computationally very efficient. In portfolio management, the
mean-variance optimization model is exactly a quadratic programming model,
meaning that it benefits from its computational power. Therefore, the success
of the Markowitz allocation model is explained by these two factors: the
quadratic utility function and the quadratic programming setup. A lot of
academics and professionals have proposed an alternative approach to the MVO
framework, but very few of these models are used in practice. The main reason
is that these competing models focus on the objective function and not on the
numerical implementation. However, we believe that any model which is not
tractable will have little success with portfolio managers. The analogy is
obvious if we consider the theory of options. The success of the Black-Scholes
model lies in the Black-Scholes analytical formula. Over the last thirty years,
many models have been created (e.g. local volatility and stochastic volatility
models), but only one can really compete with the Black-Scholes model. This is
the SABR model, and the main reason is that it has an analytical formula for
implied volatility.\smallskip

This paper focuses then on a general approach for numerically solving non-QP
portfolio allocation models. For that, we consider some algorithms that have
been successfully applied to machine learning and large-scale optimization. For
instance, the coordinate descent algorithm is the fastest method for performing
high-dimensional lasso regression, while the Dykstra's algorithm has been
created to find the solution of restricted least squares regression. Since
there is a strong link between MVO and linear regression (Scherer, 2007), this
is not a surprise if these algorithms can help solve regularized MVO allocation
models. However, these two algorithms are not sufficient for defining a general
framework. For that, we need to use the alternating direction method of
multipliers and proximal gradient methods. Finally, the combination of these
four algorithms (CD, ADMM, PO and Dykstra) allows us to consider allocation
models that cannot be cast into a QP form.\smallskip

In this paper, we have first considered allocation models with non-quadratic
objective functions. For example, we have used models based on the
diversification ratio, Shannon's entropy or the Kullback-Leibler divergence.
Second, we have solved regularized MVO models with non-linear penalty functions
such as the $\boldsymbol{\ell}_{p}$-norm penalty or the logarithmic barrier.
Third, we have discussed how to handle non-linear constraints. For instance, we
have imposed constraints on active share, volatility targeting, leverage
limits, transaction costs, etc. Most importantly, these three non-QP extensions
can be combined.\smallskip

With the development of quantitative strategies (smart beta, factor investing,
alternative risk premia, systematic strategies, robo-advisors, etc.), the asset
management industry has dramatically changed over the last five years. This is
just the beginning and we think that alternative data, machine learning methods
and artificial intelligence will massively shape investment processes in the
future. This paper is an illustration of this trend and shows how machine
learning optimization algorithms allow to move away from the traditional QP
world of portfolio management.

\clearpage

\appendix

\section*{Appendix}

\section{Mathematical results}

\subsection{QP problem when there is a benchmark}
\label{appendix:qp-benchmark}

Following Roncalli (2013), the excess return $\mathfrak{R}\left( x\mid b\right)
$\ of Portfolio $x$ with respect to Benchmark $b$ is the difference between the
return of the portfolio and the return of the benchmark:
\begin{equation*}
\mathfrak{R}\left( x\mid b\right) =\mathfrak{R}\left( x\right) -\mathfrak{R}
\left( b\right) =\left( x-b\right) ^{\top }\mathfrak{R}
\end{equation*}
It is easy to show that the expected excess return is equal to:
\begin{equation*}
\mu \left( x\mid b\right) =\mathbb{E}\left[ \mathfrak{R}\left( x\mid
b\right) \right] =\left( x-b\right) ^{\top }\mu
\end{equation*}%
whereas the volatility of the tracking error is given by:
\begin{equation*}
\sigma \left( x\mid b\right) =\sigma \left( \mathfrak{R}\left( x\mid
b\right) \right) =\sqrt{\left( x-b\right) ^{\top }\Sigma \left( x-b\right) }
\end{equation*}%
The objective function is then:%
\begin{eqnarray*}
f\left( x\mid b\right)  &=&\frac{1}{2}\left( x-b\right) ^{\top }\Sigma
\left( x-b\right) -\gamma \left( x-b\right) ^{\top }\mu  \\
&=&\frac{1}{2}x^{\top }\Sigma x-x^{\top }\left( \gamma \mu +\Sigma b\right)
+\left( \frac{1}{2}b^{\top }\Sigma b+\gamma b^{\top }\mu \right)  \\
&=&\frac{1}{2}x^{\top }Qx-x^{\top }R+C
\end{eqnarray*}%
where $C$ is a constant which does not depend on Portfolio $x$. We
recognize a QP problem where $Q=\Sigma $ and $R=\gamma \mu +\Sigma
b$.

\subsection{Augmented QP formulation of the turnover management problem}
\label{appendix:augmented-qp-turnover}

The augmented QP problem is defined by:
\begin{eqnarray*}
X^{\star } &=&\arg \min_X \frac{1}{2}X^{\top }QX-X^{\top }R \\
&\text{s.t.}&\left\{
\begin{array}{l}
AX=B \\
CX\leq D \\
\mathbf{0}_{3n}\leq X\leq \mathbf{1}_{3n}%
\end{array}%
\right.
\end{eqnarray*}%
where $X=\left( x_{1},\ldots ,x_{n},x_{1}^{-},\ldots
,x_{n}^{-},x_{1}^{+},\ldots ,x_{n}^{+}\right) $ is a $3n\times 1$ vector, $Q$
is a $3n\times 3n$ matrix:
\begin{equation*}
Q=\left(
\begin{array}{lll}
\Sigma  & \mathbf{0}_{n\times n} & \mathbf{0}_{n\times n} \\
\mathbf{0}_{n\times n} & \mathbf{0}_{n\times n} & \mathbf{0}_{n\times n} \\
\mathbf{0}_{n\times n} & \mathbf{0}_{n\times n} & \mathbf{0}_{n\times n}%
\end{array}%
\right)
\end{equation*}%
$R=\left( \gamma \mu ,\mathbf{0}_{n},\mathbf{0}_{n}\right) $ is a $3n\times 1 $
vector, $A$ is a $\left( n+1\right) \times 3n$ matrix:
\begin{equation*}
A=\left(
\begin{array}{rrr}
\mathbf{1}_{n}^{\top } & \mathbf{0}_{n}^{\top } & \mathbf{0}_{n}^{\top } \\
I_{n} & I_{n} & -I_{n}%
\end{array}%
\right)
\end{equation*}%
$B=\left( 1,\bar{x}\right) $ is a $\left( n+1\right) \times 1$ vector, $%
C=\left(
\begin{array}{ccc}
\mathbf{0}_{n}^{\top } & \mathbf{1}_{n}^{\top } & \mathbf{1}_{n}^{\top }%
\end{array}%
\right) $ is a $1\times 3n$ matrix and $D=\turnover ^{+}$.

\subsection{Augmented QP formulation of the MVO problem with transaction costs}
\label{appendix:augmented-qp-transaction-cost}

The augmented QP problem of dimension $3n$ is defined by:%
\begin{eqnarray*}
X^{\star } &=&\arg \min_X \frac{1}{2}X^{\top }QX-X^{\top }R \\
&\text{s.t.}&\left\{
\begin{array}{l}
AX=B \\
\mathbf{0}_{3n}\leq X\leq \mathbf{1}_{3n}%
\end{array}%
\right.
\end{eqnarray*}%
where $X=\left( x_{1},\ldots ,x_{n},x_{1}^{-},\ldots
,x_{n}^{-},x_{1}^{+},\ldots ,x_{n}^{+}\right) $ is a $3n\times 1$ vector, $Q$
is a $3n\times 3n$ matrix:
\begin{equation*}
Q=\left(
\begin{array}{lll}
\Sigma  & \mathbf{0}_{n\times n} & \mathbf{0}_{n\times n} \\
\mathbf{0}_{n\times n} & \mathbf{0}_{n\times n} & \mathbf{0}_{n\times n} \\
\mathbf{0}_{n\times n} & \mathbf{0}_{n\times n} & \mathbf{0}_{n\times n}%
\end{array}%
\right)
\end{equation*}%
$R=\left( \gamma \mu ,-c^{-},-c^{+}\right) $ is a $3n\times 1$ vector, $A$ is a
$\left( n+1\right) \times 3n$ matrix:
\begin{equation*}
A=\left(
\begin{array}{rrr}
\mathbf{1}_{n}^{\top } & \left( c^{-}\right) ^{\top } & \left( c^{+}\right)
^{\top } \\
I_{n} & I_{n} & -I_{n}%
\end{array}%
\right)
\end{equation*}%
and $B=\left( 1,\bar{x}\right) $ is a $\left( n+1\right) \times 1$ vector.

\subsection{QP problem with a hyperplane constraint}
\label{appendix:qp-hyperplane}

We consider the following QP problem:%
\begin{eqnarray*}
x^{\star } &=&\arg \min_{x}\frac{1}{2}x^{\top }Qx-x^{\top }R \\
&\text{s.t.}&a^{\top }x=b
\end{eqnarray*}%
The associated Lagrange function is:%
\begin{equation*}
\mathcal{L}\left( x;\lambda \right) =\frac{1}{2}x^{\top }Qx-x^{\top
}R+\lambda \left( a^{\top }x-b\right)
\end{equation*}%
The first-order conditions are then:%
\begin{equation*}
\left\{
\begin{array}{l}
\partial _{x}\,\mathcal{L}\left( x;\lambda \right) =Qx-R+\lambda a=\mathbf{0}%
_{n} \\
\partial _{\lambda }\,\mathcal{L}\left( x;\lambda \right) =a^{\top }x-b=0%
\end{array}%
\right.
\end{equation*}%
We obtain $x=Q^{-1}\left( R+\lambda a\right) $. Because $a^{\top }x-b=0$, we
have $a^{\top }Q^{-1}R+\lambda a^{\top }Q^{-1}a=b$ and:%
\begin{equation*}
\lambda ^{\star }=\frac{b-a^{\top }Q^{-1}R}{a^{\top }Q^{-1}a}
\end{equation*}%
The optimal solution is then:%
\begin{equation*}
x^{\star }=Q^{-1}\left( R+\frac{b-a^{\top }Q^{-1}R}{a^{\top }Q^{-1}a}%
a\right)
\end{equation*}

\subsection{Derivation of the soft-thresholding operator}
\label{appendix:soft-thresholding}

\begin{figure}[tbh]
\centering
\caption{Soft-thresholding operator $\mathcal{S}\left(v; \lambda\right)$}
\label{fig:soft1}
\figureskip
\includegraphics[width = \figurewidth, height = \figureheight]{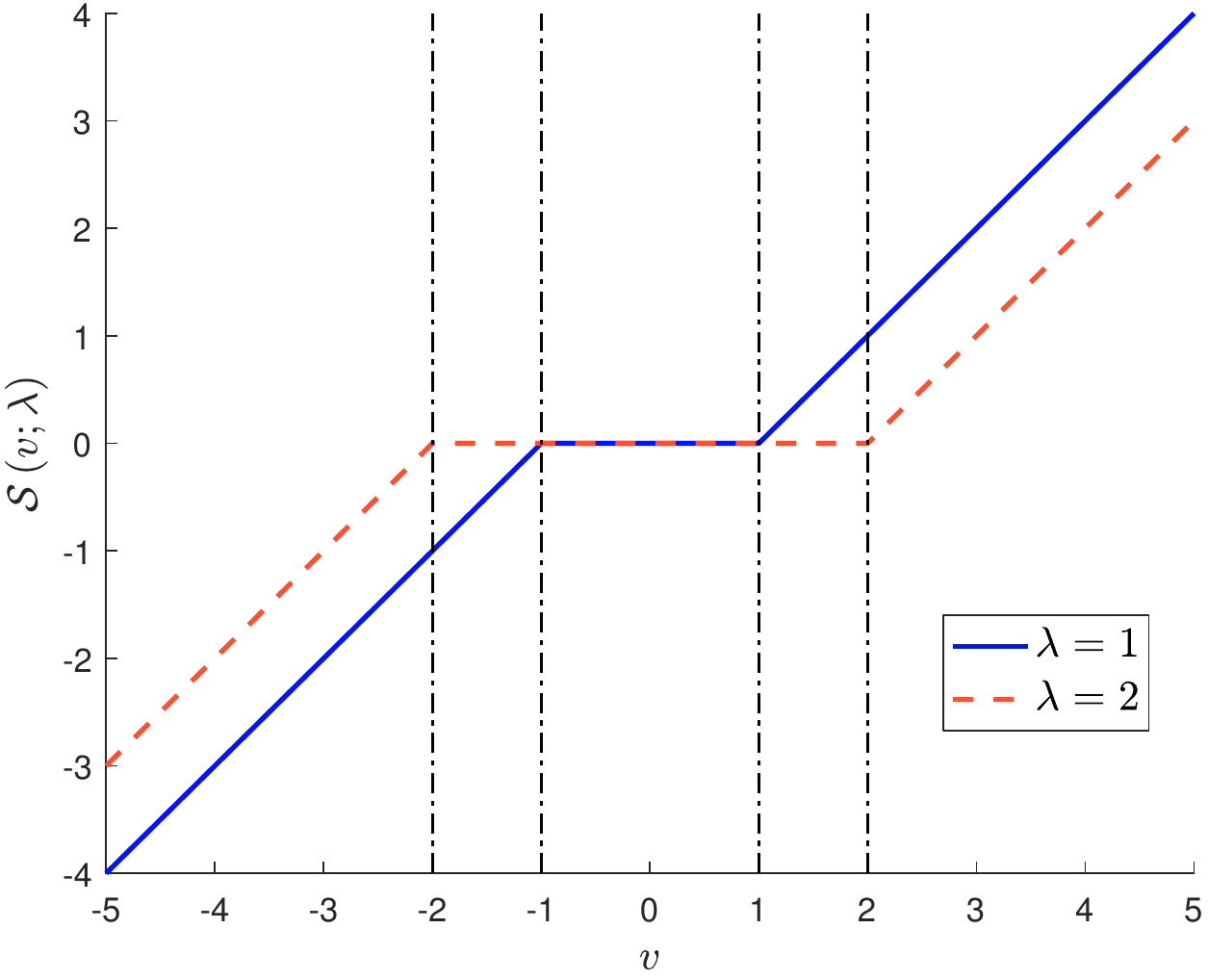}
\end{figure}

We consider the following equation:%
\begin{equation*}
cx-v+\lambda \partial \,\left\vert x\right\vert \in 0
\end{equation*}%
where $c>0$ and $\lambda >0$. Since we have $\partial \left\vert
x\right\vert =\limfunc{sign}\left( x\right) $, we deduce that:%
\begin{equation*}
x^{\star }=\left\{
\begin{array}{ll}
c^{-1}\left( v+\lambda \right)  & \text{if }x^{\star }<0 \\
0 & \text{if }x^{\star }=0 \\
c^{-1}\left( v-\lambda \right)  & \text{if }x^{\star }>0%
\end{array}%
\right.
\end{equation*}%
If $x^{\star }<0$ or $x^{\star }>0$, then we have $v+\lambda <0$ or $%
v-\lambda >0$. This is equivalent to set $\left\vert v\right\vert >\lambda >0
$. The case $x^{\star }=0$ implies that $\left\vert v\right\vert \leq
\lambda $. We deduce that:%
\begin{equation*}
x^{\star }=c^{-1}\cdot \mathcal{S}\left( v; \lambda\right)
\end{equation*}%
where $\mathcal{S}\left( v; \lambda\right)$ is the soft-thresholding operator:%
\begin{eqnarray*}
\mathcal{S}\left( v; \lambda\right)  &=&\left\{
\begin{array}{ll}
0 & \text{if }\left\vert v\right\vert \leq \lambda  \\
v-\lambda \limfunc{sign}\left( v\right)  & \text{otherwise}%
\end{array}%
\right.  \\
&=&\limfunc{sign}\left( v\right) \cdot \left( \left\vert v\right\vert
-\lambda \right) _{+}
\end{eqnarray*}
In Figure \ref{fig:soft1}, we have represented the function $\mathcal{S}\left(
v; \lambda\right)$ when $\lambda$ is respectively equal to $1$ and $2$.

\begin{remark}
The soft-thresholding operator is the proximal operator of the $\boldsymbol{\ell}_{1}$%
-norm $f\left( x\right) =\left\Vert x\right\Vert _{1}$. Indeed, we have $%
\mathbf{prox}_{f}\left( v\right) =\mathcal{S}\left( v;1\right) $ and $%
\mathbf{prox}_{\lambda f}\left( v\right) =\mathcal{S}\left( v;\lambda
\right) $.
\end{remark}

\subsection{The box-constrained QP problem}
\label{appendix:qp-ccd}

We consider the box-constrained QP problem:
\begin{eqnarray}
x^{\star } &=&\arg \min_x \frac{1}{2}x^{\top }Qx-x^{\top }R
\label{eq:appendix-ccd-qp1} \\
& \text{s.t.} & x^{-} \leq  x \leq x^{+} \notag
\end{eqnarray}%
The objective function is equal to:%
\begin{eqnarray*}
f\left( x\right)  &=&\frac{1}{2}x^{\top }Qx-x^{\top }R \\
&=&\frac{1}{2}\sum_{i=1}^{n}x_{i}\sum_{j=1}^{n}Q_{i,j}x_{j}-%
\sum_{i=1}^{n}x_{i}R_{i} \\
&=&\frac{1}{2}\sum_{i=1}^{n}x_{i}\left( Q_{i,i}x_{i}+\sum_{j\neq
i}Q_{i,j}x_{j}\right) -\sum_{i=1}^{n}x_{i}R_{i}
\end{eqnarray*}%
We deduce that:
\begin{equation*}
\frac{\partial \,f\left( x\right) }{\partial \,x_{i}}=\frac{1}{2}\left(
2x_{i}Q_{i,i}+\sum_{j\neq i}x_{j}\left( Q_{i,j}+Q_{j,i}\right) \right) -R_{i}
\end{equation*}%
We notice that:%
\begin{equation*}
\frac{\partial \,f\left( x\right) }{\partial \,x_{i}}=0\Leftrightarrow x_{i}=%
\frac{1}{Q_{i,i}}\left( R_{i}-\sum_{j\neq i}x_{j}\left( \frac{Q_{i,j}+Q_{j,i}%
}{2}\right) \right)
\end{equation*}%
The Lagrange function associated to Problem (\ref{eq:appendix-ccd-qp1}) is
equal to:%
\begin{equation*}
\mathcal{L}\left( x;\lambda ^{-},\lambda ^{+}\right) =f\left( x\right)
-\sum_{i=1}^{n}\lambda _{i}^{-}\left( x_{i}-x_{i}^{-}\right)
-\sum_{i=1}^{n}\lambda _{i}^{+}\left( x_{i}^{+}-x_{i}\right)
\end{equation*}
The first-order condition is then:%
\begin{equation*}
\frac{\partial \,\mathcal{L}\left( x;\lambda ^{-},\lambda ^{+}\right) }{%
\partial \,x_{i}}=\frac{\partial \,f\left( x\right) }{\partial \,x_{i}}%
-\lambda _{i}^{-}+\lambda _{i}^{+}=0
\end{equation*}%
Since the Kuhn-Tucker conditions are:%
\begin{equation*}
\left\{
\begin{array}{c}
\min \left( \lambda _{i}^{-},x_{i}-x_{i}^{-}\right) =0 \\
\min \left( \lambda _{i}^{+},x_{i}^{+}-x_{i}\right) =0%
\end{array}%
\right.
\end{equation*}%
we obtain three cases:

\begin{enumerate}
\item If no bound is reached, we have $\lambda _{i}^{-}=\lambda _{i}^{+}=0$
and the solution is equal to:%
\begin{equation*}
x_{i}^{\star }=\frac{1}{Q_{i,i}}\left( R_{i}-\sum_{j\neq i}x_{j}\left( \frac{%
Q_{i,j}+Q_{j,i}}{2}\right) \right)
\end{equation*}

\item If the lower bound is reached, we have $\lambda _{i}^{-}>0$, $\lambda
    _{i}^{+}=0$ and $x_{i}^{\star }=x_{i}^{-}$.

\item If the upper bound is reached, we have $\lambda _{i}^{-}=0$, $\lambda
    _{i}^{+}>0$ and $x_{i}^{\star }=x_{i}^{+}$.
\end{enumerate}

\subsection{ADMM algorithm}
\label{appendix:admm}

The optimization problem is defined as:
\begin{eqnarray}
\left\{ x^{\star },y^{\star }\right\}  &=&\arg \min_{\left( x,y\right)
}f_{x}\left( x\right) +f_{y}\left( y\right)   \label{eq:appendix-admm1} \\
&\text{s.t.}&Ax+By=c  \notag
\end{eqnarray}%
The derivation of the algorithm is fully explained in Boyd
\textsl{et al.} (2011). For that, they consider the augmented
Lagrange function:
\begin{equation}
\mathcal{L}\left( x,y;\lambda ,\varphi \right) =f_{x}\left( x\right)
+f_{y}\left( y\right) +\lambda ^{\top }\left( Ax+By-c\right) +\frac{\varphi
}{2}\left\Vert Ax+By-c\right\Vert _{2}^{2}  \label{eq:appendix-admm2}
\end{equation}%
where $\varphi >0$. According to Boyd \textsl{et al.} (2011), the
$\boldsymbol{\ell }_{2}$-norm penalty adds robustness to the dual ascent method
and accelerates its convergence. The ADMM algorithm uses the property that the
objective function is separable, and consists of the following iterations:
\begin{eqnarray*}
x^{\left( k+1\right) } &=&\arg \min_{x}\mathcal{L}\left( x,y^{\left(
k\right) };\lambda ^{\left( k\right) },\varphi \right)  \\
&=&\arg \min_{x}\left\{ f_{x}\left( x\right) +\lambda ^{\left( k\right)
^{\top }}\left( Ax+By^{\left( k\right) }-c\right) +\frac{\varphi }{2}%
\left\Vert Ax+By^{\left( k\right) }-c\right\Vert _{2}^{2}\right\}
\end{eqnarray*}%
and:%
\begin{eqnarray*}
y^{\left( k+1\right) } &=&\arg \min_{y}\mathcal{L}\left( x^{\left(
k+1\right) },y;\lambda ^{\left( k\right) },\varphi \right)  \\
&=&\arg \min_{y}\left\{ f_{y}\left( y\right) +\lambda ^{\left( k\right)
^{\top }}\left( Ax^{\left( k+1\right) }+By-c\right) +\frac{\varphi }{2}%
\left\Vert Ax^{\left( k+1\right) }+By-c\right\Vert _{2}^{2}\right\}
\end{eqnarray*}%
The update for the dual variable $\lambda $ is then:
\begin{equation*}
\lambda ^{\left( k+1\right) }=\lambda ^{\left( k\right) }+\varphi \left(
Ax^{\left( k+1\right) }+By^{\left( k+1\right) }-c\right)
\end{equation*}%
We repeat the iterations until convergence.\smallskip

Boyd \textsl{et al.} (2011) notice that the previous algorithm can be
simplified. Let $r=Ax+By-c$ be the (primal) residual. By combining linear
and quadratic terms, we have:
\begin{equation*}
\lambda ^{\top }r+\frac{\varphi }{2}\Vert r \Vert_{2}^{2}=\frac{\varphi }{2}\left\Vert
r+u\right\Vert ^{2}-\frac{\varphi }{2}\left\Vert u\right\Vert ^{2}
\end{equation*}%
where $u=\varphi ^{-1}\lambda $ is the \textit{scaled} dual variable. We can
then write the Lagrange function (\ref{eq:appendix-admm2}) as follows:
\begin{eqnarray*}
\mathcal{L}\left( x,y;u,\varphi \right)  &=&f_{x}\left( x\right)
+f_{y}\left( y\right) +\frac{\varphi }{2}\left\Vert Ax+By-c+u\right\Vert
_{2}^{2}-\frac{\varphi }{2}\left\Vert u\right\Vert ^{2}  \\
&=&f_{x}\left( x\right) +f_{y}\left( y\right) +\frac{\varphi }{2}\left\Vert
Ax+By-c+u\right\Vert _{2}^{2}-\frac{1}{2\varphi }\left\Vert \lambda
\right\Vert ^{2}
\end{eqnarray*}%
Since the last term is a constant, we deduce that the $x$- and $y$-updates
become:
\begin{eqnarray}
x^{\left( k+1\right) } &=&\arg \min_{x}\mathcal{L}\left( x,y^{\left(
k\right) };u^{\left( k\right) },\varphi \right)   \notag \\
&=&\arg \min_{x}\left\{ f_{x}\left( x\right) +\frac{\varphi }{2}\left\Vert
Ax+By^{\left( k\right) }-c+u^{\left( k\right) }\right\Vert _{2}^{2}\right\}
\label{eq:appendix-admm3a}
\end{eqnarray}%
and:%
\begin{eqnarray}
y^{\left( k+1\right) } &=&\arg \min_{y}\mathcal{L}\left( x^{\left(
k+1\right) },y;u^{\left( k\right) },\varphi \right)   \notag \\
&=&\arg \min_{y}\left\{ f_{y}\left( y\right) +\frac{\varphi }{2}\left\Vert
Ax^{\left( k+1\right) }+By-c+u^{\left( k\right) }\right\Vert
_{2}^{2}\right\}   \label{eq:appendix-admm3b}
\end{eqnarray}%
For the scaled dual variable $u$, we have:
\begin{eqnarray}
u^{\left( k+1\right) } &=&u^{\left( k\right) }+r^{\left( k+1\right) }  \notag
\\
&=&u^{\left( k\right) }+\left( Ax^{\left( k+1\right) }+By^{\left( k+1\right)
}-c\right)   \label{eq:appendix-admm3c}
\end{eqnarray}%
where $r^{\left( k+1\right) }=Ax^{\left( k+1\right) }+By^{\left(
k+1\right) }-c$ is the primal residual at iteration $k+1$. Boyd
\textsl{et al.} (2011) also define the variable $s^{\left(
k+1\right) }=\varphi A^{\top }B\left( y^{\left( k+1\right)
}-y^{\left( k\right) }\right) $ and refer to $s^{\left( k+1\right)
}$ as the dual residual at iteration $k+1$.\smallskip

Under the assumption that the traditional Lagrange function $\mathcal{L}\left(
x,y;\lambda ,0\right) $ has a saddle point, one can prove that the residual
$r^{\left( k\right) }$ converges to zero, the objective function $f_{x}\left(
x^{\left( k\right) }\right) +f_{y}\left( y^{\left( k\right) }\right) $
converges to the optimal value $f_{x}\left( x^{\star }\right) +f_{y}\left(
y^{\star }\right)$, and the dual variable $\lambda ^{\left( k\right) } =
\varphi u^{\left( k\right) }$ converges to a dual optimal point. However, the
rate of convergence is not known and the primal variables $x^{\left( k\right)
}$ and $y^{\left( k\right) }$ do not necessarily converge to the optimal values
$x^{\star }$ and $y^{\star }$. In general, the stopping criterion is defined
with respect to the residuals:
\begin{equation*}
\left\{
\begin{array}{l}
\left\Vert r^{\left( k+1\right) }\right\Vert _{2}\leqslant \varepsilon  \\[0.1cm]
\left\Vert s^{\left( k+1\right) }\right\Vert _{2}\leqslant \varepsilon
^{\prime }%
\end{array}%
\right.
\end{equation*}%
Typical values when implementing this stopping criterion are
$\varepsilon =\varepsilon ^{\prime }=10^{-15}$ (Bourgeron \textsl{et
al.}, 2018).\smallskip

From a theoretical point of view, the convergence holds regardless of the
choice of the penalization parameter $\varphi > 0$. But the choice of $ \varphi
$ affects the convergence\label{appendix:admm-varphi} rate (Ghadimi \textsl{et
al.}, 2015; Giselsson and Boyd, 2017). In practice, the penalization parameter
$\varphi $ may be changed at each iteration, implying that $\varphi $ is
replaced by $\varphi ^{\left( k\right) }$ and the scaled dual variable $u^{k}$
is equal to $\lambda ^{\left( k\right) }/\varphi ^{\left( k\right) }$. This may
improve the convergence and make the performance independent of the initial
choice $\varphi ^{\left( 0\right) }$. To update $\varphi ^{\left( k\right) }$
in practice, He \textsl{et al.} (2000) and Wang and Liao (2001) provide a
simple and efficient scheme. On the one hand, the $x$- and $y$-updates in ADMM
essentially come from placing a penalty on $\left\Vert r^{\left( k\right)
}\right\Vert _{2}^{2}$. As a consequence, if $\varphi ^{\left( k\right) }$ is
large, $\left\Vert r^{\left( k\right) }\right\Vert _{2}^{2}$ tends to be small.
On the other hand, $s^{\left( k\right) }$ depends linearly on $\varphi $. As a
consequence, if $\varphi ^{\left( k\right) }$ is small, $\left\Vert s^{\left(
k\right) }\right\Vert _{2}^{2}$ is small. To keep $\left\Vert r^{\left(
k\right) }\right\Vert _{2}^{2}$ and $\left\Vert s^{\left( k\right) }\right\Vert
_{2}^{2}$ within a factor $\mu $, one may consider\label{appendix:admm-varphi}:
\begin{equation*}
\varphi ^{\left( k+1\right) }=\left\{
\begin{array}{ll}
\tau \varphi ^{\left( k\right) } & \text{if }\left\Vert r^{\left( k\right)
}\right\Vert _{2}^{2}>\mu \left\Vert s^{\left( k\right) }\right\Vert _{2}^{2}
\\
\varphi ^{\left( k\right) }/\tau ^{\prime } & \text{if }\left\Vert s^{\left(
k\right) }\right\Vert _{2}^{2}>\mu \left\Vert r^{\left( k\right)
}\right\Vert _{2}^{2} \\
\varphi ^{\left( k\right) } & \text{otherwise}%
\end{array}%
\right.
\end{equation*}%
where $\mu $, $\tau $ and $\tau ^{\prime }$ are parameters that are greater
than one. In practice, we use $\varphi ^{\left( 0\right) }=1$, $u^{\left(
0\right) }=\mathbf{0}_{p}$, $\mu =10^{3}$ and $\tau =\tau ^{\prime }=2$.

\begin{remark}
The constant case $\varphi ^{\left( k+1\right) } = \varphi ^{\left( k\right) }
= \varphi ^{\left( 0\right) }$ is obtained by setting $\tau = \tau^{\prime} =
1$.
\end{remark}

\subsection{Proximal operators}
\label{appendix:proximal}

\subsubsection{Pointwise maximum function}
\label{appendix:proximal-max-function}

The unit simplex is the generalization of the triangle:%
\begin{equation*}
\mathbb{S}_{n}=\left\{ x\in \left[ 0,1\right] ^{n},\theta _{i}\geq
0:x=\sum_{i=0}^{n}\theta _{i}e_{i},\sum_{i=0}^{n}\theta _{i}=1,\mathbf{1}%
_{n}^{\top }x\leq 1\right\}
\end{equation*}%
where $e_{0}$ is the zero vector and $e_{i}$ are the unit vectors for $i\geq
1$. Beck (2017) shows that $\mathcal{P}_{\mathbb{S}_{n}}\left( v\right) =\left(
v-\mu ^{\star }\mathbf{1}_{n}\right) _{+}$ where $\mu ^{\star }$ is
the root of the equation $\mathbf{1}_{n}^{\top }\left( v-\mu ^{\star }%
\mathbf{1}_{n}\right) _{+}=1$. In the case of the pointwise maximum function
$f\left( x\right) =\max x$, the Moreau decomposition gives:
\begin{eqnarray*}
\mathbf{prox}_{\lambda \max x}\left( v\right)  &=&v-\lambda \mathcal{P}_{%
\mathbb{S}_{n}}\left( \frac{v}{\lambda }\right)  \\
&=&v-\lambda \left( \frac{v}{\lambda }-\mu ^{\star }\mathbf{1}_{n}\right)
_{+}
\end{eqnarray*}%
where $\mu ^{\ast }$ is the root of the equation:
\begin{eqnarray*}
\mathbf{1}_{n}^{\top }\left( \frac{v}{\lambda }-\mu ^{\star }\mathbf{1}%
_{n}\right) _{+}=1 &\Leftrightarrow &\sum_{i=1}^{n}\left( \frac{v_{i}}{%
\lambda }-\mu ^{\star }\right) _{+}=1 \\
&\Leftrightarrow &\sum_{i=1}^{n}\left( v_{i}-s^{\star }\right) _{+}=\lambda
\end{eqnarray*}%
where $s^{\star }=\lambda \mu ^{\star }$. It follows that:
\begin{eqnarray*}
\mathbf{prox}_{\lambda \max x}\left( v\right)  &=&v-\left( v-s^{\star }%
\mathbf{1}_{n}\right) _{+} \\
&=&\min \left( v,s^{\star }\right)
\end{eqnarray*}

\subsubsection{$\boldsymbol{\ell }_{2}$-norm function}

The projection onto the unit ball $\mathcal{B}_{2}\left( \mathbf{0},1\right)
=\left\{ x\in \mathbb{R}^{n}:\left\Vert x\right\Vert _{2}\leq 1\right\} $ is
equal to:
\begin{equation*}
\mathcal{P}_{\mathcal{B}_{2}\left( \mathbf{0},1\right) }\left( v\right)
=\left\{
\begin{array}{ll}
v & \text{if }\left\Vert v\right\Vert _{2}\leq 1 \\
\dfrac{v}{\left\Vert v\right\Vert _{2}} & \text{if }\left\Vert v\right\Vert
_{2}>1%
\end{array}%
\right.
\end{equation*}%
Since we have:
\begin{equation*}
\mathbf{prox}_{\lambda \left\Vert x\right\Vert _{2}}\left( v\right) +\lambda
\mathcal{P}_{\mathcal{B}_{2}\left( \mathbf{0},1\right) }\left( \frac{v}{%
\lambda }\right) =v
\end{equation*}%
we deduce that:%
\begin{eqnarray*}
\mathbf{prox}_{\lambda \left\Vert x\right\Vert _{2}}\left( v\right)
&=&v-\lambda \mathcal{P}_{\mathcal{B}_{2}\left( \mathbf{0},1\right) }\left(
\frac{v}{\lambda }\right)  \\
&=&\left\{
\begin{array}{ll}
v-\lambda \left( \dfrac{v}{\lambda }\right)  & \text{if }\left\Vert
v\right\Vert _{2}\leq \lambda  \\[0.25cm]
v-\lambda \dfrac{\dfrac{v}{\lambda }}{\left\Vert \dfrac{v}{\lambda }\right\Vert
_{2}} & \text{if }\left\Vert v\right\Vert _{2}>\lambda
\end{array}%
\right.  \\
&=&\left\{
\begin{array}{ll}
0 & \text{if }\left\Vert v\right\Vert _{2}\leq \lambda  \\
v-\dfrac{\lambda v}{\left\Vert v\right\Vert _{2}} & \text{if }\left\Vert
v\right\Vert _{2}>\lambda
\end{array}%
\right.  \\
&=&\left( 1-\frac{\lambda }{\max \left( \lambda ,\left\Vert v\right\Vert
_{2}\right) }\right) v
\end{eqnarray*}

\subsubsection{Scaling and translation}
\label{appendix:proximal-scaling}

Let $g\left( x\right) =f\left( ax+b\right) $ where $a\neq 0$. Using the
change of variable $y=ax+b$, we have:%
\begin{eqnarray*}
\mathbf{prox}_{g}\left( v\right)  &=&\arg \min_{x}\left\{ g\left( x\right) +%
\frac{1}{2}\left\Vert x-v\right\Vert _{2}^{2}\right\}  \\
&=&\arg \min_{x}\left\{ f\left( ax+b\right) +\frac{1}{2}\left\Vert
x-v\right\Vert _{2}^{2}\right\}  \\
&=&\arg \min_{y}\left\{ f\left( y\right) +\frac{1}{2}\left\Vert \frac{y-b}{a}%
-v\right\Vert _{2}^{2}\right\}
\end{eqnarray*}%
We deduce that:%
\begin{eqnarray*}
f\left( y\right) +\frac{1}{2}\left\Vert \frac{y-b}{a}-v\right\Vert _{2}^{2}
&=&f\left( y\right) +\frac{1}{2a^{2}}\left\Vert y-b-av\right\Vert _{2}^{2} \\
&=&\frac{1}{a^{2}}\left( a^{2}f\left( y\right) +\frac{1}{2}\left\Vert
y-\left( av+b\right) \right\Vert _{2}^{2}\right)
\end{eqnarray*}%
We conclude that $y^{\star }=\mathbf{prox}_{a^{2}f}\left( av+b\right) $ and:%
\begin{eqnarray*}
\mathbf{prox}_{g}\left( v\right)  &=&\frac{y^{\star }-b}{a} \\
&=&\frac{\mathbf{prox}_{a^{2}f}\left( av+b\right) -b}{a}
\end{eqnarray*}

\subsubsection{Projection onto the $\boldsymbol{\ell }_{1}$ ball}

We have:%
\begin{eqnarray*}
x &=&\mathcal{P}_{\mathcal{B}_{1}\left( c,r\right) }\left( v\right)  \\
&=&\mathcal{P}_{\mathcal{B}_{1}\left( \mathbf{0}_{n},r\right) }\left(
v-c\right) +c \\
&=&\left( v-c\right) -\limfunc{sign}\left( v-c\right) \odot \mathbf{prox}%
_{r\max x}\left( \left\vert v-c\right\vert \right) +c \\
&=&v-\limfunc{sign}\left( v-c\right) \odot \min \left( \left\vert
v-c\right\vert ,s^{\star }\right)
\end{eqnarray*}%
where $s^{\star }$ is the solution of the following equation:%
\begin{equation*}
s^{\star }=\left\{ s\in \mathbb{R}:\sum_{i=1}^{n}\left( \left\vert
v_{i}-c_{i}\right\vert -s\right) _{+}=r\right\}
\end{equation*}

\begin{remark}
$\mathcal{P}_{\mathcal{B}_{1}\left( c,r\right) }\left( v\right) $ is
sometimes expressed using the soft-thresholding operator (Beck, 2017, page
151), but the two formulas are equivalent.
\end{remark}

\subsubsection{Projection onto the $\boldsymbol{\ell }_{2}$ ball}

We have:%
\begin{eqnarray}
x &=&\mathcal{P}_{\mathcal{B}_{2}\left( c,r\right) }\left( v\right) \notag \\
&=&\mathcal{P}_{\mathcal{B}_{2}\left( \mathbf{0}_{n},r\right) }\left(v-c\right) +c \notag \\
&=&\left( v-c\right) -\mathbf{prox}_{r\left\Vert x\right\Vert _{2}}\left(v-c\right) +c \notag \\
&=&v-\left( 1-\frac{r}{\max \left( r,\left\Vert v-c\right\Vert _{2}\right) }\right) \left( v-c\right) \notag \\
&=&c+\frac{r}{\max \left( r,\left\Vert v-c\right\Vert _{2}\right) }\left(v-c\right) \label{eq:proximal-ball2}
\end{eqnarray}

\subsubsection{$\boldsymbol{\ell }_{2}$-penalized logarithmic barrier function}
\label{appendix:proximal-penalized-logarithmic}

We note $f_{1}\left( x\right) =-\lambda \sum_{i=1}^{n}b_{i}\ln x_{i}$ and
$f_{2}\left( x\right) =\mathbf{\mathds{1}}_{\Omega }\left( x\right) $ where
$\Omega =\mathcal{B}_{2}\left( c,r\right) $. The Dykstra's algorithm becomes:
\begin{equation*}
\left\{
\begin{array}{l}
v_{x}^{\left( k\right) }=y^{\left( k\right) }+z_{1}^{\left( k\right) } \\
x^{\left( k+1\right) }=\mathbf{prox}_{f_{1}}\left( v_{x}^{\left( k\right)
}\right)  \\
z_{1}^{\left( k+1\right) }=y^{\left( k\right) }+z_{1}^{\left( k\right)
}-x^{\left( k+1\right) } \\
v_{y}^{\left( k\right) }=x^{\left( k+1\right) }+z_{2}^{\left( k\right) } \\
y^{\left( k+1\right) }=\mathbf{prox}_{f_{2}}\left( v_{y}^{\left( k\right)
}\right)  \\
z_{2}^{\left( k+1\right) }=x^{\left( k+1\right) }+z_{2}^{\left( k\right)
}-y^{\left( k+1\right) }%
\end{array}%
\right.
\end{equation*}%
It follows that:%
\begin{equation*}
x^{\left( k+1\right) }=\frac{v_{x}^{\left( k\right) }+\sqrt{v_{x}^{\left(
k\right) }\odot v_{x}^{\left( k\right) }+4\lambda b}}{2}
\end{equation*}%
and:%
\begin{eqnarray*}
y^{\left( k+1\right) } &=&\mathcal{P}_{\mathcal{B}_{2}\left( c,r\right)
}\left( v_{y}^{\left( k\right) }\right)  \\
&=&c+\frac{r}{\max \left( r,\left\Vert v_{y}^{\left( k\right) }-c\right\Vert
_{2}\right) }\left( v_{y}^{\left( k\right) }-c\right)
\end{eqnarray*}

\subsubsection{Quadratic function}
\label{appendix:proximal-quadratic}

Let $f\left( x\right) =\dfrac{1}{2}x^{\top }Qx-x^{\top }R$. We have:%
\begin{eqnarray*}
\mathbf{prox}_{f}\left( v\right)  &=&\arg \min_{x}\left\{ \frac{1}{2}x^{\top
}Qx-x^{\top }R+\frac{1}{2}\left\Vert x-v\right\Vert _{2}^{2}\right\}  \\
&=&\arg \min_{x}\left\{ \frac{1}{2}x^{\top }\left( Q+I_{n}\right) x-x^{\top
}\left( R+v\right) +\frac{1}{2}v^{\top }v\right\}  \\
&=&\left( Q+I_{n}\right) ^{-1}\left( R+v\right)
\end{eqnarray*}

\subsubsection{Projection onto the intersection of a $\boldsymbol{\ell }_{2}$
ball and a box} \label{appendix:proximal-ball-box}

We note $f_{1}\left( x\right) =\mathbf{\mathds{1}}_{\Omega _{1}}\left( x\right)
$ and $f_{2}\left( x\right) =\mathbf{\mathds{1}}_{\Omega _{2}}\left( x\right) $
where $\Omega _{1}=\mathcal{B}_{2}\left( c,r\right) $ and $\Omega
_{2}=\mathcal{B}_{ox}\left[ x^{-},x^{+}\right] =\left\{ x\in
\mathbb{R}^{n}:x^{-}\leq x\leq x^{+}\right\} $. The Dykstra's algorithm
becomes:
\begin{equation*}
\left\{
\begin{array}{l}
x^{\left( k+1\right) }=c+\dfrac{r}{\max \left( r,\left\Vert y^{\left(
k\right) }+z_{1}^{\left( k\right) }-c\right\Vert _{2}\right) }\left(
y^{\left( k\right) }+z_{1}^{\left( k\right) }-c\right)  \\
z_{1}^{\left( k+1\right) }=y^{\left( k\right) }+z_{1}^{\left( k\right)
}-x^{\left( k+1\right) } \\
y^{\left( k+1\right) }=\mathcal{T}\left( x^{\left( k+1\right)
}+z_{2}^{\left( k\right) };x^{-},x^{+}\right)  \\
z_{2}^{\left( k+1\right) }=x^{\left( k+1\right) }+z_{2}^{\left( k\right)
}-y^{\left( k+1\right) }%
\end{array}%
\right.
\end{equation*}%
This algorithm is denoted by $\mathcal{P}_{\mathcal{B}\mathrm{ox}-\mathcal{B}%
\mathrm{all}}\left( v;x^{-},x^{+},c,r\right) $.

\subsubsection{Shannon's entropy and Kullback-Leibler divergence}
\label{appendix:proximal-shannon} \label{appendix:proximal-kullblack_leibler}

If we consider the scalar function $f\left( x\right) =\lambda x\ln \left( x/%
\tilde{x}\right) $ where $\tilde{x}$ is a constant, we have:
\begin{eqnarray*}
\lambda f\left( x\right) +\frac{1}{2}\left\Vert x-v\right\Vert _{2}^{2} &=&\lambda
x\ln \frac{x}{\tilde{x}}+\frac{1}{2}\left( x-v\right) ^{2} \\
&=&\lambda x\ln \frac{x}{\tilde{x}}+\frac{1}{2}x^{2}-xv+\frac{1}{2}v^{2}
\end{eqnarray*}%
The first-order condition is:%
\begin{eqnarray*}
\lambda \frac{1}{\tilde{x}}+\lambda \ln \frac{x}{\tilde{x}}+x-v=0
&\Leftrightarrow &\ln x+\lambda ^{-1}x=\ln \tilde{x}+\lambda ^{-1}v-\frac{1}{%
\tilde{x}} \\
&\Leftrightarrow &e^{\ln x+\lambda ^{-1}x}=e^{\lambda ^{-1}v-\frac{1}{\tilde{%
x}}+\ln \tilde{x}} \\
&\Leftrightarrow &xe^{\lambda ^{-1}x}=\tilde{x}e^{\lambda ^{-1}v-\frac{1}{%
\tilde{x}}} \\
&\Leftrightarrow &\left( \lambda ^{-1}x\right) e^{\left( \lambda
^{-1}x\right) }=\lambda ^{-1}\tilde{x}e^{\lambda ^{-1}v-\frac{1}{\tilde{x}}}
\end{eqnarray*}%
We deduce that the root is equal to:%
\begin{equation*}
x^{\star }=\lambda W\left( \frac{\tilde{x}e^{\lambda ^{-1}v-\frac{1}{\tilde{x%
}}}}{\lambda }\right)
\end{equation*}
where $W\left( x\right) $ is the Lambert $W$ function satisfying
$W\left( x\right) e^{W\left( x\right) }=x$ (Corless \textit{et al.},
1996). In the case of the Kullback-Liebler divergence
$\mathrm{\limfunc{KL}}\left( x\right) =\sum_{i=1}^{n}x_{i}\ln
\left(x_{i}/ \tilde{x}_{i}\right)$, it follows that:
\begin{equation*}
\mathbf{prox}_{\lambda \limfunc{KL}\left( v\mid \tilde{x}\right) }\left( v\right) =\lambda
\left(
\begin{array}{c}
W\left( \lambda ^{-1}\tilde{x}_{1}e^{\lambda ^{-1}v_{1}-\tilde{x}_{1}^{-1}}\right)  \\
\vdots  \\
W\left( \lambda ^{-1}\tilde{x}_{n}e^{\lambda ^{-1}v_{n}-\tilde{x}_{n}^{-1}}\right)
\end{array}%
\right)
\end{equation*}

\begin{remark}
The proximal of Shannon's entropy $\mathrm{\limfunc{SE}}\left( x\right)
=-\sum_{i=1}^{n}x_{i}\ln x_{i}$ is a special case of the previous
result\footnote{We use the fact that $\max \func{SE}\left( x\right) =\min
-\func{SE}\left( x\right) $.} with $\tilde{x}_{i}=1$:%
\begin{equation*}
\mathbf{prox}_{\lambda \limfunc{SE}\left( x\right) }\left( v\right) =\lambda
\left(
\begin{array}{c}
W\left( \lambda ^{-1}e^{\lambda ^{-1}v_{1}-1}\right)  \\
\vdots  \\
W\left( \lambda ^{-1}e^{\lambda ^{-1}v_{n}-1}\right)
\end{array}%
\right)
\end{equation*}
This result has been first obtained by Chaux \textsl{et al.} (2007).
\end{remark}

\subsubsection{Projection onto the complement $\bar{\mathcal{B}}_{2}\left(
c,r\right) $ of the $\boldsymbol{\ell }_{2}$ ball}

We consider the following proximal problem:
\begin{equation*}
x^{\star }=\arg \min_{x}\left\{ \mathds{1}_{\Omega }\left( x\right) +\frac{1}{2}%
\left\Vert x-v\right\Vert _{2}^{2}\right\}
\end{equation*}%
where:%
\begin{equation*}
\Omega =\left\{ x\in \mathbb{R}^{n}:\left\Vert x-c\right\Vert _{2}\geq
r\right\}
\end{equation*}%
This problem is equivalent to:
\begin{eqnarray*}
x^{\star } &=&\arg \min_{x}\frac{1}{2}\left( x-v\right) ^{\top }\left(
x-v\right)  \\
&\text{s.t.}&\left( x-c\right) ^{\top }\left( x-c\right) -r^{2}\geq 0
\end{eqnarray*}%
We deduce that the Lagrange function is equal to:
\begin{equation*}
\mathcal{L}\left( x;\lambda \right) =\frac{1}{2}\left( x-v\right) ^{\top
}\left( x-v\right) -\lambda \left( \left( x-c\right) ^{\top }\left(
x-c\right) -r^{2}\right)
\end{equation*}%
The first-order condition is:%
\begin{equation*}
\frac{\partial \,\mathcal{L}\left( x;\lambda \right) }{\partial \,x}%
=x-v-2\lambda \left( x-c\right) =\mathbf{0}_{n}
\end{equation*}%
whereas the KKT condition is $\min \left( \lambda ,\left( x-c\right) ^{\top
}\left( x-c\right) -r^{2}\right) =0$. We distinguish two cases:
\begin{enumerate}
\item If $\lambda =0$, this means that $x^{\star }=v$ and $\left( x-c\right)
    ^{\top }\left( x-c\right) -r^{2}>0$.

\item If $\lambda >0$, we have $\left( x-c\right) ^{\top }\left( x-c\right)
    =r^{2}$. Then we obtain the following system:
\begin{equation*}
\left\{
\begin{array}{l}
x-v-2\lambda \left( x-c\right) =\mathbf{0}_{n} \\
\left( x-c\right) ^{\top }\left( x-c\right) =r^{2}%
\end{array}%
\right.
\end{equation*}%
We deduce that:
\begin{equation}
\left( x-c\right) -\left( v-c\right) -2\lambda \left( x-c\right) =\mathbf{0}%
_{n}  \label{eq:proximal-comp-ball1}
\end{equation}%
and:%
\begin{equation*}
\left( x-c\right) ^{\top }\left( x-c\right) -\left( x-c\right) ^{\top
}\left( v-c\right) -2\lambda \left( x-c\right) ^{\top }\left( x-c\right) =0
\end{equation*}%
It follows that $r^{2}-\left( x-c\right) ^{\top }\left( v-c\right) -2\lambda
r^{2}=0$, meaning that:%
\begin{equation*}
\lambda ^{\star }=\frac{r^{2}-\left( x-c\right) ^{\top }\left( v-c\right) }{2r^{2}}
\end{equation*}%
We notice that:%
\begin{eqnarray*}
\left( \ref{eq:proximal-comp-ball1}\right)  &\Leftrightarrow &\left(
x-c\right) -\left( v-c\right) -2\frac{r^{2}-\left( x-c\right) ^{\top }\left(
v-c\right) }{2r^{2}}\left( x-c\right) =\mathbf{0}_{n} \\
&\Leftrightarrow &-r^{2}\left( v-c\right) +\left( x-c\right) ^{\top }\left(
v-c\right) \left( x-c\right) =\mathbf{0}_{n} \\
&\Leftrightarrow &\left( x-c\right) ^{\top }\left( v-c\right) \left(
x-c\right) = r^{2}\left( v-c\right)
\end{eqnarray*}%
Because $\left( x-c\right) ^{\top }\left( v-c\right) $ is a scalar, we deduce
that $x-c$ and $v-c$ are two collinear vectors:
\begin{equation*}
x-c=r\frac{\left( v-c\right) }{\left\Vert v-c\right\Vert _{2}}
\end{equation*}%
The optimal solution is:%
\begin{equation*}
x^{\star }=c+r\frac{\left( v-c\right) }{\left\Vert v-c\right\Vert _{2}}
\end{equation*}
\end{enumerate}
Combining the two cases gives:
\begin{equation*}
\mathbf{prox}_{\mathds{1}_{\Omega }\left( x\right) }\left( v\right) =c+\frac{r}{\min \left(
r,\left\Vert v-c\right\Vert _{2}\right) }\left( v-c\right)
\end{equation*}%
This is the formula of the projection onto the $\boldsymbol{\ell }_{2}$ ball,
but the minimum function has replaced the maximum function\footnote{ See
Equation (\ref{eq:proximal-ball2}) on page \pageref{eq:proximal-ball2}.}.

\subsubsection{Projection onto the complement $\bar{\mathcal{B}}_{1}\left( c,r\right) $
of the $\boldsymbol{\ell }_{1}$ ball}
\label{appendix:proximal-l1-ball-complement}

This proximal problem associated with $\bar{\mathcal{B}}_{1}\left(
\mathbf{0}_{n},r\right) $ is:
\begin{eqnarray*}
x^{\star } &=&\arg \min_{x}\frac{1}{2}\left( x-v\right) ^{\top }\left(
x-v\right)  \\
&\text{s.t.}&\left\Vert x\right\Vert _{1}\geq r
\end{eqnarray*}%
We deduce that the Lagrange function is equal to:
\begin{equation*}
\mathcal{L}\left( x;\lambda \right) =\frac{1}{2}\left( x-v\right) ^{\top
}\left( x-v\right) -\lambda \left( \left\Vert x\right\Vert _{1}-r\right)
\end{equation*}%
The first-order condition is:%
\begin{equation*}
\frac{\partial \,\mathcal{L}\left( x;\lambda \right) }{\partial \,x}%
=x-v-\lambda \func{sign}\left( x\right) =\mathbf{0}_{n}
\end{equation*}%
whereas the KKT condition is $\min \left( \lambda ,\left\Vert x\right\Vert
_{1}-r\right) =0$. We distinguish two cases:

\begin{enumerate}
\item If $\lambda =0$, this means that $x^{\star }=v$ and $\left\Vert
    x\right\Vert _{1}\geq r$.

\item If $\lambda >0$, we have $\left\Vert x\right\Vert _{1}=r$. Then we
    obtain the following system of equations:
\begin{equation*}
\left\{
\begin{array}{l}
x-v=\lambda \func{sign}\left( x\right)  \\
\left\Vert x\right\Vert _{1}=r%
\end{array}%
\right.
\end{equation*}%
The first condition gives that $x-v$ is a vector whose elements are $%
+\lambda $ and/or $-\lambda $, whereas the second condition shows that $x$ is
on the surface of the $\boldsymbol{\ell }_{1}$ ball. Unfortunately, there is
no unique solution. This is why we assume that $\func{sign}\left(
x\right) =\func{sign}\left( v\right) $ and we modify the sign function: $%
\func{sign}\left( a\right) =1$ if $a\geq 0$ and $\func{sign}\left( a\right)
=-1$ if $a<0$. In this case, there is a unique solution $x^{\star }=v+\lambda
^{\star }\func{sign}\left( v\right) $ where $\lambda ^{\star }=n^{-1}\left(
r-\left\Vert v\right\Vert _{1}\right) $ because $\left\vert v+\lambda
\func{sign}\left( v\right) \right\vert =\left\vert v\right\vert +\lambda $.
\end{enumerate}

\noindent Combining the two cases implies that:
\begin{equation*}
\mathcal{P}_{\mathcal{\bar{B}}_{1}\left( \mathbf{0}_{n},r\right) }\left(
v\right) =\left\{
\begin{array}{ll}
v & \text{if }\left\Vert v\right\Vert _{1}\geq r \\
v+\func{sign}\left( v\right) \odot n^{-1}\left( r-\left\Vert v\right\Vert
_{1}\right)  & \text{if }\left\Vert v\right\Vert _{1}<r%
\end{array}%
\right.
\end{equation*}%
Using the translation property, we deduce that:%
\begin{equation*}
\mathcal{P}_{\mathcal{\bar{B}}_{1}\left( c,r\right) }\left( v\right) =v+%
\func{sign}\left( v-c\right) \odot \frac{\max \left( r-\left\Vert
v-c\right\Vert _{1},0\right) }{n}
\end{equation*}

\subsubsection{The bid-ask linear cost function}
\label{appendix:linear-cost}

If we consider the scalar function:%
\begin{equation*}
f\left( x\right) =\alpha \left( \gamma -x\right) _{+}+\beta \left( x-\gamma
\right) _{+}
\end{equation*}%
we have:%
\begin{eqnarray*}
f_{v}\left( x\right)  &=&\lambda f\left( x\right) +\frac{1}{2}\left\Vert
x-v\right\Vert _{2}^{2} \\
&=&\lambda \alpha \left( \gamma -x\right) _{+}+\lambda \beta \left( x-\gamma
\right) _{+}+\frac{1}{2}x^{2}-xv+\frac{1}{2}v^{2}
\end{eqnarray*}%
Following Beck (2017), we distinguish three cases:
\begin{enumerate}
\item If $f_{v}\left( x\right) =\lambda \alpha \left( \gamma -x\right) +%
\frac{1}{2}x^{2}-xv+\frac{1}{2}v^{2}$, then $f_{v}^{\prime }\left( x\right)
=-\lambda \alpha +x-v$ and $x^{\star }=v+\lambda \alpha $. This implies that
$\gamma -x^{\star }>0$ or $v<\gamma -\lambda \alpha $.

\item If $f_{v}\left( x\right) =\lambda \beta \left( x-\gamma \right) _{+}+%
\frac{1}{2}x^{2}-xv+\frac{1}{2}v^{2}$, then $f_{v}^{\prime }\left( x\right)
=\lambda \beta +x-v$ and $x^{\star }=v-\lambda \beta $. This implies that $%
x^{\star }-\gamma >0$ or $v<\gamma +\lambda \beta $.

\item If $v\in \left[ \gamma -\lambda \alpha ,\gamma +\lambda \beta \right] $%
, the minimum is not obtained at a point of differentiability. Since $\gamma
$ is the only point of non-differentiability, we obtain $x^{\star }=\gamma $.
\end{enumerate}
Therefore, we can write the proximal operator in the following compact form:
\begin{equation*}
x^{\star }=\gamma +\left( v-\gamma -\lambda \beta \right) _{+} - \left(
v-\gamma +\lambda \alpha \right) _{-}
\end{equation*}%
where $x_{-}$ and $x_{+}$ are the negative part and the positive part of $x$.
If we consider the vector-value function $f\left( x\right)
=\sum_{i=1}^{n}\alpha _{i}\left( \gamma _{i}-x_{i}\right) _{+}+\beta
_{i}\left( x_{i}-\gamma _{i}\right) _{+}$, we deduce that:
\begin{equation*}
\mathbf{prox}_{\lambda f\left( x\right) }\left( v\right) =\gamma +\mathcal{S}%
\left( v-\gamma ;\lambda \alpha ,\lambda \beta \right)
\end{equation*}%
where $\mathcal{S}\left( v;\lambda _{-},\lambda _{+}\right) =
\left( v-\lambda_{+}\right) _{+} - \left( v+\lambda
_{-}\right) _{-}$ is the two-sided soft-thresholding operator.

\subsection{The QP form of the ADMM-QP problem}
\label{appendix:admm-qp}

We have:%
\begin{eqnarray*}
f_{\mathrm{QP}}\left( x\right)  &=&f_{\mathrm{MVO}}\left( x\right)
+f_{\boldsymbol{\ell }_{2}}\left( x\right)  \\
&=&\frac{1}{2}\left( x-b\right) ^{\top }\Sigma _{t}\left( x-b\right) -\gamma
\left( x-b\right) ^{\top }\mu _{t}+\frac{1}{2}\varrho _{2}\left\Vert \Gamma
_{2}\left( x-x_{t}\right) \right\Vert _{2}^{2}+\frac{1}{2}\tilde{\varrho}%
_{2}\left\Vert \tilde{\Gamma}_{2}\left( x-\tilde{x}\right) \right\Vert
_{2}^{2} \\
&=&\frac{1}{2}x^{\top }\Sigma _{t}x-x^{\top }\Sigma _{t}b+\frac{1}{2}b^{\top
}\Sigma _{t}b-\gamma x^{\top }\mu _{t}+\gamma b^{\top }\mu _{t}+ \\
&&\frac{1}{2}x^{\top }\left( \varrho _{2}\Gamma _{2}^{\top }\Gamma
_{2}\right) x-x^{\top }\left( \varrho _{2}\Gamma _{2}^{\top }\Gamma
_{2}\right) x_{t}+\frac{1}{2}x_{t}^{\top }\left( \varrho _{2}\Gamma
_{2}^{\top }\Gamma _{2}\right) x_{t}+ \\
&&\frac{1}{2}x^{\top }\left( \tilde{\varrho}_{2}\tilde{\Gamma}_{2}^{\top }%
\tilde{\Gamma}_{2}\right) x-x^{\top }\left( \tilde{\varrho}_{2}\tilde{\Gamma}%
_{2}^{\top }\tilde{\Gamma}_{2}\right) \tilde{x}+\frac{1}{2}\tilde{x}\left(
\tilde{\varrho}_{2}\tilde{\Gamma}_{2}^{\top }\tilde{\Gamma}_{2}\right)
\tilde{x} \\
&=&\frac{1}{2}x^{\top }\left( \Sigma _{t}+\varrho _{2}\Gamma _{2}^{\top
}\Gamma _{2}+\tilde{\varrho}_{2}\tilde{\Gamma}_{2}^{\top }\tilde{\Gamma}%
_{2}\right) x-x^{\top }\left( \gamma \mu _{t}+\Sigma _{t}b+\varrho
_{2}\Gamma _{2}^{\top }\Gamma _{2}x_{t}+\tilde{\varrho}_{2}\tilde{\Gamma}%
_{2}^{\top }\tilde{\Gamma}_{2}\tilde{x}\right) + \\
&&\gamma b^{\top }\mu _{t}+\frac{1}{2}\left( b^{\top }\Sigma _{t}b+\varrho
_{2}x_{t}^{\top }\Gamma _{2}^{\top }\Gamma _{2}x_{t}+\tilde{\varrho}_{2}%
\tilde{x}\tilde{\varrho}_{2}\tilde{\Gamma}_{2}^{\top }\tilde{\Gamma}_{2}%
\tilde{x}\right)
\end{eqnarray*}

\subsection{The CCD algorithm of a QP form with a logarithmic barrier}
\label{appendix:admm-ccd}

We consider the following optimization problem:%
\begin{equation*}
x^{\star }=\arg \min_{x}\frac{1}{2}x^{\top }Qx-x^{\top
}R-\sum_{i=1}^{n}\lambda _{i}\ln x_{i}
\end{equation*}%
where $Q$ is a positive-definite matrix and $\lambda _{i}>0$. The
first-order condition with respect to coordinate $x_{i}$ is:%
\begin{equation*}
\left( Qx\right) _{i}-R_{i}-\frac{\lambda _{i}}{x_{i}}=0
\end{equation*}%
It follows that $x_{i}\left( Qx\right) _{i}-R_{i}x_{i}-\lambda _{i}=0$ or
equivalently:%
\begin{equation*}
Q_{i,i}x_{i}^{2}+\left( \sum_{j\neq i}x_{j}Q_{i,j}-R_{i}\right)
x_{i}-\lambda _{i}=0
\end{equation*}%
The polynomial function is convex because we have $Q_{i,i}>0$. Since the
product of the roots is negative\footnote{%
We have $-Q_{i,i}\lambda _{i}<0$.}, we have two solutions with opposite signs.
We deduce that the solution is the positive root of the second-degree
equation:%
\begin{equation*}
x_{i}^{\star }=\frac{R_{i}-\sum_{j\neq i}x_{j}Q_{i,j}+\sqrt{\left(
\sum_{j\neq i}x_{j}Q_{i,j}-R_{i}\right) ^{2}+4\lambda _{i}Q_{i,i}}}{2Q_{i,i}}
\end{equation*}%
It follows that CCD algorithm is:%
\begin{eqnarray*}
x_{i}^{\left( k+1\right) } &=&\frac{R_{i}-\sum_{j<i}x_{j}^{\left( k+1\right)
}Q_{i,j}-\sum_{j>i}x_{j}^{\left( k\right) }Q_{i,j}}{2Q_{i,i}}+ \\
&&\frac{\sqrt{\left( \sum_{j<i}x_{j}^{\left( k+1\right)
}Q_{i,j}+\sum_{j>i}x_{j}^{\left( k\right) }Q_{i,j}-R_{i}\right)
^{2}+4\lambda _{i}Q_{i,i}}}{2Q_{i,i}}
\end{eqnarray*}


\section{Data}

\subparagraph{Parameter set \#1}
\label{appendix:data1}

We consider a capitalization-weighted stock index, which is composed of eight
stocks. The weights of this benchmark are equal to $23\%$, $19\%$, $17\%$,
$9\%$, $8\%$, $6\%$ and $5\%$. We assume that their volatilities are $21\%$,
$20\%$, $40\%$, $18\%$, $35\%$, $23\%$, $7\%$ and $29\%$. The correlation
matrix is defined as follows:%
\begin{equation*}
\rho =\left(
\begin{array}{rrrrrrrr}
100\% &       &       &       &       &       &       &        \\
 80\% & 100\% &       &       &       &       &       &        \\
 70\% &  75\% & 100\% &       &       &       &       &        \\
 60\% &  65\% &  90\% & 100\% &       &       &       &        \\
 70\% &  50\% &  70\% &  85\% & 100\% &       &       &        \\
 50\% &  60\% &  70\% &  80\% &  60\% & 100\% &       &        \\
 70\% &  50\% &  70\% &  75\% &  80\% &  50\% & 100\% &        \\
 60\% &  65\% &  70\% &  75\% &  65\% &  70\% &  80\% & 100\%%
\end{array}%
\right)
\end{equation*}

\subparagraph{Parameter set \#2}
\label{appendix:data2}

We consider a universe of eight stocks. We assume that their volatilities are
$25\%$, $20\%$, $15\%$, $18\%$, $30\%$, $20\%$, $15\%$ and $35\%$. The
correlation
matrix is defined as follows:%
\begin{equation*}
\rho =\left(
\begin{array}{rrrrrrrr}
100\% &       &       &       &       &       &       &         \\
 20\% & 100\% &       &       &       &       &       &         \\
 55\% &  60\% & 100\% &       &       &       &       &         \\
 60\% &  60\% &  60\% & 100\% &       &       &       &         \\
 60\% &  60\% &  60\% &  60\% & 100\% &       &       &         \\
 60\% &  60\% &  60\% &  60\% &  60\% & 100\% &       &         \\
 60\% &  60\% &  60\% &  60\% &  60\% &  60\% & 100\% &         \\
 60\% &  60\% &  60\% &  60\% &  60\% &  60\% &  60\% & 100\%
\end{array}%
\right)
\end{equation*}

\clearpage

\section{Notations}
\label{appendix:notations}

\begin{itemize}
\item $\mu =\left( \mu _{1},\ldots ,\mu _{n}\right) $ is the vector of
    expected return.

\item $\Sigma =\left[ \rho _{i,j}\sigma _{i}\sigma _{j}\right]
    _{i,j=1}^{i,j=1}$ is the covariance matrix where $\sigma _{i}$ is the
    volatility of Asset $i$ and $\rho _{i,j}$ is the correlation between
    Asset $i $ and Asset $j$.

\item $b$ is the vector of benchmark weights.

\item $\mu \left( x\right) =x^{\top }\mu $ is the expected return of
    Portfolio $x$.

\item $\sigma \left( x\right) =\sqrt{x^{\top }\Sigma x}$ is the volatility of
    Portfolio $x$.

\item $\mu \left( x\mid b\right) =\left( x-b\right) ^{\top }\mu $ is the
    expected excess return of Portfolio $x$ with respect to Benchmark $b$.

\item $\sigma \left( x\mid b\right) =\sqrt{\left( x-b\right) ^{\top }\Sigma
    \left( x-b\right) }$ is the tracking error volatility of Portfolio $x$
    with respect to Benchmark $b$.

\item $\mathcal{R}\left( x\right) $ is a convex risk measure.

\item $\mathcal{RB=}\left( \mathcal{RB}_{1},\ldots ,\mathcal{RB}_{n}\right) $
    is the vector of risk budgets.

\item $\mathcal{RC}_{i}\left( x\right) $ is the risk contribution of Asset $i
    $ with respect to Portfolio $x$.

\item $\turnover\left( x\mid \tilde{x}\right) =\sum_{i=1}^{n}\left\vert
    x_{i}-\tilde{x}_{i}\right\vert $ is the turnover between Portfolios $x$
    and $\tilde{x}$. The maximum acceptable turnover is denoted by
    $\turnover^{+}$.

\item $\cost\left( x\mid \tilde{x}\right) $ is the cost function when
    rebalancing Portfolio $x$ from Portfolio $\tilde{x}$. The maximum
    acceptable cost is denoted by $\cost^{+}$.

\item $\mathcal{AS}\left( x\mid b\right) =\frac{1}{2}\sum_{i=1}^{n}\left
    \vert x_{i}-b_{i}\right\vert $ is the active share of Portfolio $x$ with
    respect to Benchmark $b$. $\mathcal{AS}^{-}$ is the minimum acceptable
    active share.

\item $\mathcal{H}\left( x\right) =\sum_{i=1}^{n}x_{i}^{2}$ is the Herfindahl
    index.

\item $\mathcal{N}\left( x\right) =1/\mathcal{H}\left( x\right) $ is the
    number of effective bets. $\mathcal{N}^{-}$ corresponds to the minimum
    acceptable number of effective bets.

\item $\mathcal{DR}\left( x\right) =\left( x^{\top }\sigma \right) /\sqrt{
    x^{\top }\Sigma x}$ is the diversification ratio of Portfolio $x$.

\item $\mathcal{LS}\left( x\right) =\left\vert \sum_{i=1}^{n}x_{i}\right\vert
    $ is the long/short exposure of Portfolio $x$.

\item $\mathcal{L}\left( x\right) =\sum_{i=1}^{n}\left\vert x_{i}\right\vert
    $ is the leverage of Portfolio $x$.

\item $\limfunc{SE}\left( x\right) =-\sum_{i=1}^{n}x_{i}\ln x_{i}$ is
    Shannon's entropy of $x$.

\item $\limfunc{KL}\left( x\right) =\sum_{i=1}^{n}x_{i}\ln \left( x_{i}/
    \tilde{x}_{i}\right) $ is the Kullback-Leibler divergence between $x$ and
    $ \tilde{x}$.

\item $W\left( x\right) $ is the Lambert $W$ function satisfying $W\left(
    x\right) e^{W\left( x\right) }=x$.

\item $\mathbf{0}_{n}$ is the vector of zeros.

\item $\mathbf{1}_{n}$ is the vector of ones.

\item $e_{i}$ is the unit vector, i.e. $\left[ e_{i}\right] _{i}=1$ and $
    \left[ e_{i}\right] _{j}=0$ for all $j\neq i$.

\item $x_{-} = \max \left( -x,0\right) =-\min \left( x,0\right) $ is the
    negative part of $x$.

\item $x_{+} = \max \left( x,0\right) $ is the positive part of $x$.

\item $\mathds{1}_{\Omega }\left( x\right) $ is the convex indicator function
    of $\Omega $: $\mathds{1}_{\Omega }\left( x\right) =0$ for $x\in \Omega $
    and $\mathds{1}_{\Omega }\left( x\right) =+\infty $ for $x\notin \Omega $.

\item $A^{\dagger }$ is the Moore-Penrose pseudo-inverse matrix of $A$.

\item $\left\Vert x\right\Vert _{p}=\left( \sum_{i=1}^{n}\left\vert
    x_{i}\right\vert ^{p}\right) ^{1/p}$ is the $\boldsymbol{\ell }_{p}$
    norm.

\item $\left\Vert x\right\Vert _{A}=\left( x^{\top }Ax\right) ^{1/2}$ is the
    weighted $\boldsymbol{\ell }_{2}$ norm.

\item $x\odot y$ is the Hadamard element-wise product: $\left[ x\odot y
    \right] _{i,j}=\left[ x\right] _{i,j}\left[ y\right] _{i,j}$.

\item $\mathbf{prox}_{f}\left( v\right) $ is the proximal operator of $
    f\left( x\right) $: $\mathbf{prox}_{f}\left( v\right) =\arg
    \min\nolimits_{x}\left\{ f\left( x\right) +\frac{1}{2}\left\Vert
    x-v\right\Vert _{2}^{2}\right\} $.

\item $\mathcal{S}\left( v;\lambda \right) =\limfunc{sign}\left( v\right)
    \cdot \left( \left\vert v\right\vert -\lambda \right) _{+}$ is the
    soft-thresholding operator.

\item $\mathcal{S}\left( v;\lambda _{-},\lambda _{+}\right) = \left(
    v-\lambda_{+}\right) _{+} - \left( v+\lambda _{-}\right) _{-}$ is the
    two-sided soft-thresholding operator. We have the following property:
    $\mathcal{S}\left( v;\lambda \right) = \mathcal{S}\left(
    v;\lambda,\lambda\right)$.

\item $\mathcal{T}\left( v,x^{-},x^{+}\right) =\max \left( x^{-},\min \left(
    x,x^{+}\right) \right) $ is the truncation operator.

\item $\mathcal{P}_{\Omega }\left( v\right) $ is the projection of $v$ onto the
    set $\Omega $: $\mathcal{P}_{\Omega }\left( v\right) =\arg
    \min\nolimits_{x\in \Omega }\frac{1}{2}\left\Vert
    x-v\right\Vert_{2}^{2}=\mathbf{prox}_{\mathds{1}_{\Omega }\left( x\right)
    }\left( v\right)$.

\item $\mathbb{S}_{n}$ is the unit simplex with dimension $n$.

\item $\mathcal{A}_{ffineset}\left[ A,B\right] $ is the affine set $\left\{
    x\in \mathbb{R}^{n}:Ax=B\right\} $.

\item $\mathcal{H}_{yperplane}\left[ a,b\right] $ is the hyperplane $\left\{
    x\in \mathbb{R}^{n}:a^{\top }x=b\right\} $.

\item $\mathcal{H}_{alfspace}\left[ c,d\right] $ is the half-space $\left\{
    x\in \mathbb{R}^{n}:c^{\top }x\leq d\right\} $.

\item $\mathcal{B}_{ox}\left[ x^{-},x^{+}\right] $ is the box $\left\{ x\in
    \mathbb{R}^{n}:x^{-}\leq x\leq x^{+}\right\} $.

\item $\mathcal{B}_{p}\left( c,r\right) $ is the $\boldsymbol{\ell
    }_{p}$-ball $\left\{ x\in \mathbb{R}^{n}:\left\Vert x-c\right\Vert
    _{p}\leq r\right\} $.

\item $\mathfrak{D}$ is the weight diversification set $\left\{ x\in
    \mathbb{R}^{n}:\mathcal{D}\left( x\right) \geq \mathcal{D}^{-}\right\} $
    where $\mathcal{D}\left( x\right) $ is the diversification measure and
    $\mathcal{D}^{-}$ is the minimum acceptable diversification.
\end{itemize}



\end{document}